\pdfoutput=1
\documentclass[cernpreprint, atlasdraft=false, USenglish, texmf, orcidlogo]{atlasdoc}
\usepackage{atlaspackage}
\usepackage{atlasbiblatex}

\usepackage{atlasphysics}
\graphicspath{{logos/}{figures/}}

\usepackage{ANA-HION-2023-04-PAPER-defs}

\AtlasTitle{Azimuthal anisotropies of charged particles with high transverse
momentum in \mbox{Pb+Pb} collisions
at $\sqrt{s_{_\text{NN}}}=5.02$~\ensuremath{\text{T\text{e\kern-0.1em V}}}\xspace with the ATLAS detector}

\AtlasAbstract{%
A measurement is presented of elliptic ($v_2$) and triangular ($v_3$) azimuthal anisotropy coefficients for charged particles produced in \mbox{Pb+Pb} collisions at $\sqrt{s_{_\text{NN}}} = 5.02$~\ensuremath{\text{T\text{e\kern-0.1em V}}}\xspace
using a dataset corresponding to an integrated luminosity of $0.44$~\mbox{$\text{nb}^{-1}$}\xspace collected with the ATLAS detector at the LHC in 2018.
The values of $v_2$ and $v_3$ are measured
for charged particles over a wide range of transverse momentum ($\ensuremath{p_{\text{T}}}$), 1--400~\ensuremath{\text{G\text{e\kern-0.1em V}}}\xspace,
and \mbox{Pb+Pb} collision centrality, 0--60\%, using the scalar-product and multiparticle cumulant methods.
These methods are sensitive to event-by-event fluctuations and nonflow effects in the measurements of azimuthal anisotropies.
Positive values of $v_2$ are observed up to a \ensuremath{p_{\text{T}}}\xspace\ of approximately 100~\ensuremath{\text{G\text{e\kern-0.1em V}}}\xspace from both methods across all centrality intervals.
Positive values of $v_3$ are observed up to approximately 25~\ensuremath{\text{G\text{e\kern-0.1em V}}}\xspace using both methods,
though the application of the three-subevent technique to the multiparticle cumulant method leads to significant changes at the highest \ensuremath{p_{\text{T}}}\xspace.
At high \ensuremath{p_{\text{T}}}\xspace\ ($\ensuremath{p_{\text{T}}}\xspace \gtrapprox 10$~\ensuremath{\text{G\text{e\kern-0.1em V}}}\xspace), charged particles are dominantly from jet fragmentation.
These jets, and hence the measurements presented here, are sensitive to the path-length dependence of parton energy loss in the quark-gluon plasma produced in \mbox{Pb+Pb} collisions.
}

\AtlasRefCode{HION-2023-04}

\PreprintIdNumber{CERN-EP-2024-334}

\AtlasJournal{Phys.\ Rev.\ C.}
\AtlasJournalRef{\PRC 112 (2025) 024910}
\AtlasDOI{10.1103/d46f-yl4n}

\AtlasCoverEgroupAnalysisTeam{atlas-hion-2023-04-analysis-team@cern.ch}

\hypersetup{pdftitle={ATLAS document},pdfauthor={The ATLAS Collaboration}}

\begin{document}

\maketitle

\section{Introduction}
%

% The next lines are included from the .//ANA-HION-2023-04-PAPER-intro.tex input file
The primary aim of the heavy-ion program at the Large Hadron Collider (LHC) is to produce
and study the quark-gluon plasma (QGP), the high-temperature state of
matter in which quarks and gluons are no longer confined within protons and neutrons (for a recent review, see
Ref.~\cite{Busza:2018rrf}).
Measurements of jets originating from hard parton scatterings in the early stages of
heavy-ion collisions provide information about the short-distance-scale
interactions of high-energy partons with the QGP (for a recent review, see Ref.~\cite{Cunqueiro:2021wls}).
The overall rate of jets at a given transverse momentum, \pt,
is found to be reduced by approximately a factor of two
in central \pbpb\ collisions compared to \pp\ collisions scaled to account for the increased partonic luminosity in \pbpb\
collisions~\cite{ATLAS:2018gwx, CMS:2021vui, ATLAS:2023hso, ALICE:2019qyj}.
This suppression can be explained by the energy loss of partons propagating through the QGP, %
and the magnitude of this energy loss depends on the amount of QGP that
the parton travels through.

The geometry of the overlap of the two nuclei
leads to a shorter average path length if the jet is oriented along the direction of the collision impact
parameter\footnote{The impact parameter is defined as the distance between the centers of the two colliding nuclei.} than if the jet is oriented in the perpendicular direction.
This is expected to
lead to a dependence of the jet yield on the azimuthal angle~\cite{Wang:2000fq,Gyulassy:2000gk,Shuryak:2001me}.
Charged particles with \pt\ greater than 10~\GeV, hereinafter referred to as high-\pt\ charged particles,
are likely to come from jet fragmentation.
Therefore, the measurements of azimuthal anisotropies of high-\pt\ charged particles are useful
for investigating the path-length dependence of jet energy loss.
An azimuthal modulation of the same direction is also present in the low-\pt\ charged particles due to the hydrodynamic flow of the QGP.
The amplitudes of these anisotropies can be used to constrain the bulk properties of the QGP (for a review, see Ref.~\cite{Heinz:2013th}).

The azimuthal anisotropies are quantified via the values of Fourier coefficients describing the
azimuthal angular distribution of charged particles with respect to the
event planes \cite{Poskanzer:1998yz}:
\begin{equation}\nonumber
\frac{{d}N}{{d}\phi} \propto 1 + 2 \sum_{n=1}^\infty \vn{n} \cos(n(\phi - \Psi_n)),
\end{equation}
where $n$ is the order of harmonics, \Psin\ is the \nth\ order event plane angle, and $\phi$ is the azimuthal angle of charged particles.
The order of harmonics in \vn{n} corresponds to the order of eccentricities in the initial geometry of the QGP, such as the ellipticity for \vn{2} and the triangularity for \vn{3}.
Measurements of \vn{n} also include contributions from nonflow effects,
defined as the correlations unrelated to the initial geometry of the QGP,
such as resonance decays, global momentum conservation, jet fragmentation, and dijet production.
Therefore,
various methods have been developed
to suppress nonflow effects and applied in \vn{n} measurements at the LHC ~\cite{CMS:2017xgk,ATLAS:2021ktw,ATLAS:2013ssy,ATLAS:2018ezv, ALICE:2019zfl}
and the Relativistic Heavy Ion Collider (RHIC) ~\cite{STAR:2002hbo, PHENIX:2018lfu}.
The \textit{scalar-product} (SP) method~\cite{STAR:2002hbo,Voloshin:2008dg,Luzum:2012da} provides an estimate of $\sqrt{\langle\vn{n}^2\rangle}$ that is independent of the detector resolution.
In addition, the nonflow contributions are mitigated if a pseudorapidity gap is imposed between the correlated particles.
Alternatively, lower-order short-distance correlations from particle decays can be effectively suppressed by utilizing genuine multiparticle correlations within the \textit{multiparticle cumulant} (MPC) framework ~\cite{Borghini:2001vi, Borghini:2000sa}, which can be efficiently implemented using the so-called $Q$ cumulants \cite{Bilandzic:2010jr, Bilandzic:2013kga}.
Following the convention from this framework, in this paper, correlations, cumulants, and \vn{n} measurements that are integrated in \pt\ are termed \textit{reference},
in contrast to the \textit{differential} quantities that are differential in \pt.
The three-subevent $Q$-cumulant method extends the standard method
by requiring a pseudorapidity gap between some of the correlated particles.
This has been shown to reduce further the short-range nonflow effects in measurements of reference \vn{n} \cite{Jia:2017hbm, Huo:2017nms, ATLAS:2019peb}.
The measurements of MPCs used to obtain \vn{n} are conducted over an ensemble of events of similar centralities.
It has been observed that the results obtained with the MPC method are also sensitive to the centrality resolution of the event ensemble chosen.
Therefore, different strategies of constructing the \vn{n} values in a centrality interval have been investigated~\cite{ATLAS:2019peb}.

Measurements of the azimuthal anisotropies of jets~\cite{ATLAS:2013ssy,ALICE:2015efi, ATLAS:2021ktw}
and charged particles at high-\pt~\cite{ATLAS:2014qxy,CMS:2017xgk,ATLAS:2018ezv} in \pbpb\ collisions have
been previously performed.
For jets with $\pT > 70$~\GeV, positive values of \vn{2} are measured for all centrality intervals, except the most central,
and positive values of \vn{3} are measured for mid-central collisions \cite{ATLAS:2021ktw}.
Measurements of \vn{n}\ using the SP method and the measurements of \vn{2} using the MPC method have also been performed for charged particles with a \pT\ up to 100~\GeV at the LHC~\cite{CMS:2017xgk,ATLAS:2014qxy}.
It has been observed that \vn{2} values measured with both the SP and the MPC method decrease with \pT\ for charged particles with a \pt\ greater than 20~\GeV\
while remaining positive up to a \pT\ of approximately $80$~\GeV\ for the 60\% most central collisions.
Meanwhile, the values of \vn{3} measured with the SP method remain positive up to a \pT\ of approximately $20$~\GeV\ for the 40\% most central collisions \cite{CMS:2017xgk}.
The positive values of \vn{n} observed in the high-\pt\ sector suggest that the energy loss of hard-scattered partons is affected by the event-by-event initial geometry of the QGP.

This analysis extends the measurement of \vn{n}\ values using the SP and MPC methods to higher \pt\ for charged particles in $\sqrtsnn=5.02~\TeV$ \pbpb\ collisions.
The dataset used corresponds to an integrated luminosity of \PEBlum\ collected by the ATLAS detector using the innovative partial event building technique.
Measurements of \vn{n} were performed in intervals of \pt\ up to 400~\GeV\ and intervals of centrality spanning the 0--60\% most central collisions, and compared between different acceptance ranges and strategies.
Measurements of \vn{n} with the SP method, denoted by \vnsp{n}, were carried out for tracks in three pseudorapidity ranges\footnote{ATLAS uses a right-handed
coordinate system with its origin at the nominal IP in the center of the detector,
and the $z$-axis along the beam pipe.
The $x$-axis points from the IP to the center of the LHC ring, and the $y$-axis points upward.
Cylindrical coordinates $(r,\phi)$ are used in the transverse plane,
$\phi$ being the azimuthal angle around the $z$-axis.
The pseudorapidity is defined in terms of the polar angle $\theta$ as $\eta=-\ln\tan(\theta/2)$.
The rapidity is defined as $y = 0.5\text{ln}[(E + p_z)/(E-p_z)]$
where $E$ and $p_z$ are the energy and $z$-component of the momentum along the beam direction respectively.
Transverse momentum and transverse energy are defined as
$\pt = p \sin\theta$ and $\ET = E \sin\theta$, respectively.} $|\eta| < 1.1$, $1.1 < |\eta| < 2.5$ and $|\eta| < 2.5$.
And measurements of \vn{n} using the MPC method were performed for four-particle cumulants, denoted by \vnm{n}{4}, for charged particles over the pseudorapidity range of $|\eta|<2.5$.
Furthermore, the three-subevent $Q$-cumulant method used for reference \vn{n} measurements is expanded to include \pt-differential \vnm{2}{4} and
\vnm{3}{4}.
For the MPC method, the cumulants are firstly computed within narrower centrality intervals before combining into wider ones.
The differences in the MPC measurements by using different centrality intervals before combining are shown.
Finally, results from the SP method and the MPC method are compared with each other.

The paper is organized as follows. Section~\ref{sec:detector} describes the detector, trigger and datasets, and Section~\ref{sec:mc_data} the event and track selections, as well as event combination procedures. Section~\ref{sec:analysis} provides the mathematical framework for the scalar-product and multiparticle cumulant methods. The systematic uncertainties are described in Sections~\ref{sec:systematics}. Section~\ref{sec:result} presents the \vn{n} measurements using the SP and MPC methods. Comparisons of these measurements with existing measurements and between using different strategies and kinematic ranges are also presented. Finally, the conclusions are included in Section~\ref{sec:conclusion}.
%
%
%

% End of text imported from the .//ANA-HION-2023-04-PAPER-intro.tex input file

%
\section{ATLAS detector}
\label{sec:detector}
%

% The next lines are included from the .//ANA-HION-2023-04-PAPER-detector.tex input file
The ATLAS detector~\cite{PERF-2007-01} at the LHC covers nearly the entire solid angle around the collision point.
It consists of an inner tracking detector surrounded by a thin superconducting solenoid, electromagnetic and hadronic calorimeters,
and a muon spectrometer incorporating three large superconducting toroidal magnets.

The inner-detector system (ID) is immersed in a \SI{2}{\tesla} axial magnetic field
and provides charged-particle tracking in the pseudorapidity range \(|\eta| < 2.5\).
The high-granularity silicon pixel detector covers the vertex region,
and is composed of four layers including
the insertable B-layer~\cite{ATLAS-TDR-19,PIX-2018-001}.
It is followed by the silicon microstrip tracker (SCT), which usually provides eight measurements per track.
These silicon detectors are complemented by the transition radiation tracker (TRT),
which enables radially extended track reconstruction up to \(|\eta| = 2.0\).
The TRT also provides electron identification information based on
the fraction of hits (typically 30 in total) above a higher energy-deposit threshold corresponding to transition radiation.

The calorimeter system covers the pseudorapidity range \(|\eta| < 4.9\).
Within the region \(|\eta|< 3.2\), electromagnetic calorimetry is provided by barrel and
endcap high-granularity lead/liquid-argon (LAr) electromagnetic calorimeters,
with an additional thin LAr presampler covering \(|\eta| < 1.8\)
to correct for energy loss in material upstream of the calorimeters.
The hadronic calorimeters have three sampling layers longitudinal in
shower depth in \(|\eta| < 1.7\) and four sampling layers in \(1.5 < |\eta| < 3.2\),
with a slight overlap in \(\eta\).
The solid angle coverage is completed with forward copper/LAr and tungsten/LAr calorimeter modules (FCal)
optimized for electromagnetic and hadronic measurements respectively.
The FCals cover a pseudorapidity range of $3.2 < |\eta| < 4.9$.

The zero-degree calorimeters (ZDCs) are located symmetrically at $z = \pm 140$~m and cover $|\eta| > 8.3$ during the \pbpb\ data-taking period.
The ZDCs use tungsten plates as absorbers and quartz rods sandwiched between the tungsten plates as the active medium.
In Pb+Pb collisions the ZDCs primarily measure spectator neutrons.
A ZDC coincidence trigger is implemented by requiring the pulse height from each ZDC to be above a threshold set to accept the energy of a single neutron.
The luminosity is measured mainly by the LUCID--2~\cite{LUCID2}
detector that records Cherenkov light produced in the quartz windows of photomultipliers located close to the beampipe.

A two-level trigger system is used to select interesting events~\cite{TRIG-2016-01, ATLAS:2020esi}.
The first-level (L1) trigger is implemented in hardware and uses a subset of detector information,
including ZDC coincidences in \pbpb\ collisions,
to reduce the event rate to a design value of at most 100~kHz.
This is followed by a software-based high-level trigger (HLT) which reduces the event rate to several kHz.
A software suite~\cite{SOFT-2022-02} is used in the reconstruction and analysis of real
and simulated data, in detector operations, and in the trigger and data acquisition systems of the experiment.

% End of text imported from the .//ANA-HION-2023-04-PAPER-detector.tex input file

%
%
%
%

% The next lines are included from the .//ANA-HION-2023-04-PAPER-Monte-Carlo-and-Data-Selection.tex input file
\section{Monte Carlo simulation and data selection}
\label{sec:mc_data}
\subsection{Monte Carlo samples}

Monte Carlo (MC) simulations are used in this analysis to evaluate the track reconstruction performance.
A minimum-bias sample of \pbpb\ MC events at 5.02 TeV was generated with \hijing~\cite{Gyulassy:1994ew}.
After the generation,
the flow harmonics were added to the simulated events
using an “afterburner”~\cite{Poskanzer:1998yz} procedure, which implements the \pt, $\eta$, and centrality dependence of the \vn{n}, as measured
in the $\sqrtsnn = 2.76$~\TeV\ \pbpb\ data~\cite{ATLAS:2012at}, by artificially rearranging the $\phi$ positions of the
generated particles.
The detector response is simulated with \textsc{AtlFast-II}~\cite{SOFT-2010-01} for calorimeters and \textsc{Geant4}~\cite{Agostinelli:2002hh} for the ID.
The simulated events are then reconstructed using the same algorithms as data events.

\subsection{Event selection}
The dataset of \pbpb\ collisions at $\sqrtsnn=5.02$~\TeV\ used in this analysis was collected by the ATLAS detector in 2018 and corresponds to an integrated luminosity of \PEBlum\ after data-quality requirements~\cite{DAPR-2018-01}.
Events were required to satisfy one of the three L1 minimum-bias (MB) triggers.
The first MB trigger requires an event to have a total transverse energy (\ETcal) measured in the calorimeter system above 600 GeV.
The second MB trigger requires an event to have a \ETcal\ above 50 GeV and less than 600 GeV.
The third MB trigger will fire when an event is rejected by both of the previous two triggers and satisfies a ZDC coincidence trigger at L1 and have at least one track in the HLT.
The MB triggers above have a small prescale\footnote{A prescale is a number that defines what fraction of events are recorded out of all possible events that would have passed the trigger requirements.}
of approximately $3.95$.
The small prescale is enabled by a novel strategy known as the partial event building~\cite{TRIG-2016-01},
in which data from selected ATLAS subdetectors are used to compose an event, thus enabling higher recording rate and optimized event size.
These events contain the ID, FCal and ZDC information required for this analysis as in other events; other information has not been recorded.

For this analysis, events are required to have the $z$-coordinate of the primary vertex~\cite{ ATL-PHYS-PUB-2015-026} within 100~mm of the nominal interaction point.
Events with more than one hadronic interaction from the same bunch crossing are estimated to be less than 0.5\% of collisions.
These events are suppressed by utilizing the observed anti-correlation, expected from the nuclear geometry,
between the total transverse energy deposited in both of the forward calorimeters, \ETfcal,
and the energy in the ZDC, with the latter proportional to the number of observed spectator neutrons.
Additional hadronic interactions from just before or right after the collision
of interest could interfere with calorimeter performance and centrality determinations.
Criteria for rejecting these events are determined by the expected strong and linear correlation between charged particle multiplicity and measured \ETfcal.
Events with a charged particle multiplicity that falls out of this correlation are rejected.
A very stringent criteria is chosen to remove such events,
excluding approximately $15\%$ of events nearly independent of centrality.
This rejection is made necessary due to the focus of this analysis on measuring correlations of small magnitudes and a smaller bunch spacing used in the ATLAS 2018 \pbpb\ dataset.
Whereas the ATLAS 2015 dataset for \pbpb\ collisions at \sqrtsnn=5.02~\TeV\ used a bunch spacing interval of \SI{100}{ns}, this analysis uses the ATLAS 2018 dataset where a bunch spacing interval of \SI{75}{ns} was used for approximately half of the dataset for higher event rate.

The centrality of an event in this analysis is obtained using its \ETfcal\ energy, and the centrality percentile requirements are determined by using a MC Glauber analysis~\cite{Miller:2007ri, ATLAS:2011ag}, starting from the most central events,
where the collisions have the smallest impact parameter and the highest \ETfcal.
In this analysis, results are obtained in seven centrality intervals: 0--5\%, 5--10\%, 10--20\%, 20--30\%, 30--40\%, 40--50\%, and 50--60\%.

\subsection{Track selection}
\label{subsec:trk_sel}
Tracks of charged particles are reconstructed from hits in the ID using a track reconstruction algorithm
that was optimized for the high hit density in heavy-ion collisions~\cite{ATLAS:2017kyn}.
Tracks used in this analysis have $|\eta| < 2.5$ and are required to have at least 10 hits in the silicon detectors.
Additionally, a track must have no more than one hole in the SCT,
where a hole is defined as the absence of a hit predicted by the track trajectory.
All charged-particle tracks used in this analysis are required to have reconstructed transverse momentum $\pT > 1.0 $~\GeV.
In order to suppress the contribution from secondary particles,\footnote{Primary particles are defined as particles with a mean
lifetime $\tau>0.3\times 10^{-10}$ s either directly produced in the collisions or from subsequent decays of particles with a shorter lifetime.
All other particles are considered to be secondary.}
the distance of the closest approach of the track to the primary vertex is required to be less
than a \pt-dependent value varying from 0.6~mm at $\pT=1$~\GeV\ to 0.2~mm at
$\pT=20$~\GeV\ in the transverse plane and less than 1.0~mm in the longitudinal direction \cite{ATLAS:2018bvp}.
The transverse and longitudinal distances of the closest approach to the primary vertex are denoted by $d_0$ and $z_0$ respectively. The significances $|d_0|/\sigma_{d_0}$ and $|z_0 \sin{\theta}|/\sigma_{z_0\sin{\theta}}$ are required to be smaller than $3.0$, where $\theta$ is the polar angle of the track, and $\sigma_{d_0}$ and $\sigma_{z_0\sin{\theta}}$ are the uncertainties in $d_0$ and $z_0\sin{\theta}$, respectively.
The primary vertex is determined using vertex finding and fitting algorithms described in Ref.~\cite{ATLAS:2015jge}.

The track weight factor in this analysis is denoted by $\omega_j$ where $j$ is the track index, and is calculated from the track reconstruction inefficiency and the ID acceptance non-uniformity.
The track reconstruction efficiency $\epsilon$ for tracks that pass the
selections is evaluated as a function of
\pt, in intervals of centrality and absolute pseudorapidity \abseta.
The charged particle tracks are weighted by $1/\epsilon$ to correct for reconstruction inefficiency in this analysis.
In the most central collision interval 0--5\% for $\abseta < 1.1$, $\epsilon$ increases from approximately $63\%$ to
approximately $70\%$ in the \pt\ range of $1$--$20$~\GeV\
and reaches a plateau for $\pt>20$~\GeV.
For $1.1 < \abseta < 2.5$ in the same centrality interval,
$\epsilon$ increases from approximately $43\%$ to approximately $48\%$ in the \pt\ range of $1$--$20$~\GeV,
and reaches a plateau for $\pt>20$~\GeV.
The efficiency values in other centrality intervals follow the same trend and increase toward peripheral events nearly independent of \pt.
The difference in $\epsilon$ between the most peripheral centrality interval 50--60\% and
the most central centrality interval 0--5\% is approximately $4\%$ for $\abseta < 1.1$ and approximately $8\%$ for $1.1 < \abseta < 2.5$.
Reconstructed tracks that are not matched to a generated primary particle in the
MC samples are considered “fake” tracks.
The fake rate is negligible for tracks that pass the selection criteria for all centrality, \pt, and $\eta$ ranges used in this analysis,
and thus, it is not taken into account.
In addition, to correct for non-uniformity in the ID acceptance along azimuthal direction, the tracks are weighted by a factor proportional to $N_{\mathrm{trk}}(\eta)/N_{\mathrm{trk}}(\eta,\phi)$, where $N_{\mathrm{trk}}(\eta)$ is the total number of tracks within a given pseudorapidity interval of 0.1 and $N_{\mathrm{trk}} (\eta,\phi)$ is the number of tracks for a small given pseudorapidity and azimuthal angle interval of $0.1\times0.1$~\cite{ATLAS:2018ezv}.
This factor yields a weighted track distribution that is uniform along the azimuthal direction for any given pseudorapidity interval.

% End of text imported from the .//ANA-HION-2023-04-PAPER-Monte-Carlo-and-Data-Selection.tex input file

% The next lines are included from the .//ANA-HION-2023-04-PAPER-analysis-method.tex input file
\section{Analysis method}
\label{sec:analysis}
\subsection{Scalar-product method}
The SP method is defined in Ref.~\cite{STAR:2002hbo}, further discussed in Ref.~\cite{Voloshin:2008dg},
and results using this method for \pbpb\ and \xexe\ collisions have been published by ATLAS in Refs.~\cite{ATLAS:2018ezv, ATLAS:2019dct}.
It uses flow vectors \unj\ and \Qn, where the \unj\ for each object of interest $j$, for example, a charged-particle track or energy deposited at a single calorimeter tower,
is defined as
\begin{equation}
\unj =  \ephij,
\end{equation}
and the average flow vector \Qn\ of a subevent, for example, one side of the FCal, is defined as
\begin{equation}
\label{flow_vector}
\Qn =\frac{1}{\sum_j\wj}\sum_{j}\wj\unj,
\end{equation}
where the summation goes over all objects of interest in the subevent.
In this analysis, flow vectors are evaluated separately
for the two sides of the FCal and are denoted \QnNP, where the N and P
correspond to a pseudorapidity range of $-4.9 < \eta < -3.2$ and $3.2 < \eta < 4.9$, respectively.
In this case, the sum in Eq.~\ref{flow_vector} runs over the calorimeter
towers with approximate granularity of $\Delta\eta \times \Delta\phi =0.1\times0.1$
and the weights \wj\ are the transverse energies \Et\ measured in the towers.

The azimuthal anisotropy coefficients using  this method, \vnsp{n}, are defined for each \pt\ and centrality interval as,
\begin{equation}
\vnsp{n}\equiv Re \frac{\left\langle\unj\QnNP^{\ast}\right\rangle}{\sqrt{\left\langle\QnN\QnP^{\ast}\right\rangle}}
= \frac{\left\langle|\unj||\QnNP|\cos\left[n\left(\phi_j-\Psi_n^{N|P}\right)\right]\right\rangle}{\sqrt{\left\langle|\QnN||\QnP|\cos\left[n\left(\Psi_n^{N}-\Psi_n^{P}\right)\right] \right\rangle}}.
\end{equation}
The numerator calculates the
scalar product of the flow vector of the reference subevent, \QnNP, and that
of each charged-particle track, \unj, where the $\langle\rangle$ denotes an average over all tracks within the particular \pt\ interval from all events of the given centrality interval.
This average is weighted for each track to correct for azimuthal non-uniformity of the detector and track reconstruction inefficiency, as detailed in Section~\ref{subsec:trk_sel}.
The denominator calculates the estimated detector resolutions,
as determined using two reference subevents, positive and negative ends of the FCal, of the same event, %
and the $\langle\rangle$ denotes an average over all events in the given centrality interval.
The flow vector \QnP\ (\QnN) of the calorimeter is
correlated with tracks with $\eta <$~0 ($\eta >$~0).
This implementation of the SP method imposes a pseudorapidity gap of at least $3.2$ between the reference \Qn\ and
tracks, thus suppressing short-range nonflow correlations,
such as those arising from resonance decays
and same-jet correlations ~\cite{Luzum:2012da}.
The \vnsp{n}\ values are measured for ID pseudorapidity ranges of $|\eta| < 2.5$, $|\eta| < 1.1$, and $1.1 < |\eta| < 2.5$.

\subsection{Multi-particle cumulant method}
The MPC method using standard $Q$ cumulants, as applied in this analysis, is based on the generic framework described in Ref.~\cite{Bilandzic:2013kga}.
The three-subevent $Q$ cumulants method extends this framework and was introduced in Ref.~\cite{Jia:2017hbm} for reference \vnm{n}{4} measurements.
This analysis extends the three-subevent cumulants to \pt-differential \vnm{n}{4} measurements for high-\pt\ charged particles in \pbpb\ collisions.

\subsubsection{Standard $Q$ cumulants}
\label{subsub:std_qc}
For a single event, the \nth\ order two- and four-particle azimuthal correlators are denoted as \corr{n}{2} and \corr{n}{4}, respectively, and defined as \cite{Bilandzic:2010jr,Bilandzic:2013kga},
\begin{equation}
\begin{aligned}
\corr{n}{2}\equiv&\langle e^{in(\phi_1-\phi_2)}\rangle, \\
\corr{n}{4}\equiv&\langle e^{in(\phi_1+\phi_2-\phi_3-\phi_4)}\rangle, \\
\end{aligned}
\end{equation}
where $\phii$ represents the azimuthal angles of distinct charged-particle tracks, and the $\langle\rangle$ denotes an average over all two- or four-particle combinations in the given event.
In the standard $Q$ cumulants, all tracks within the pseudorapidity range of $|\eta|<2.5$ are used.
As it is computationally challenging to calculate four-particle correlations using nested loops,
the correlators in this analysis are computed using the $Q$ cumulants, following the generic framework from Ref.~\cite{Bilandzic:2013kga},
with the $k$-th power weighted flow vector \Qnk,\footnote{The notation \Qnk\ defined for the MPC method by Eq.~\ref{qnk} should be distinguished from the notation \Qn\ defined for the SP method by Eq.~\ref{flow_vector}.
The notation \Qnk\ is a weighted summation of \unj, whereas \Qn\ is a weighted average of \unj\ computed by the weighted summation of \unj\ normalized by the total weight of \unj.} defined as
\begin{equation}
\label{qnk}
Q_{n,k}=\sum_{j=1}^{M}\omega_{j}^k\unj,
\end{equation}
where $M$ is the charged-particle multiplicity of the event and \wj\ is the track weight.

The \pt-integrated reference four-particle cumulants \cn{n}{4}, computed over reference particles (REF), defined as all charged particles in the soft \pt\ range of 1--5~\GeV,
are calculated as,
\begin{equation}
\begin{aligned}
\cn{n}{4}\equiv&\corrs{n}{4}-2\corrs{n}{2}^2=\langle v_n^4\rangle-2\langle v_n^2\rangle^2=2\sigma^2(v_n^2)-\langle v_n^4\rangle=-(\vnm{n}{4})^4, \\
\end{aligned}
\end{equation}
where \corrs{n}{2} and \corrs{n}{4} are respectively the event-averaged two- and four-particle correlators, with the $\llangle\rrangle$ denoting an average of $\langle\rangle$ over all events, and $\sigma^2(v_n^2)$ is the variance of $v_n^2$.
The corresponding differential cumulants \dn{n}{4} are defined similarly as,
\begin{equation}
\begin{aligned}
\dn{n}{4}=&\corrspt{n}{4}-2\corrspt{n}{2}\corrs{n}{2},
\end{aligned}
\end{equation}
where \corrspt{n}{2} and \corrspt{n}{4} are calculated using one particle of interest (POI) in a specific \pt\ bin, while the other particles are REF.
The \pt\ differential \vnm{n}{4} is computed as,
\begin{equation}
\label{eqn:formula_vnpt_mpc}
\begin{aligned}
\vnmpt{n}{4}=&\frac{-\dn{n}{4}}{(-\cn{n}{4})^{3/4}}.
\end{aligned}
\end{equation}
Depending on the shape of the underlying reference \vnm{n}{4} distributions,
\cn{n}{4}\ could change sign across different centrality ranges.
Values of \cn{2}{4} have been observed to become positive in the 2\% most central collisions,
giving an imaginary reference \vnm{2}{4} \cite{ATLAS:2019peb}.
Therefore for this analysis, the centrality interval 0--5\% is omitted for the differential \vnm{2}{4} measurements to ensure that the values of reference \vnm{2}{4},
thus values of the denominator in the differential \vnm{2}{4} formula, are always real-valued.

\subsubsection{Three-subevent $Q$ cumulants}
In the standard $Q$ cumulants, all of the two or four particles correlated come from the full pseudorapidity range of the ID, $|\eta|<2.5$.
To further suppress the nonflow correlations,
rapidity gaps are applied in the MPC method using the three-subevent cumulants \cite{Jia:2017hbm,Huo:2017nms,ATLAS:2019peb} by requiring the correlated particles in an event to come from different pseudorapidity ranges.
The ID acceptance is divided equally into
three subevents of non-overlapping pseudorapidity ranges, indexed as $a$, $b$ and $c$.
The pseudorapidity ranges of these three subevents are as follows,
\begin{equation}
-2.5 < \eta_a < -\frac{2.5}{3},\text{  }|\eta_b| < \frac{2.5}{3},\text{  }\frac{2.5}{3}<\eta_c<2.5.
\end{equation}
The $k$-th power weighted flow vector for subevent $a$, $Q^a_{n,k}$, is defined as
\begin{equation}
\label{eqn:qnk3}
Q^a_{n,k}=\sum_{j=1}^{M_a}\omega_{j}^k\unj,
\end{equation}
where $M_a$ is the charged-particle multiplicity within subevent $a$.
The three-subevent single-event reference correlators and their corresponding event weights $W_{(2)}$ and $W_{(4)}$, calculated also using flow vectors where $n=0$, are defined as
\begin{equation}
\begin{aligned}
\corrsub{n}{2}{\ab}\equiv&\langle e^{in(\phi_a-\phi_b)}\rangle=Re\frac{Q_{n,1}^aQ_{-n,1}^b}{\Wt{2}{\ab}},\\
\Wt{2}{\ab}=&Q_{0,1}^aQ_{0,1}^b,\\
\end{aligned}
\end{equation}
\begin{equation}
\begin{aligned}
\corrsub{n}{4}{\af}\equiv&\langle e^{in(\phi_a+\phi'_a-\phi_b-\phi_c)}\rangle=Re\frac{[(Q_{n,1}^a)^2-Q_{2n,2}^a]Q_{-n,1}^bQ_{-n,1}^c}{\Wt{4}{\af}},\\
\Wt{4}{\af}=&[(Q_{0,1}^a)^2-Q_{0,2}^a]Q_{0,1}^bQ_{0,1}^{c}.\\
\end{aligned}
\end{equation}
where the superscript in event weights $W_{(2)}$ and $W_{(4)}$ corresponds to the track configuration of correlators.
Here the prime label on the azimuthal angle $\phi'_a$ indicates that a second distinct particle is selected from subevent $a$.
The reference cumulants \cnsub{n}{4}{\af} are calculated from the event-averaged correlators as,
\begin{equation}
\label{eqn:formula_cn_3sub}
\begin{aligned}
\cnsub{n}{4}{\af}=&\corrssub{n}{4}{\af}-2\corrssub{n}{2}{\ab}\corrssub{n}{2}{\ac}.\\
\end{aligned}
\end{equation}
To boost the statistics and reduce $\eta$-dependent detector bias, the subevent index $a$ is interchanged with $b$ and $c$ to create two additional configurations.
The weighted average of these three
measurements is used to obtain three-subevent reference cumulants \cnsub{n}{4}{\text{3-sub}}.
The formulae for these reference correlators, presented with different notations, are detailed in Ref.~\cite{Jia:2017hbm}.

Each subevent configuration of reference correlator \corr{n}{2} is subdivided into two separate configurations of the differential flow correlators by selecting each of the two correlated particles as the POI.
Similarly, each reference correlator configuration of \corr{n}{4} is subdivided into four differential correlator configurations.
For subevent $a$, the \nth\ order harmonic $k$-th power weighted flow vector using only POIs is denoted by $p^a_{n,k}$, which is defined for a given \pt\ interval.
Similarly, the flow vector made from particles that are both REF and POI for a given \pT\ interval is denoted by $q^a_{n,k}$ for subevent $a$.
These two flow vectors are calculated similarly to Eq.~\ref{eqn:qnk3}.
For \corr{n}{4}, two formulae were provided: one for when the POI comes from a subevent within which two particles are selected, as shown in Eq.~\ref{eqn:formula_corrpt_3sub:4a}, and the other for when the POI is the only particle from its subevent, as shown in Eq.~\ref{eqn:formula_corrpt_3sub:4b}.
A prime in the subevent index indicates that the subevent from which the POI of this correlator is selected.
\begin{align}
\centering
\begin{split}
\corrptsub{n}{2}{\Ab}=Re&\frac{p_{n,1}^aQ_{-n,1}^b}{\Wtpt{2}{\Ab}},\\
\Wtpt{2}{\Ab}=&p_{0,1}^aQ_{0,1}^b,\\
\end{split}\\
\begin{split}\label{eqn:formula_corrpt_3sub:4a}
\corrptsub{n}{4}{\Af}=&\corrptsub{n}{4}{\aAf}=Re\frac{(p_{n,1}^aQ_{n,1}^a-q_{2n,2}^a)Q_{-n,1}^bQ_{-n,1}^c}{\Wt{4}{\Af}},\\
\Wt{4}{\Af}=&(p_{0,1}^aQ_{0,1}^a-q_{0,2}^a)Q_{0,1}^bQ_{0,1}^{c},\\
\end{split}\\
\begin{split}\label{eqn:formula_corrpt_3sub:4b}
\corrptsub{n}{4}{\aF}=&Re\frac{[(Q_{n,1}^a)^2-Q_{2n,2}^a]p_{-n,1}^bQ_{-n,1}^c}{\Wt{4}{\aF}},\\
\Wt{4}{\aF}=&[(Q_{0,1}^a)^2-Q_{0,2}^a]p_{0,1}^bQ_{0,1}^{c}.\\
\end{split}
\end{align}
Other configurations can be written similarly by permuting the subevent indices.
Similar to Eq.~\ref{eqn:formula_cn_3sub}, the differential cumulants are defined as
\begin{equation}
\begin{aligned}
\dnsub{n}{4}{\Af}=&\corrsptsub{n}{4}{\Af}-2\corrsptsub{n}{2}{\Ab}\corrssub{n}{2}{\ac}.\\
\end{aligned}
\end{equation}
Each of the three configurations for the reference cumulant corresponds to four configurations for the differential cumulant.
The three-subevent differential cumulants \dnsub{n}{4}{\text{3-sub}} are calculated as the weighted average from these 12 configurations,
and then used for the calculation of three-subevent cumulants \vnm{n}{4} (as defined in Eq.~\ref{eqn:formula_vnpt_mpc}), denoted by \vnmsub{n}{4}{\text{3-sub}}.

\subsubsection{Event combination procedure}
In this analysis, the event-averaged cumulants \cn{n}{4} and \dn{n}{4} are firstly computed in narrower centrality intervals, and then averaged toward wider centrality intervals in which the final results of \vn{n} are presented.
The choice of initial centrality intervals in which the cumulants are calculated defines an event combination procedure.
The choice of event combination procedure can affect the measured \vn{n} in different ways.
High-\pt\ particle production is biased toward more central events, and therefore, for quantities measured using high-\pt\ charged particles, an average over a wide centrality interval is biased toward the more central edge of the interval.
Also, the event-by-event fluctuations in the underlying azimuthal anisotropy distributions are sensitive to the event combination procedure.
Therefore, the choice of the initial centrality intervals in an event combination procedure can be used to test the sensitivity of an observable to the underlying fluctuations and to identify the source of this fluctuation.
These fluctuations can stem from the initial state geometry or nonflow effects.
It has been shown in measurements of reference \vn{n}
that the centrality resolution of the initial centrality intervals affects the MPC measurements of \vn{n} due to fluctuations in nonflow effects~\cite{ATLAS:2019peb,Jia:2017hbm}.

By default, cumulants are calculated within 1\% centrality percentiles before combining.
For comparison, cumulants are also calculated within 2\% centrality percentiles before combining and directly calculated within the quoted centrality intervals for results.
Results obtained following these procedures are referred to later as \ec{1}, \ec{2} and \ec{10}, respectively.
The final results are presented in 5\% centrality intervals for the most central 10\% of events, and in 10\% centrality intervals for all other events.

% End of text imported from the .//ANA-HION-2023-04-PAPER-analysis-method.tex input file

\section{Systematic uncertainties}
\label{sec:systematics}

% The next lines are included from the .//ANA-HION-2023-04-PAPER-sys.tex input file
The systematic uncertainties in this analysis
arise from the choice of the pseudorapidity ranges used in reference \vn{n}
definitions, variation of track selection criteria, detector material
uncertainties affecting reconstruction efficiency and overall imperfections in
the detector response.
These uncertainties are evaluated by first repeating the analysis with
alternative pseudorapidity ranges, the varied selection criteria on tracks, or
different track reconstruction efficiency corrections, and then calculating the absolute differences in the resulted
\vn{n}\ values from the default results.
For each alteration, track weights and other corrections are re-calculated accordingly.

For a detector with perfect azimuthal symmetry, the imaginary part of the
scalar product should be zero.
Therefore, the imaginary part of the scalar product, referred to as the
residual sine term,
is used as a systematic uncertainty.
The sensitivity of the result to the pseudorapidity range of the FCals used to
determine
the flow vector is evaluated by using only the inner or outer
halves of the FCals, corresponding to pseudorapidity ranges of $|\eta_\mathrm{FCal}| <
4.0$ and $|\eta_\mathrm{FCal}| > 4.0$, respectively.
Additionally, \vn{n}\ values are expected to be consistent
between the positive and negative halves of ID
due to the symmetry of the collision system.
Any asymmetry of the results in $\eta$
is included in the systematic uncertainty, evaluated using $\eta_\mathrm{trk} < 0$
and $\eta_\mathrm{trk} > 0$.
The
sensitivity of the result to the track selection is assessed
by tightening the track selection criteria, requiring at least 12 hits in the
silicon detectors and no holes in the SCT.
The default \ETfcal\ requirements used for the centrality percentile is determined by
matching $85\%$ of the Glauber-like dataset to MC events using the Glauber
Model~\cite{Miller:2007ri, ATLAS:2011ag}.
The centrality determination uncertainty is accounted for by up and down varying
$1\%$ of data used to match the simulations for alternative \ETfcal\ requirements.
The track reconstruction efficiency used in this analysis is fitted as a
function of \pt\ in order to obtain a smooth correction.
The difference in the measured \vn{n} between the fitted and binned efficiency
is used as a systematic uncertainty.
The same systematic uncertainties are evaluated for \vnsp{n} results for
the three ranges of charged-particle pseudorapidity $|\eta| < 1.1$, $1.1 <
|\eta| < 2.5$ and $|\eta| < 2.5$.

In addition to \vnsp{n} values, the difference between two \vnsp{n}
measurements using tracks with non-overlapping pseudorapidity ranges is also
calculated.
Only contributions of systematic uncertainties that are uncorrelated between
the two \vnsp{n} measurements of non-overlapping ranges are included in this
quantity.
For this analysis, the charged-particle $\eta$-asymmetry
and residual sine term uncertainties use non-overlapping parts of sub-detectors
and are therefore considered uncorrelated.

For the MPC method, only a subset of systematic uncertainties that are
considered for the SP method is included.
No systematic uncertainties related to $\eta$-symmetry are included,
as evaluating multiparticle correlations within half of the default pseudorapidity range
would significantly
affect the sensitivity of measurements to physical effects such as nonflow correlations.
Systematics related to the FCal are irrelevant for the MPC method.
Therefore, the list of systematic uncertainty items used for the MPC method is
as follow:
the sine residual term in the differential cumulant \dn{n}{4},
the centrality determination variations, and
the systematic uncertainty from the track reconstruction efficiency.

\begin{figure}
\centering
\subfloat[]{
\centering
\includegraphics[width=0.48\textwidth]{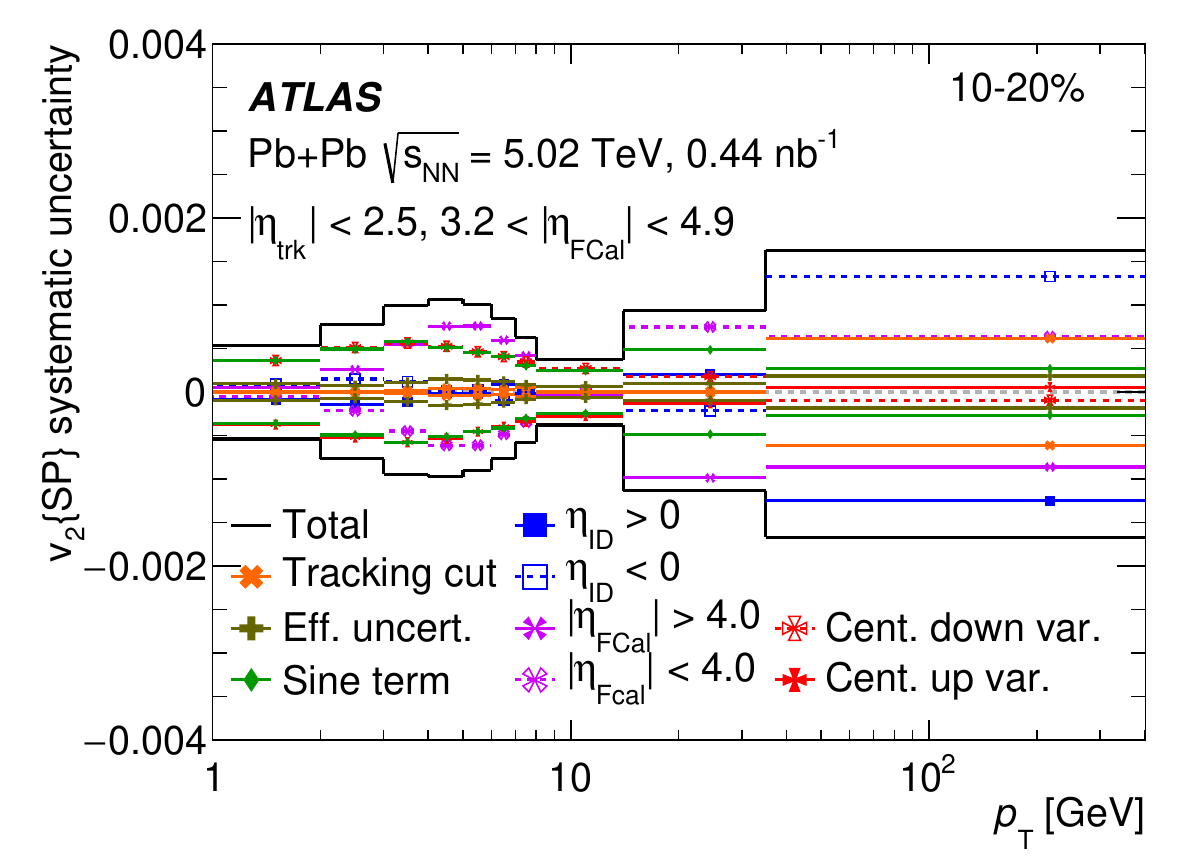}
}
\subfloat[]{
\centering
\includegraphics[width=0.48\textwidth]{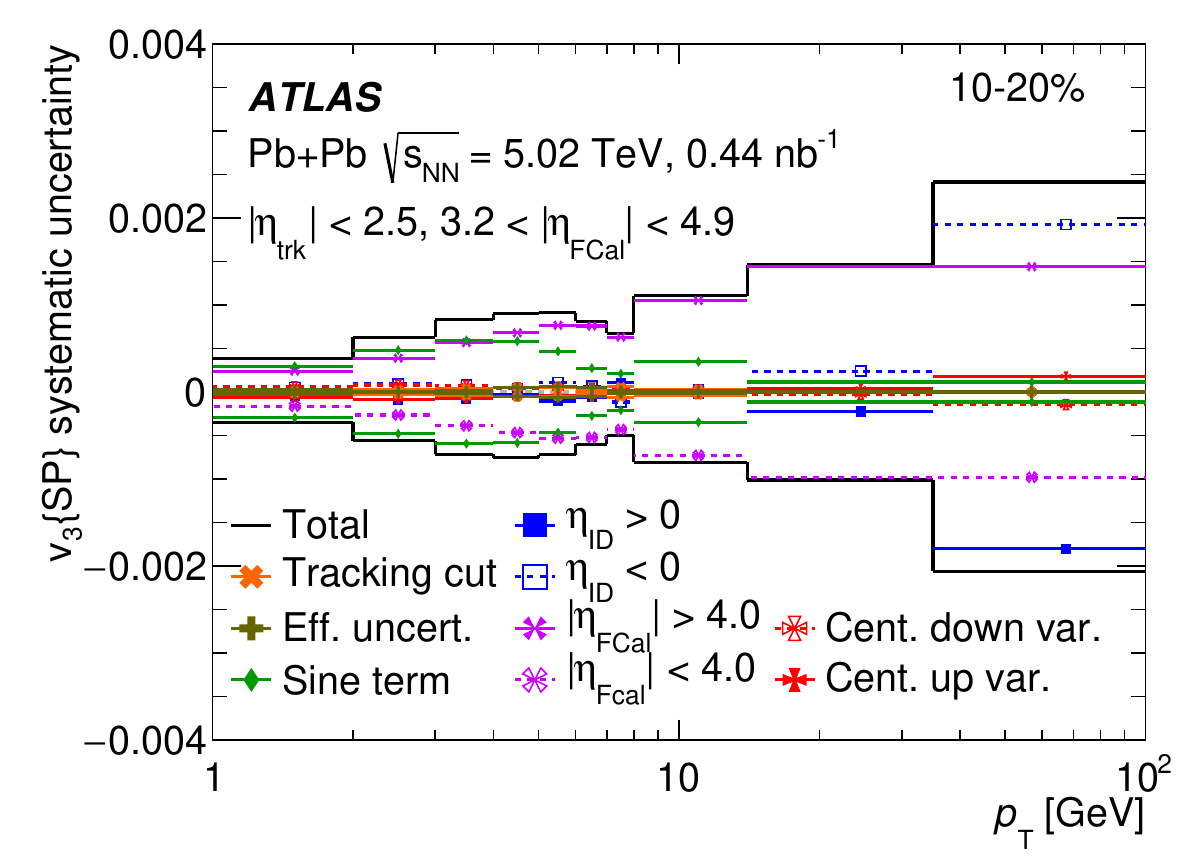}

}
\hfill
\caption{Breakdown of absolute systematic uncertainties as a function of \pt\ for (a) \vnsp{2} and
(b) \vnsp{3} values
in 10--20\% central collisions.}
\label{fig:syssp}
\end{figure}

\begin{figure}
\centering
\subfloat[]{
\centering

\includegraphics[width=0.48\textwidth]{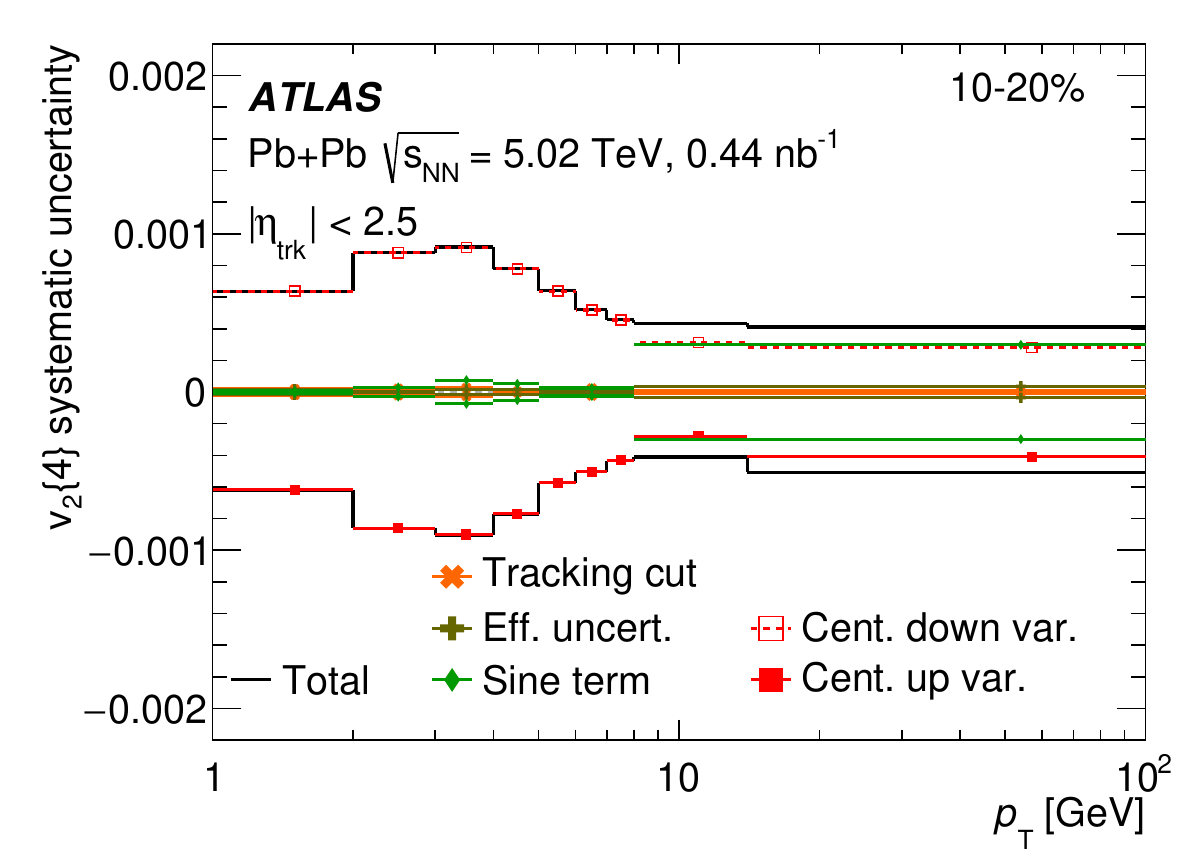}}
\subfloat[]{
\centering

\includegraphics[width=0.48\textwidth]{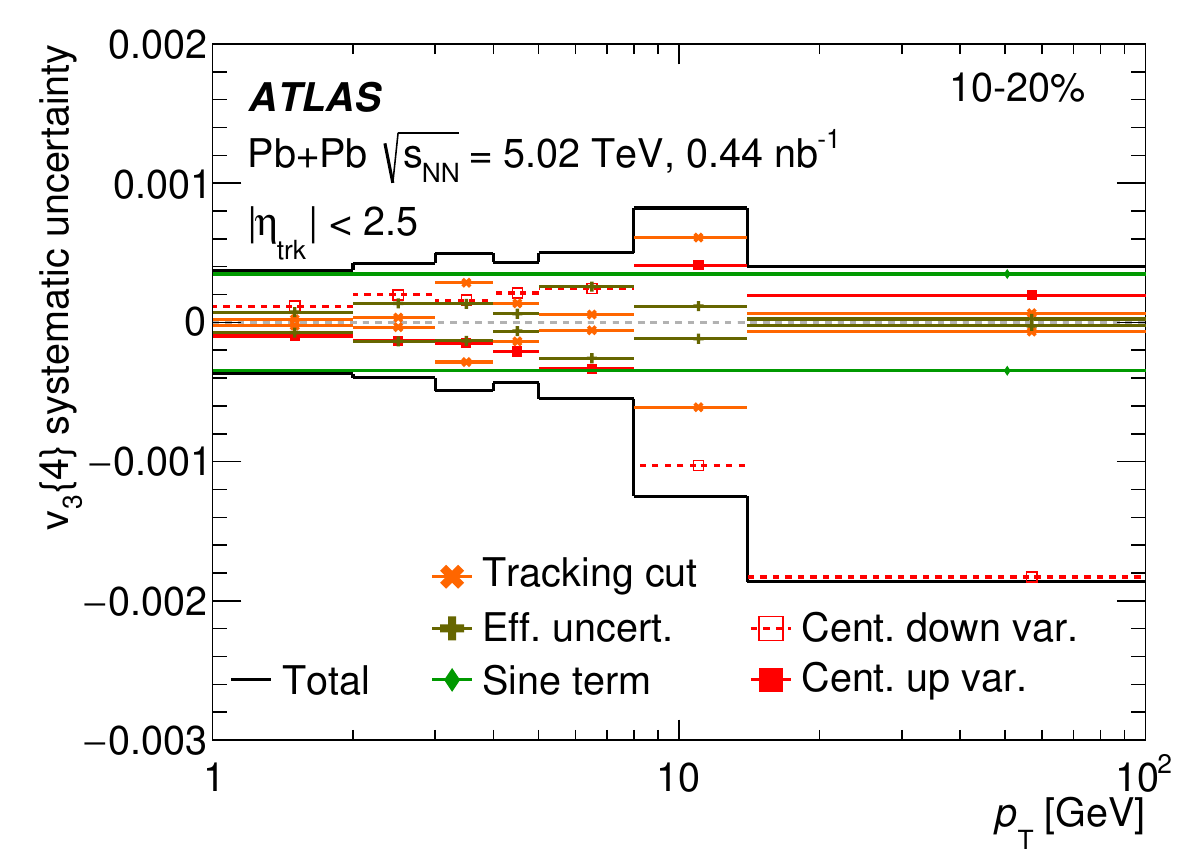}}
\hfill
\caption{Breakdown of absolute systematic uncertainties as a function of \pt\ for (a) \vnm{2}{4}
and (b) \vnm{3}{4} values
in 10--20\% central collisions.}
\label{fig:sysmpc}
\end{figure}

All sources of systematic uncertainties are taken as uncorrelated, and the
total
systematic uncertainty is calculated as the quadrature sum of the individual
components, separately for positive and negative terms.
The systematic uncertainty sources with no opposite-side alternations, residual sine terms,
track reconstruction efficiency, and tightened track selection criteria, are symmetrized to account
for the possibility of opposite-side uncertainties.
The magnitudes of systematic uncertainties are summarized in
Figure~\ref{fig:syssp} for the SP method and Figure~\ref{fig:sysmpc} for the
MPC method for
the selected centrality interval of 10--20\%.
Systematic uncertainties of the SP method are dominated by the choice of the
pseudorapidity range for ID or FCal as well as sine terms, whereas the
systematic uncertainties of the MPC method are dominated by the variation
in the centrality determination.
Similar relative magnitudes of systematic uncertainties are seen in other
centrality intervals.
The magnitudes of total systematic uncertainties increase toward peripheral
events.
Systematic uncertainties in \vnsp{n} using alternative pseudorapidity ranges
$0<|\eta|<1.1$ and $1.1<|\eta|<2.5$ show similar relative magnitudes among
different terms and increased total magnitudes in comparison to the
default pseudorapidity range of $0<|\eta|<2.5$.
The centrality and \pt\ dependence of the total systematic uncertainty are
similar for both \vnm{n}{4} and \vnmsub{n}{4}{\text{3-sub}}, with the latter
having a larger magnitude especially toward more peripheral events.
In both the SP and the MPC methods, the statistical uncertainties dominate over systematic ones,
especially toward the high-\pt\ region.

\FloatBarrier

% End of text imported from the .//ANA-HION-2023-04-PAPER-sys.tex input file

%
\section{Results}
\label{sec:result}

% The next lines are included from the .//ANA-HION-2023-04-PAPER-result.tex input file
This section presents measurements of \vn{n}\ using the SP method and the MPC method as a function of the
charged-particle \pt\ for different centrality intervals in \pbpb\ collisions.
Comparisons in this section are done for both central and peripheral events, the interval 10--20\% for ``central events'' and the interval 40--50\% or 50--60\% for ``peripheral events''. The former interval is the most central 10\%-wide centrality interval, and the latter is the most peripheral centrality interval that is statistically available for the variables concerned.
The values are measured over intervals of \pT, and the \pt-position of each data point corresponds to the bin center of each \pt\ interval.
In each centrality interval, the \vn{n}\ values reach a maximum at charged-particle
\pt\ between 3--4~\GeV\ and then decrease with increasing \pt.
Additional centrality intervals for some plots are included in Appendix~\ref{sec:appendix}.

\subsection{Scalar-product method}

\begin{figure*}[!ht]
\centering
\subfloat[]{
\centering
\includegraphics[width=0.48\textwidth]{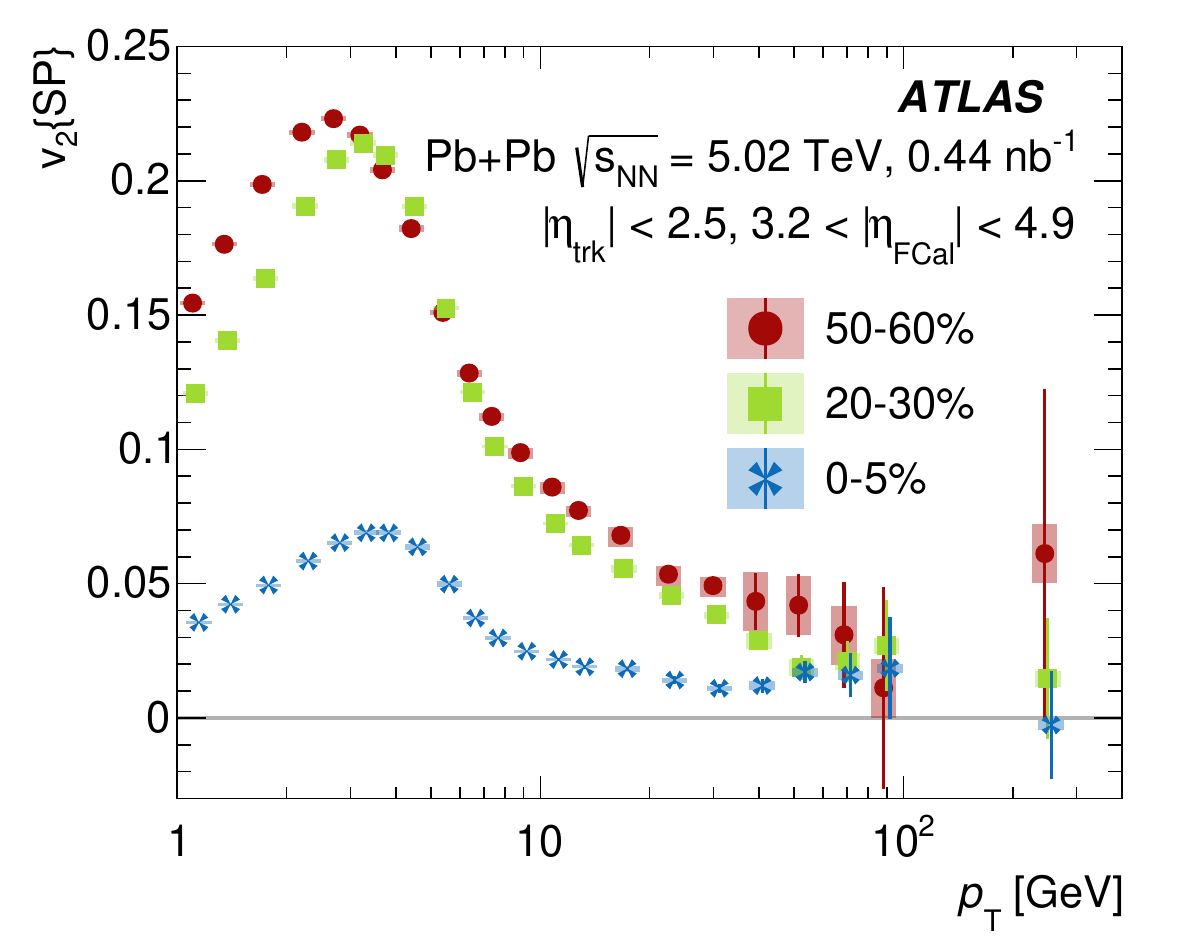}
}
\subfloat[]{
\centering
\includegraphics[width=0.48\textwidth]{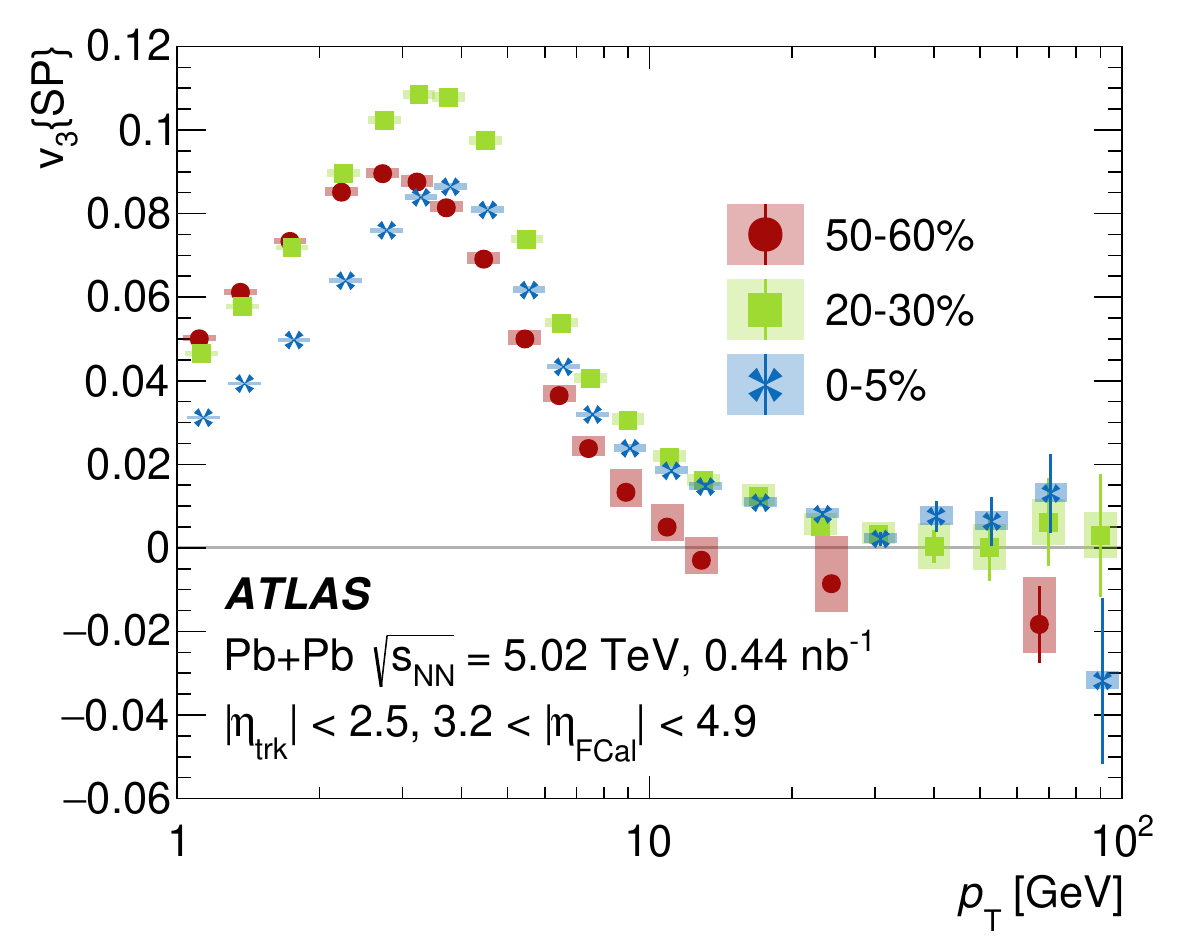}
}
\hfill
\caption{The (a) \vnsp{2} and (b) \vnsp{3} values as a function of charged-particle \pt\
for centrality intervals 0--5\%, 20--30\%, and 50--60\%.
The statistical uncertainties are shown as error bars and the systematic uncertainties are shown as boxes.
}
\label{fig:vn_sys_sp}
\end{figure*}

\begin{figure}
\centering
\subfloat[]{
\centering
\includegraphics[width=0.48\textwidth]{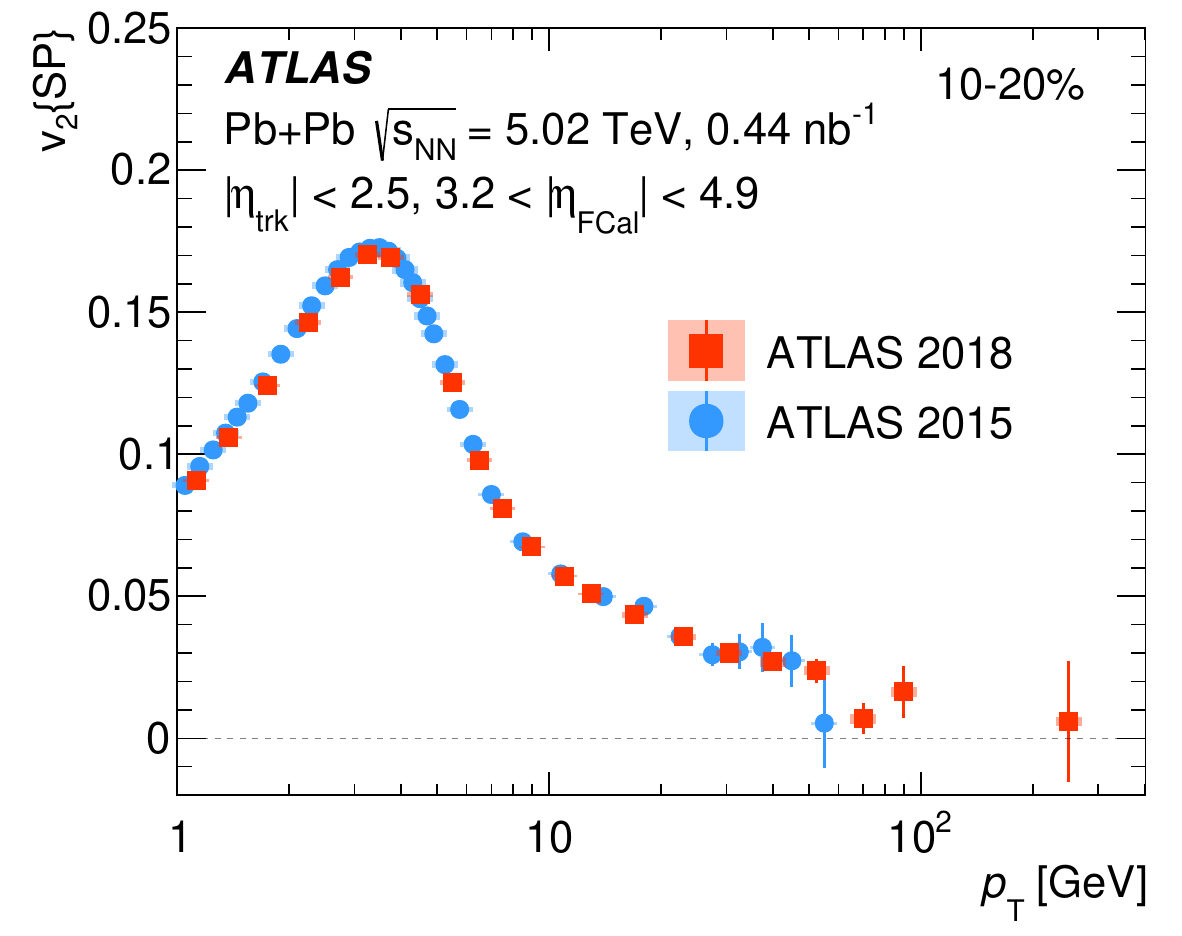}}
\subfloat[]{
\centering
\includegraphics[width=0.48\textwidth]{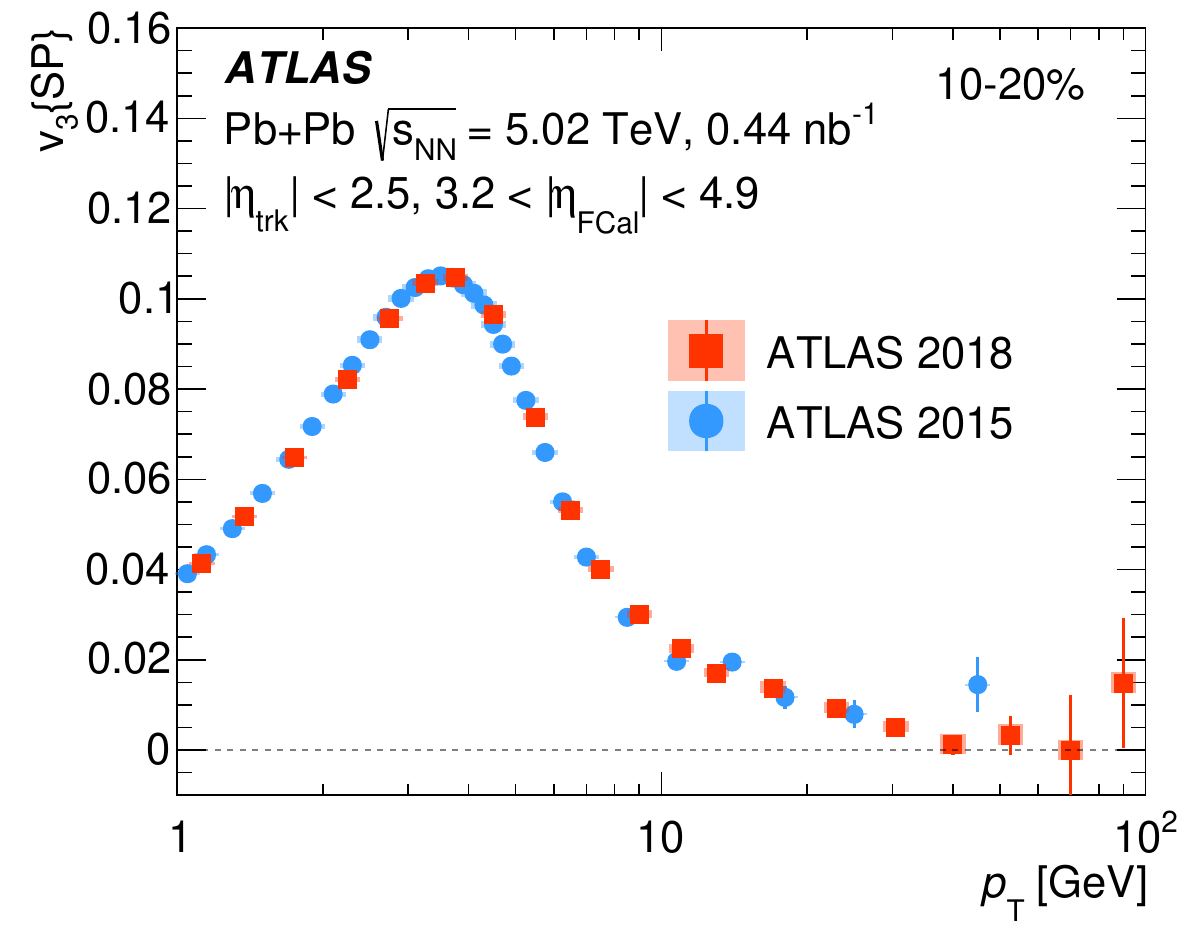}}
\caption{The (a) \vnsp{2} and (b) \vnsp{3} values as a function of charged-particle \pt, for the 10--20\% centrality interval,
compared with
ATLAS 2015 measurements from Ref.~\cite{ATLAS:2018ezv} in the same charged particle and calorimeter pseudorapidity ranges.
The statistical uncertainties are shown as error bars and the systematic uncertainties are shown as boxes.
}
\label{fig:comp_atlas}
\end{figure}

\begin{figure}
\centering
\subfloat[]{
\centering
\includegraphics[width=0.48\textwidth]{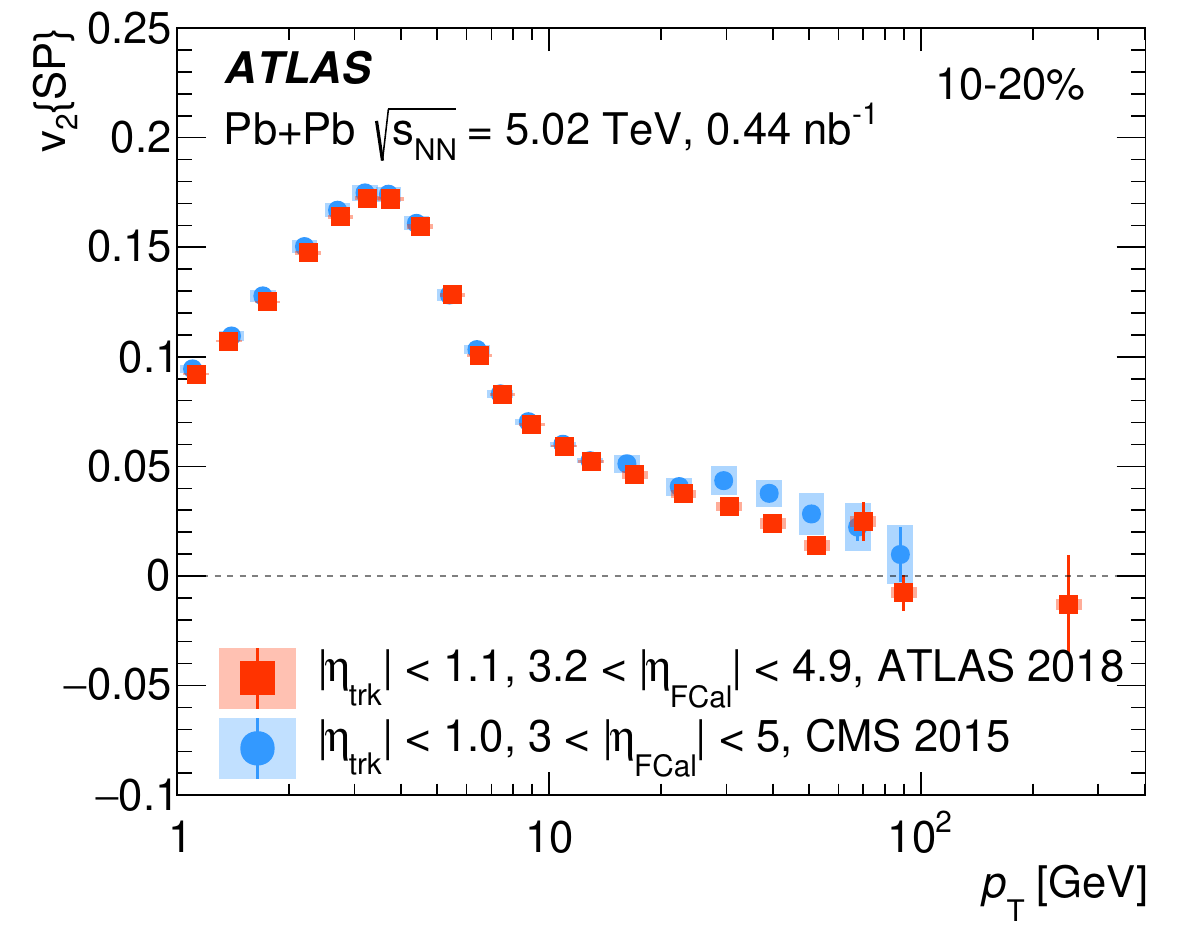}}
\subfloat[]{
\centering
\includegraphics[width=0.48\textwidth]{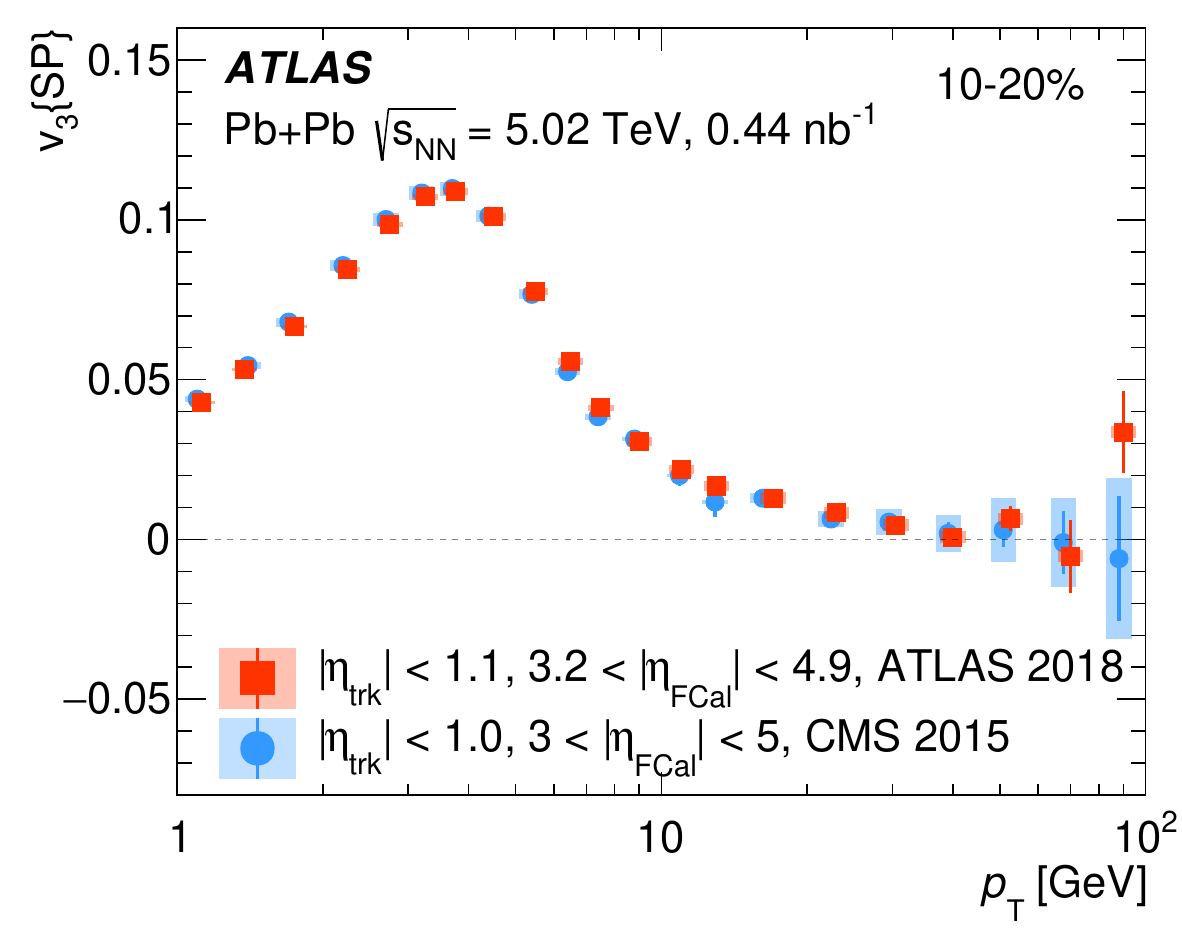}}
\caption{The (a) \vnsp{2} and (b) \vnsp{3} values as a function of charged-particle \pt, for the 10--20\% centrality interval,
compared with CMS 2015 measurements from Ref.~\cite{CMS:2017xgk}, which uses charged particles with pseudorapidity range $|\eta| < 1.0$ and calorimeter in pseudorapidity range $3 < |\eta| < 5$.
The statistical uncertainties are shown as error bars and the systematic uncertainties are shown as boxes.}
\label{fig:comp_cms}
\end{figure}

Figure~\ref{fig:vn_sys_sp} shows the \vnsp{2} and \vnsp{3} values as functions of \pt\ for the selected
centrality intervals.
The measured \vnsp{2} values are positive for the selected centrality intervals up to a \pt\ of 100~\GeV, and become constant over the \pt\ range of approximately 50 to 100~\GeV\ for the 5--10\% and 20--30\% centrality intervals. Note that due to decreased statistics, \vnsp{3} results in the 50--60\% centrality interval use wider \pt\ intervals toward high \pt\ than other centrality intervals.
The measured \vnsp{3} values are positive up to approximately $25$~\GeV\ for the selected centrality intervals, except for the 50--60\% centrality interval, where values of \vnsp{3} become negative and have a downward trend for $\pt>20$~\GeV.
Figures~\ref{fig:comp_atlas} and~\ref{fig:comp_cms} show the consistency of the \vnsp{n} measured in \pbpb\ collision at $\sqrtsnn=5.02~\TeV$ for the centrality interval 10--20\%.
Note that in the two figures, two different pseudorapidity ranges are used for the ATLAS 2018 measurements from this analysis.
In Figure~\ref{fig:comp_atlas}, the reported ATLAS 2015 measurements~\cite{ATLAS:2018ezv} use the same pseudorapidity ranges of ID and FCal with the ATLAS 2018 measurements, and the two results show good agreement.
In Figure~\ref{fig:comp_cms}, the CMS 2015 measurements~\cite{CMS:2017xgk} shown use tracks with pseudorapidity range $|\eta| < 1.0$ and calorimeters in pseudorapidity range $3 < |\eta| < 5$.
Overall, good agreement is found between the ATLAS 2018 and the CMS 2015 measurements, except for \vnsp{2} in the \pt\ range of 14--50~\GeV, where the CMS 2015 measurements yield higher values than the measurements from this analysis.

\begin{figure}
\centering
\subfloat[]{
\centering
\includegraphics[width=0.48\textwidth]{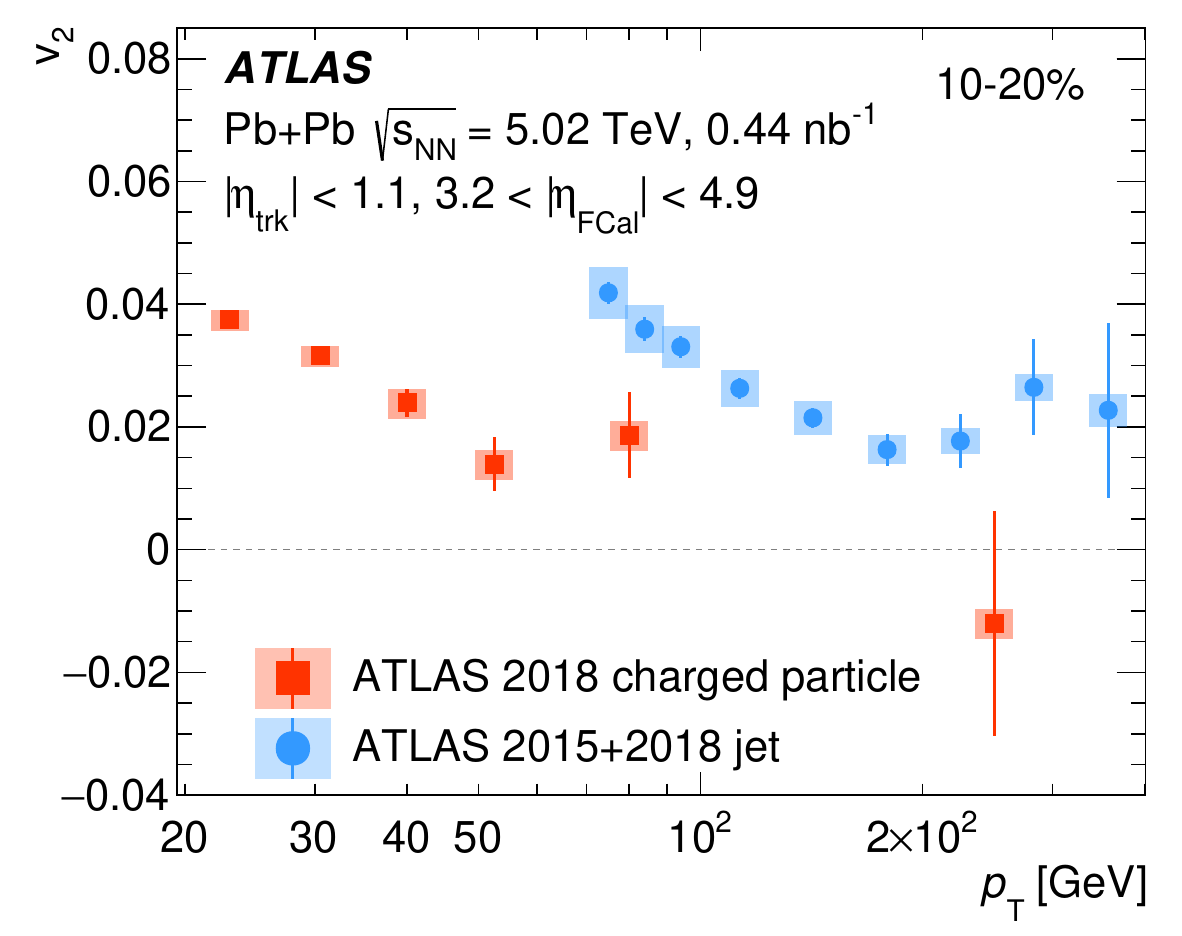}
}
\subfloat[]{
\centering
\includegraphics[width=0.48\textwidth]{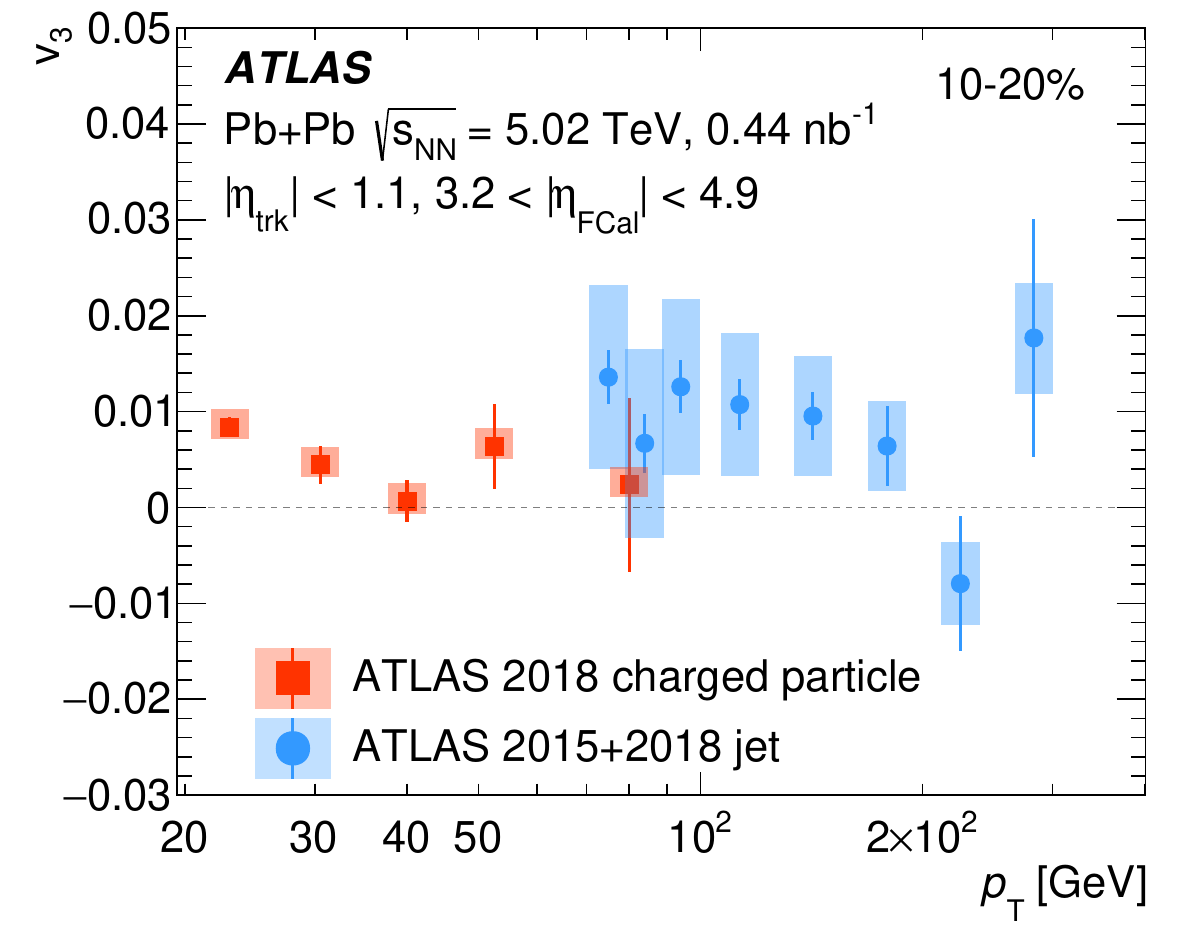}
}
\caption{The (a) \vn{2} and (b) \vn{3} values as a function of charged particle or jet \pt, for the 10--20\% centrality interval,
compared with
ATLAS jet \vn{n} measurements from Ref.~\cite{ATLAS:2021ktw}.
Due to decreased statistics, the measurements of charged particle \vnsp{3} in (b) covers up to 100~\GeV.
The statistical uncertainties are shown as error bars and the systematic uncertainties are shown as boxes.}
\label{fig:jet_comp}
\end{figure}

\begin{figure}
\centering
\subfloat[]{
\centering
\includegraphics[width=0.48\textwidth]{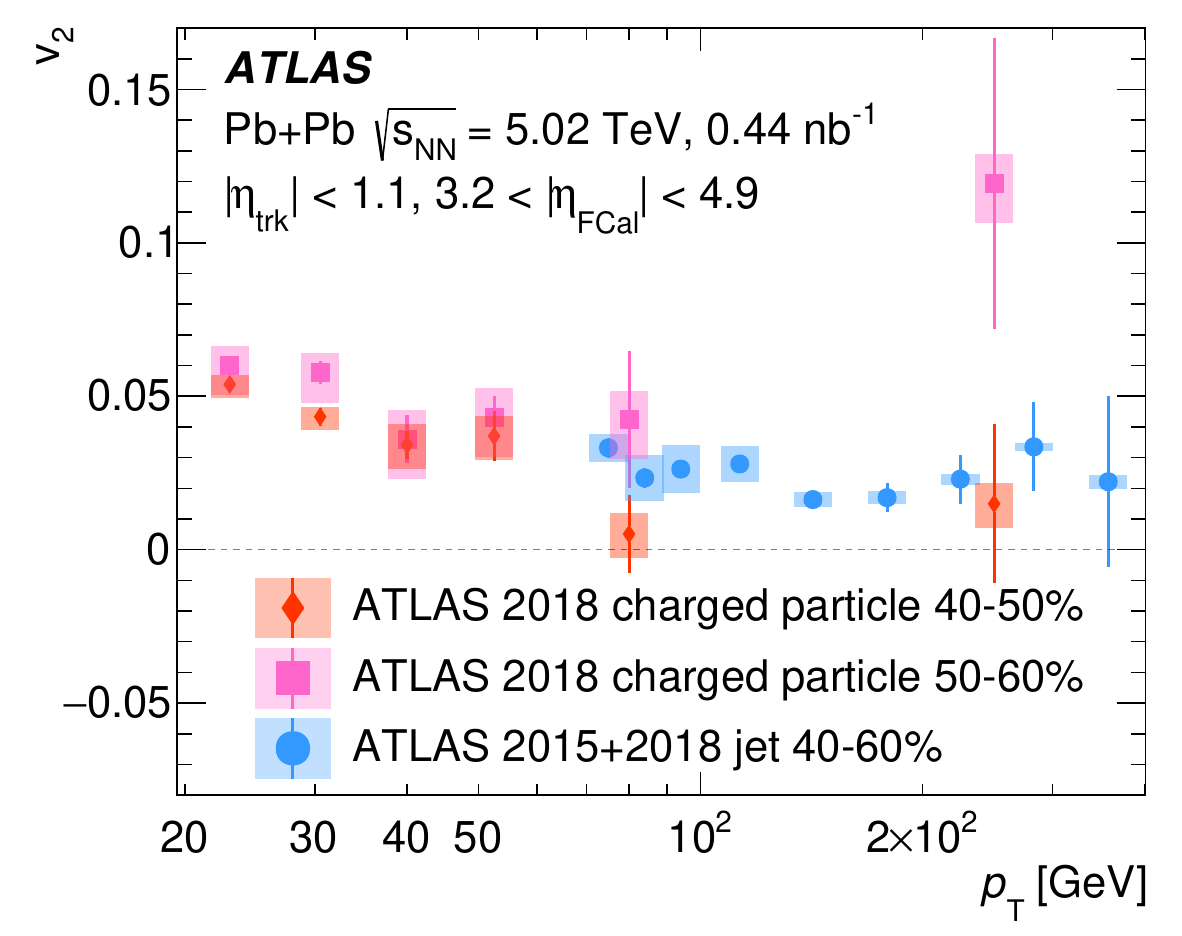}
}
\subfloat[]{
\centering
\includegraphics[width=0.48\textwidth]{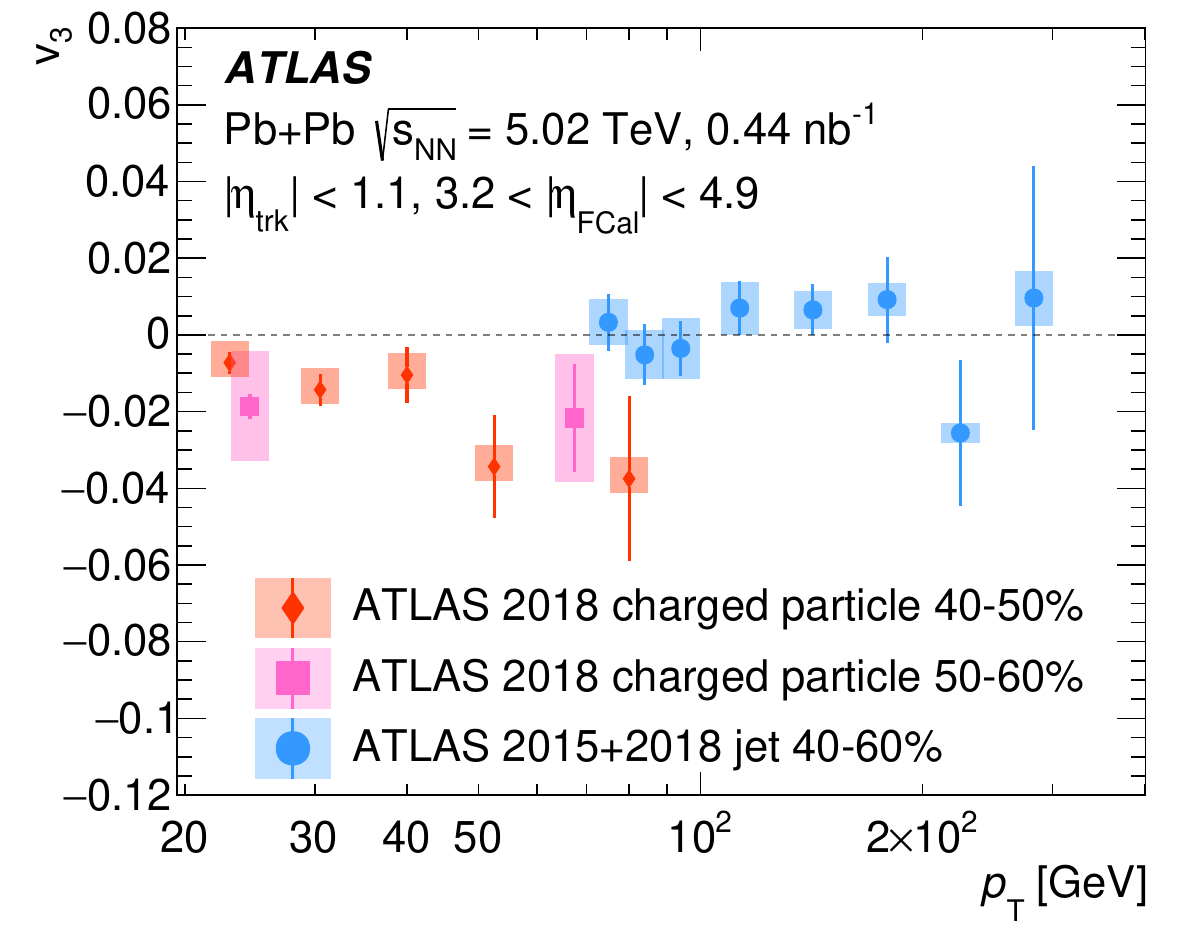}
}
\caption{The (a) \vn{2} and (b) \vn{3} values as a function of charged particle or jet \pt, for the centrality intervals within 40--60\%,
compared with ATLAS jet \vn{n} measurements from Ref.~\cite{ATLAS:2021ktw}.
Due to decreased statistics, the measurements of charged particle \vnsp{3} in (b) covers up to 100~\GeV.
The statistical uncertainties are shown as error bars and the systematic uncertainties are shown as boxes.}
\label{fig:jet_comp2}
\end{figure}

In Figures~\ref{fig:jet_comp} and~\ref{fig:jet_comp2}, a comparison is made between the charged-particle \vnsp{n} measurements from this analysis and the jet \vn{n} measurements from ATLAS using the event plane method from Ref.~\cite{ATLAS:2021ktw}.
High-\pt\ charged particles are from jet fragmentation but carry only a fraction of a given jet's \pt. The measured \vn{n} values are qualitatively similar between the jets and charged particles in the quoted high-\pt\ range of 20--400~\GeV.

\begin{figure*}[!ht]
\centering
\subfloat[]{
\centering
\includegraphics[width=0.46\textwidth]{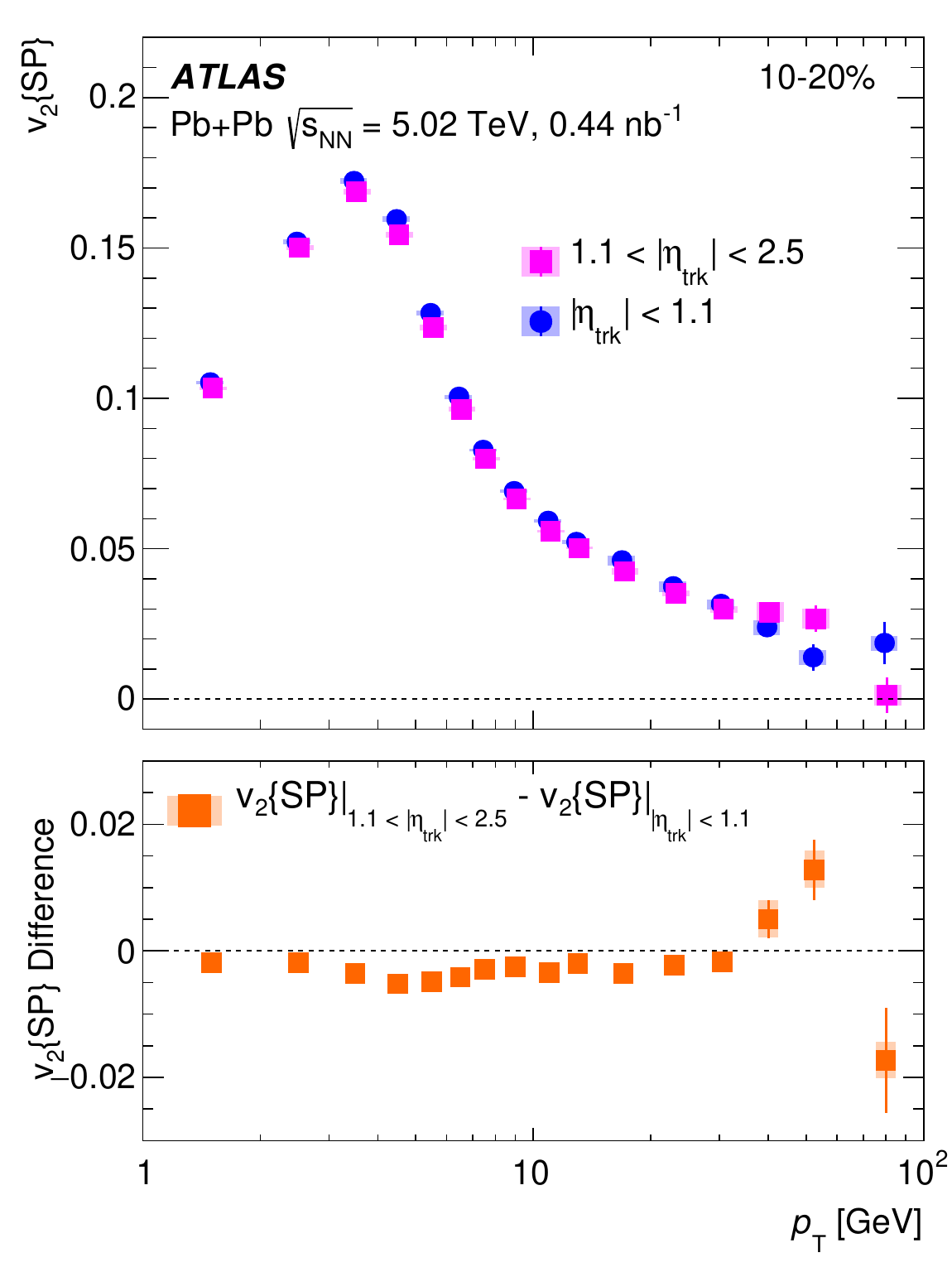}
}
\subfloat[]{
\centering
\includegraphics[width=0.46\textwidth]{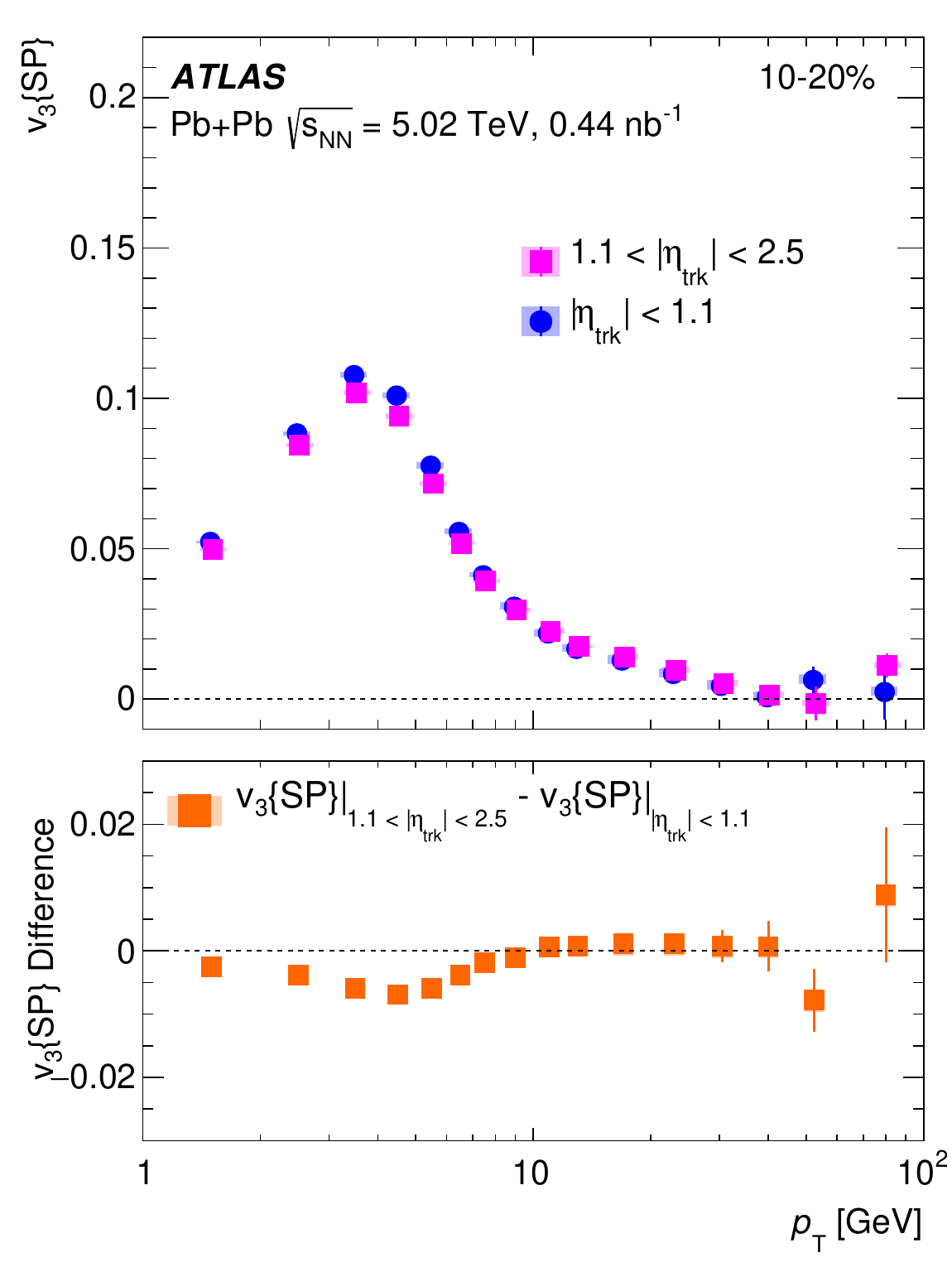}
}
\hfill
\caption{The (a) \vnsp{2} and (b) \vnsp{3} values as a function of charged-particle \pt\
in the 10--20\%\ centrality interval for different pseudorapidity ranges.
The bottom panel shows the difference in the \vnsp{n} measured between the two ranges.
The statistical uncertainties are shown as error bars and the systematic uncertainties are shown as boxes.}
\label{fig:eta_dep}
\end{figure*}

\begin{figure*}[!ht]
\centering
\subfloat[]{
\centering
\includegraphics[width=0.46\textwidth]{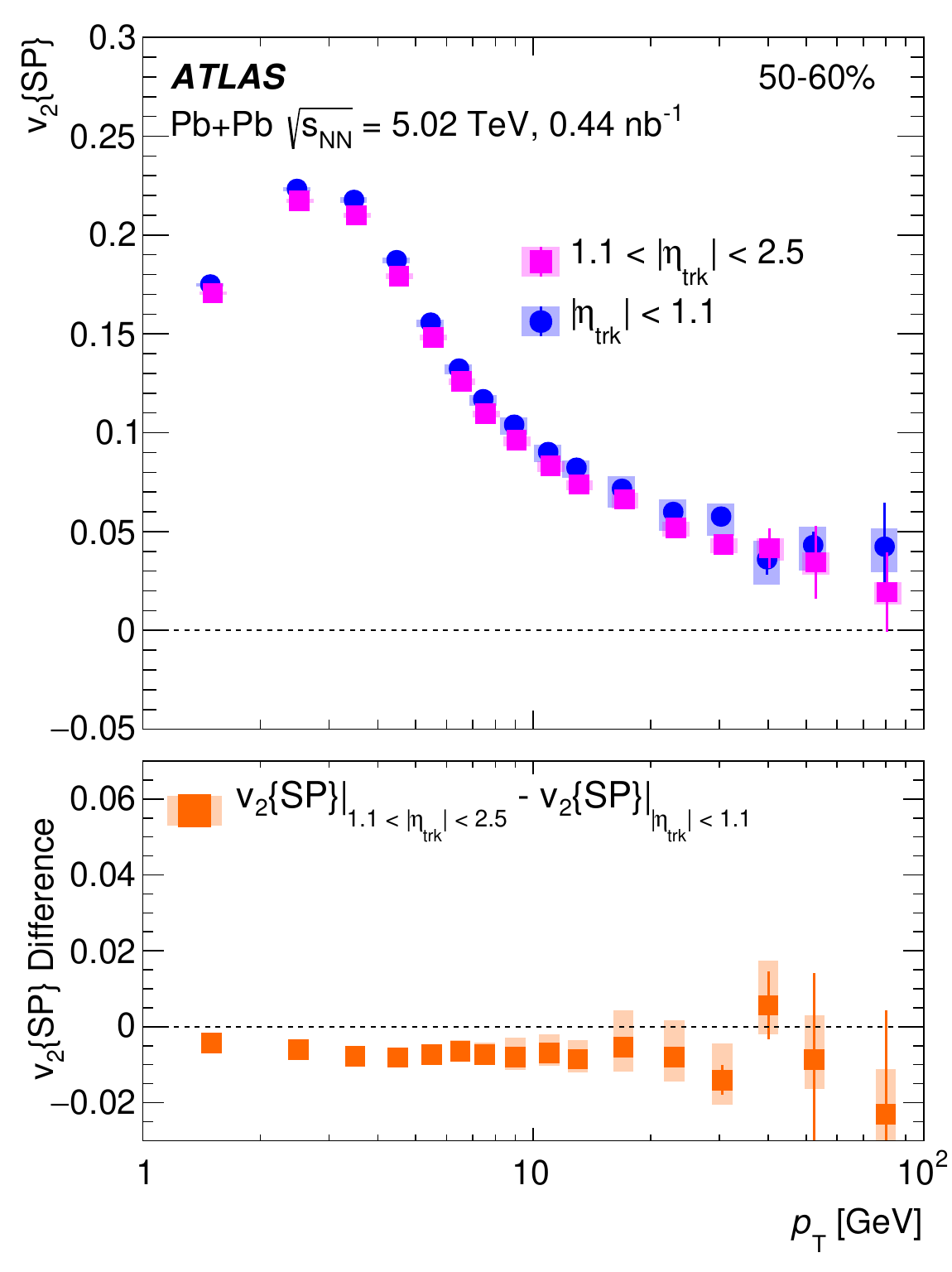}
}
\subfloat[]{
\centering
\includegraphics[width=0.46\textwidth]{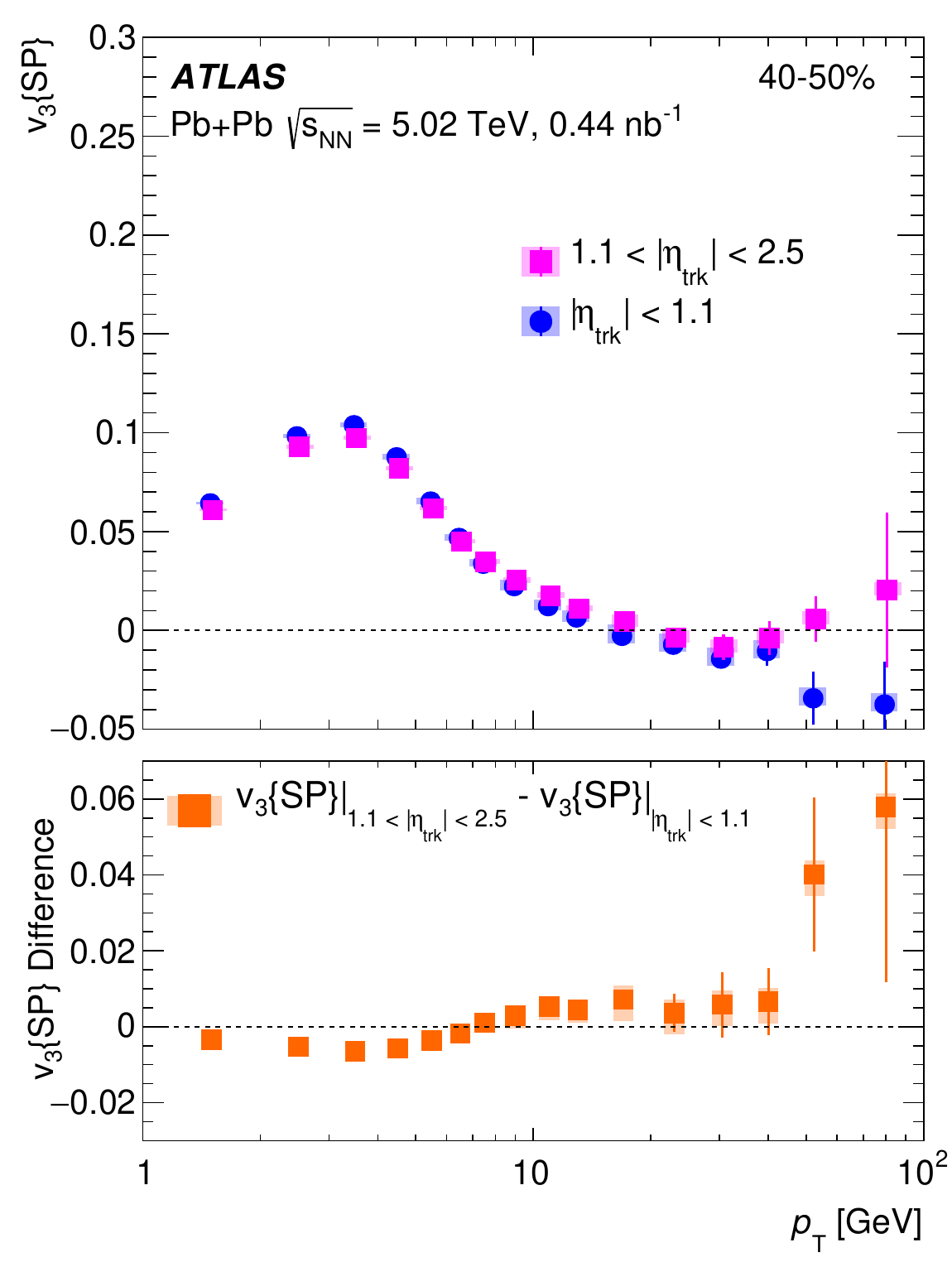}
}
\hfill
\caption{The \vnsp{n} values as a function of charged-particle \pt\ for different pseudorapidity ranges in (a) the 50--60\% centrality interval for \vnsp{2} and (b) the 40--50\% centrality interval for \vnsp{3}.
The bottom panel shows the difference in the \vnsp{n} measured between the two ranges.
The statistical uncertainties are shown as error bars and the systematic uncertainties are shown as boxes.
}
\label{fig:eta_dep3}
\end{figure*}

Measurements of \vnsp{n} using two non-overlapping pseudorapidity ranges of tracks are compared in Figures~\ref{fig:eta_dep} and~\ref{fig:eta_dep3} in two centrality intervals.
The pseudorapidity range $1.1 < |\eta| < 2.5$ corresponds to a greater average pseudorapidity gap between the tracks and correlated reference \Qn\ than the pseudorapidity range $|\eta| < 1.1$.
At $\pt<10$~\GeV\ for both centrality intervals shown, tracks with the pseudorapidity range $1.1 < |\eta| < 2.5$ yield smaller values of both \vnsp{2} and \vnsp{3} than tracks with the pseudorapidity range $|\eta| < 1.1$.
With a larger pseudorapidity gap imposed, certain nonflow effects are better suppressed, while the longitudinal decorrelation effects in the event plane become stronger~\cite{ATLAS:2017rij}, both of which reduce the measured values of \vn{n}.
For the 10--20\% centrality interval, the difference in \vnsp{2} between the two pseudorapidity ranges at $10 < \pt < 20$~\GeV\ is similar to that at low \pt, while the difference in \vnsp{3} between the two pseudorapidity ranges becomes consistent with zero.
For the most peripheral centrality intervals, 50--60\% for \vnsp{2} and 40--50\% for \vnsp{3}, tracks with the pseudorapidity range $1.1<|\eta|<2.5$ yield smaller \vnsp{2} values but larger \vnsp{3} values than tracks with the pseudorapidity range $|\eta| < 1.1$.
In this \pt\ range, particle production is dominated by jets.
The observed contrasting behavior of \vnsp{2} and \vnsp{3} at high \pt, especially in peripheral events, is consistent with a back-to-back dijet contamination,
which contributes positively to even-order harmonics and negatively to odd-order harmonics~\cite{Huo:2017nms}.
At $\pt>20$~\GeV, similar conclusions cannot be drawn due to increased statistical uncertainties and fluctuations.

\FloatBarrier

\subsection{Multi-particle cumulant method}
\begin{figure*}[ht!]
\centering
\subfloat[]{
\centering
\includegraphics[width=0.48\textwidth]{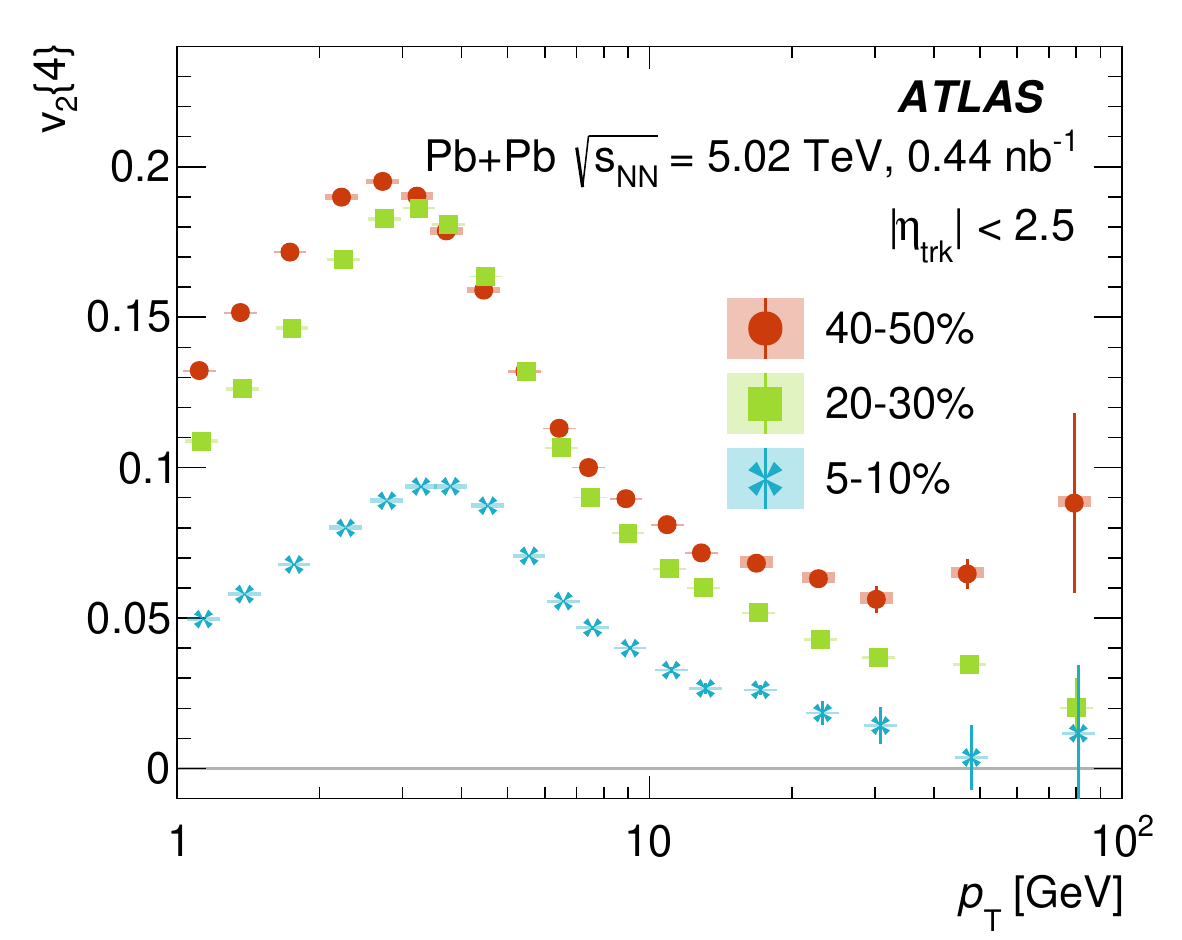}
}
\subfloat[]{
\centering
\includegraphics[width=0.48\textwidth]{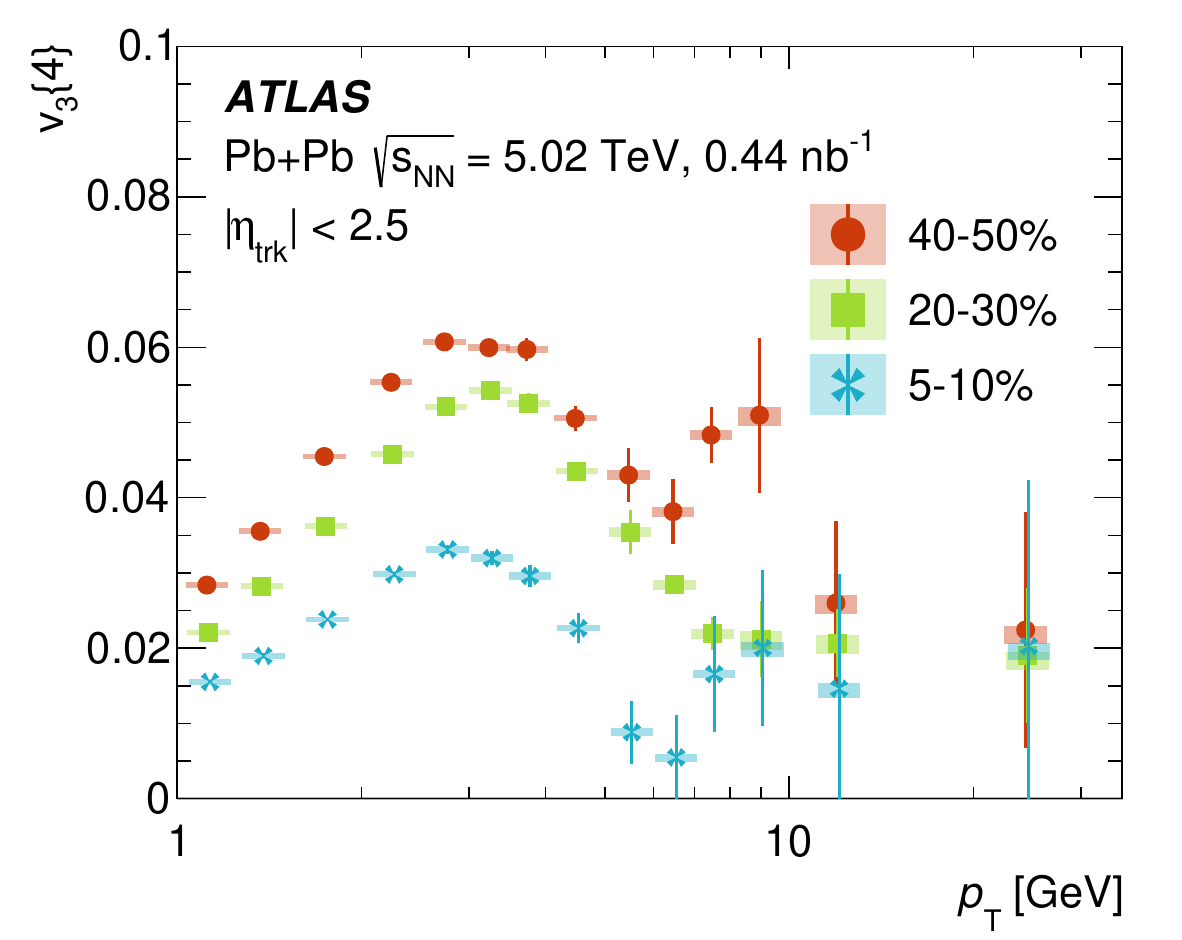}
}
\hfill
\caption{(a) \vnm{2}{4} and (b) \vnm{3}{4} using standard $Q$ cumulants for the centrality intervals 5--10\%, 20--30\%, and 40--50\%.
The statistical uncertainties are shown as error bars and the systematic uncertainties are shown as boxes.}
\label{fig:vn4}
\end{figure*}

Measurements of \vnm{n}{4} with the MPC method with standard $Q$ cumulants using the \ec{1} event combination procedure are shown in Figure~\ref{fig:vn4}.
Due to statistical limitations, \vnm{3}{4} is measured only up to the 50\% most central events.
As explained in Section~\ref{subsub:std_qc}, 0--5\% central events are omitted for \vnm{2}{4} to avoid imaginary-valued reference \vnm{2}{4}.
At high \pt, above 10 GeV, the \vnm{2}{4} values for centrality intervals within the 5--40\% most central events decrease with \pt\ across the measured range.
In contrast, for the 40--50\%~(peripheral events), values of \vnm{2}{4} decrease with \pT\ up to approximately 30~\GeV\
and then become approximately constant with \pt.
In the centrality interval 20--30\%, the values of \vnm{3}{4} become approximately constant above 8~\GeV, while with decreased statistics, bigger fluctuations are observed for 5--10\% and 40--50\% centrality intervals above approximately 8~\GeV.

\begin{figure*}[ht!]
\centering
\subfloat[]{
\centering
\includegraphics[width=0.48\textwidth]{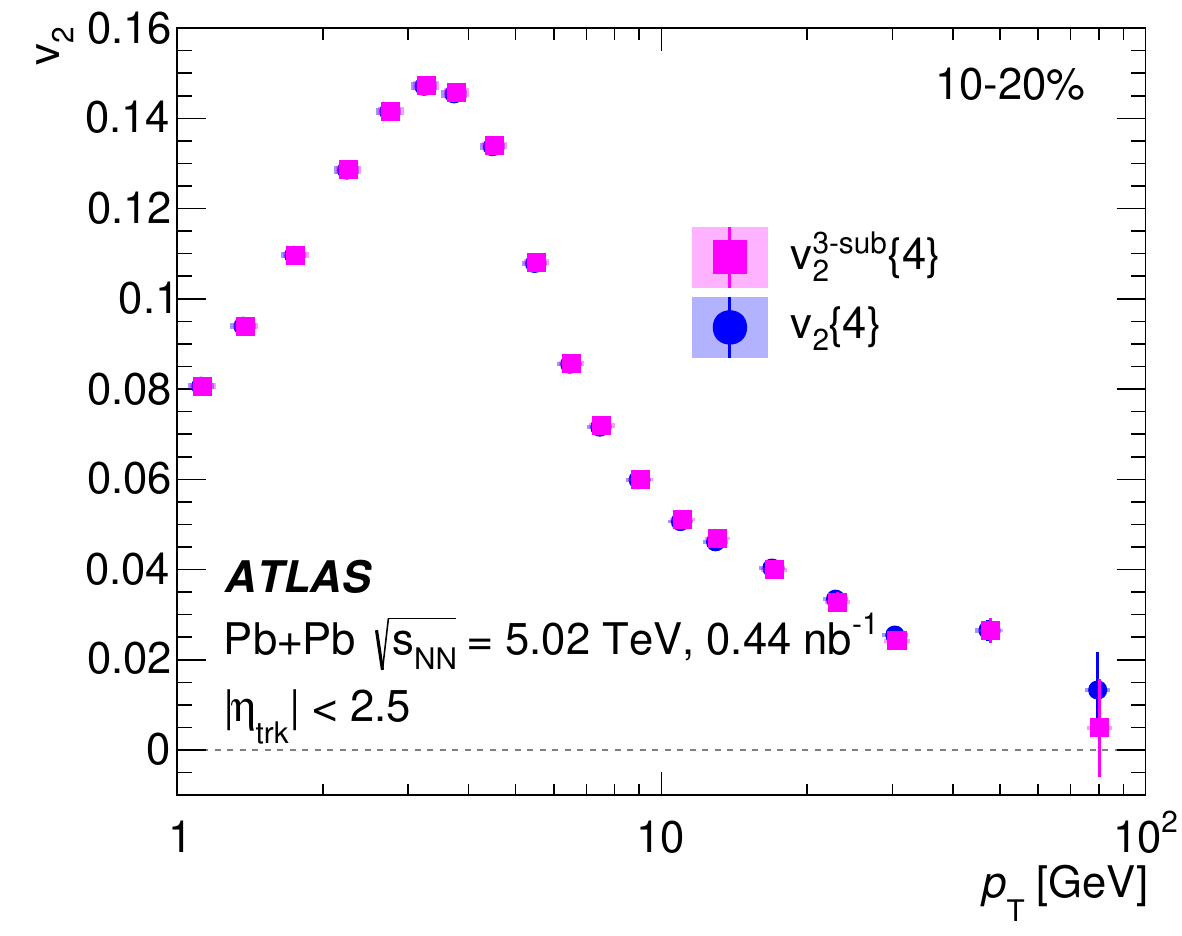}
}
\subfloat[]{
\centering
\includegraphics[width=0.48\textwidth]{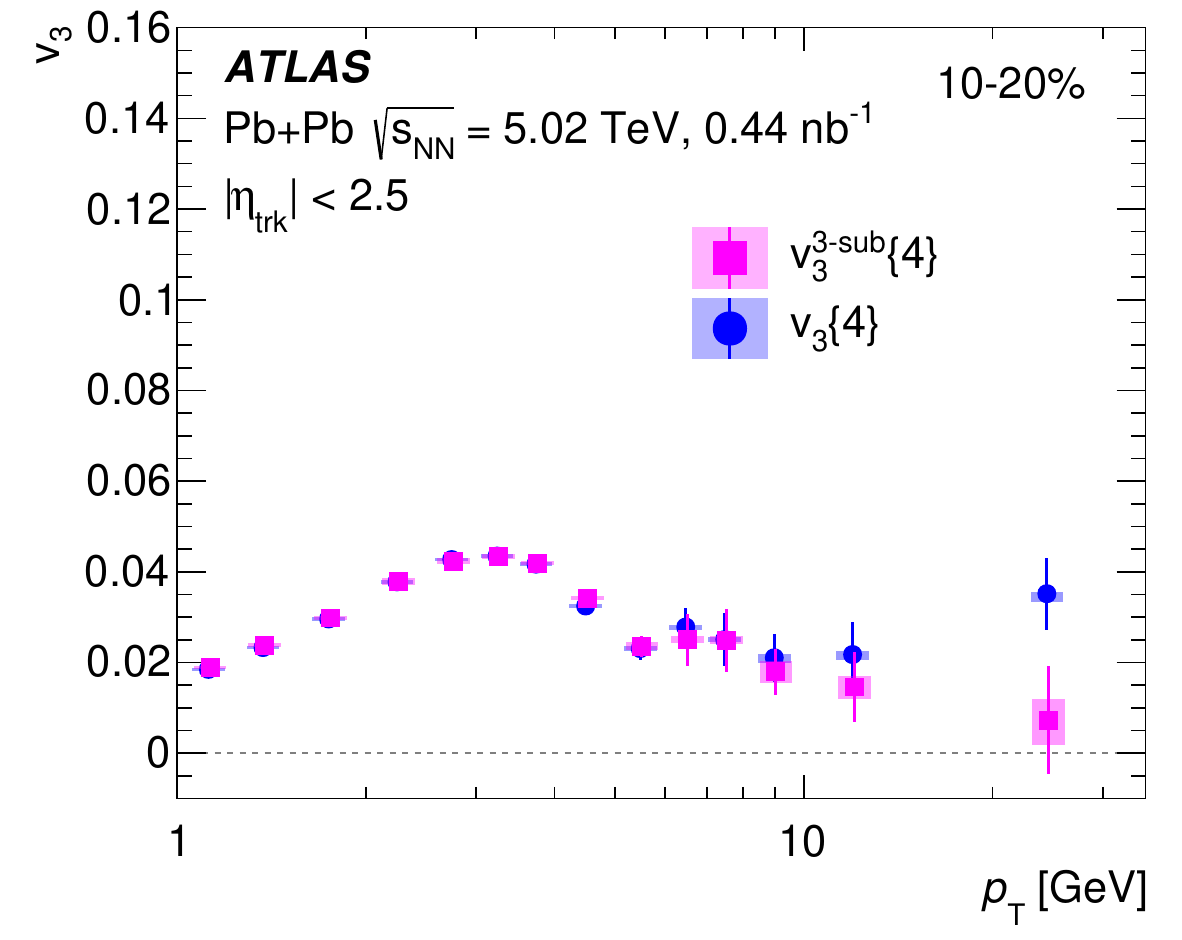}
}
\hfill
\centering
\subfloat[]{
\centering
\includegraphics[width=0.48\textwidth]{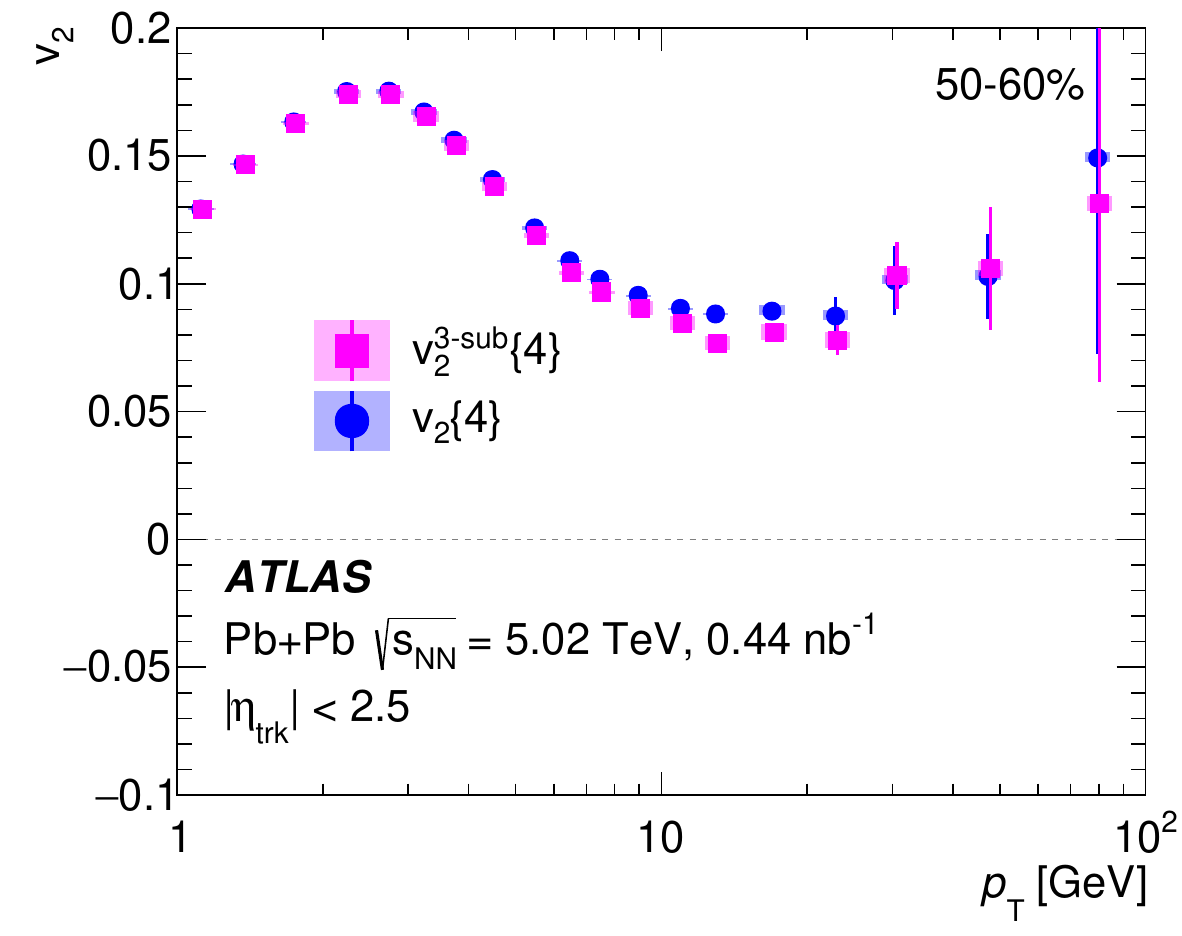}
}
\subfloat[]{
\centering
\includegraphics[width=0.48\textwidth]{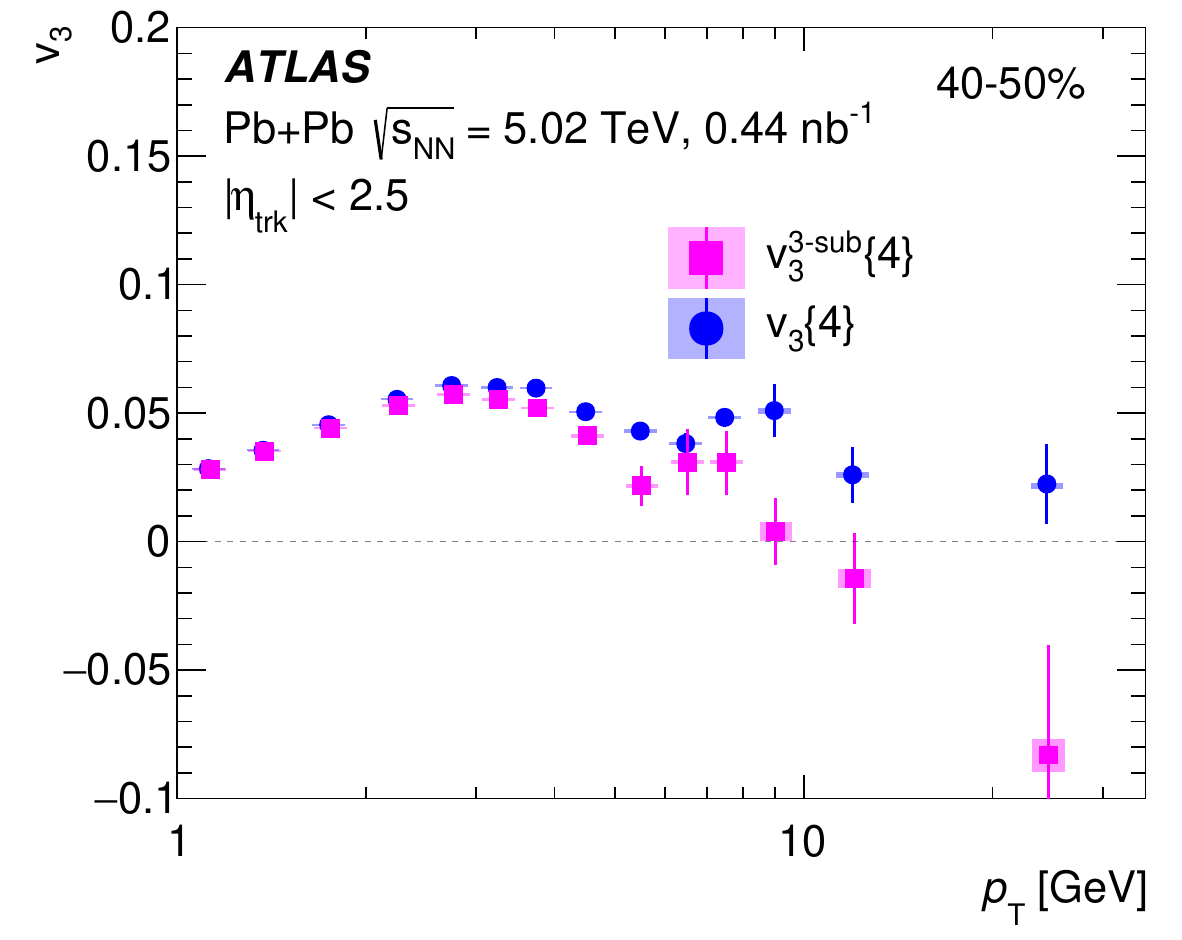}
}
\hfill
\caption{Comparison between standard and three-subevent $Q$ cumulants for (a) \vnm{2}{4} and (b) \vnm{3}{4} in the centrality interval 10--20\%, and for (c) \vnm{2}{4} in the centrality interval 50--60\% and (d) \vnm{3}{4} in the centrality interval 40--50\%. The statistical uncertainties are shown as error bars and the systematic uncertainties are shown as boxes.}
\label{fig:3-sub}
\end{figure*}

The flattening and slight increase of \vn{n} with \pt\ in peripheral events could arise from nonflow effects.
A comparison is made between the standard $Q$-cumulant result \vnm{n}{4} and three-subevent $Q$-cumulant results $\vnmsub{n}{4}{\text{3-sub}}$ to further understand the role of nonflow effects.
Note that $\vnmsub{3}{4}{\text{3-sub}}$ values are only measured up to $\pt=35$~\GeV and the most central 50\% collisions due to statistical limitations.
As shown in Figure~\ref{fig:3-sub}, values of $\vnmsub{n}{4}{\text{3-sub}}$ agree with \vnm{n}{4} for central events,
and deviate from \vnm{n}{4} in peripheral events at high \pt.
The \pt\ and centrality dependence of this deviation is consistent with a suppression of short-range nonflow correlations by the three-subevent cumulants.
As has been shown in Ref.~\cite{Huo:2017nms}, these nonflow correlations have a more significant effect in events of smaller charged-particle multiplicity.
Different behaviors are observed for $\vnmsub{2}{4}{\text{3-sub}}$ and $\vnmsub{3}{4}{\text{3-sub}}$ in the most peripheral centrality intervals available.
For $\pt>20$~\GeV, the values of $\vnmsub{2}{4}{\text{3-sub}}$ become nearly constant,
whereas the values of $\vnmsub{3}{4}{\text{3-sub}}$ continue to decrease for higher \pT.
As in the cases of \vnsp{n}, the contrasting behavior between \vn{2} and \vn{3} in the high-\pT\ region is consistent with contamination from back-to-back dijet production,
which contributes positively to even-order harmonics and negatively to odd-order harmonics and has a bigger impact in more peripheral events.
In these measurements, dijet contributions are more pronounced in the MPC method than the SP method, and can arise from the narrower pseudorapidity range used for the MPC method.

\begin{figure*}[ht!]
\centering
\subfloat[]{
\centering
\includegraphics[width=0.48\textwidth]{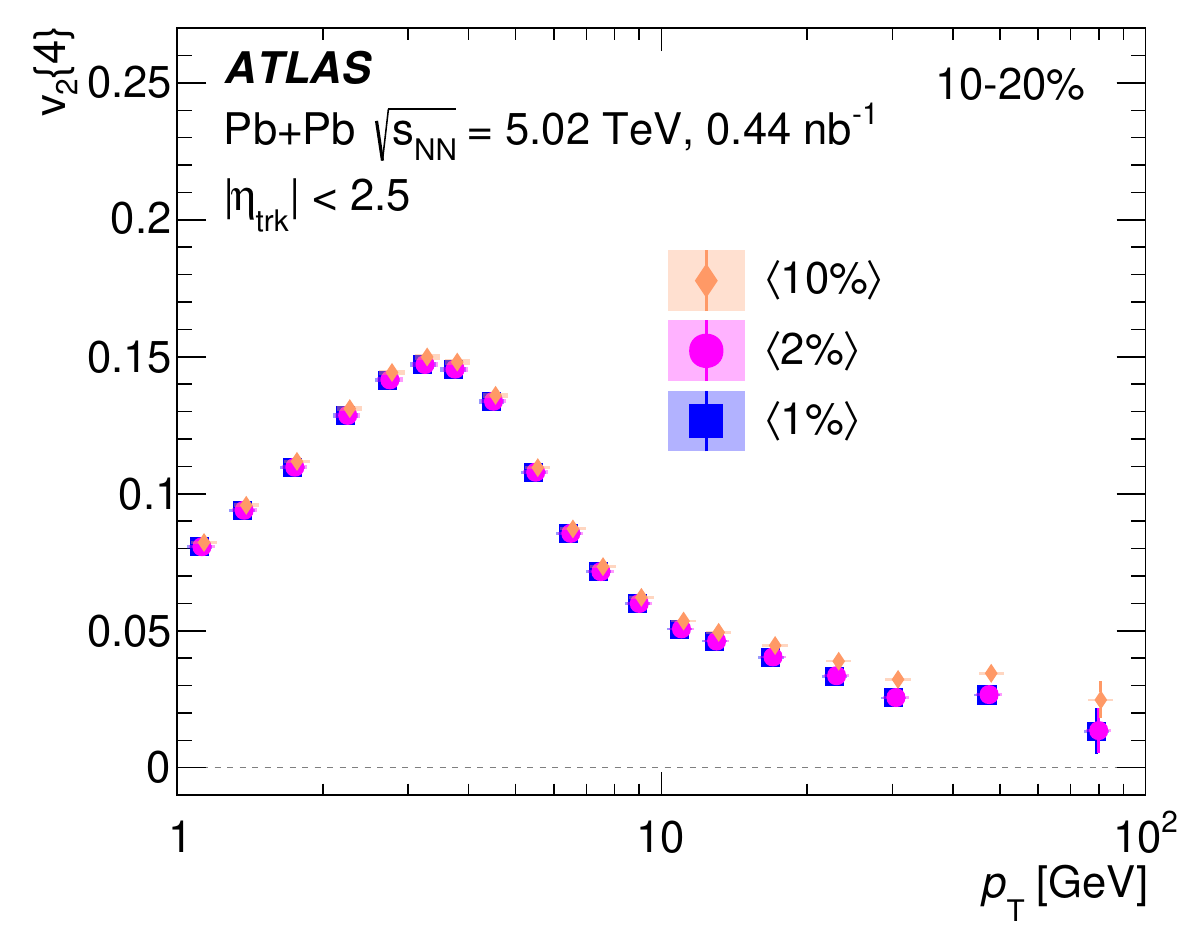}
}
\subfloat[]{
\centering
\includegraphics[width=0.48\textwidth]{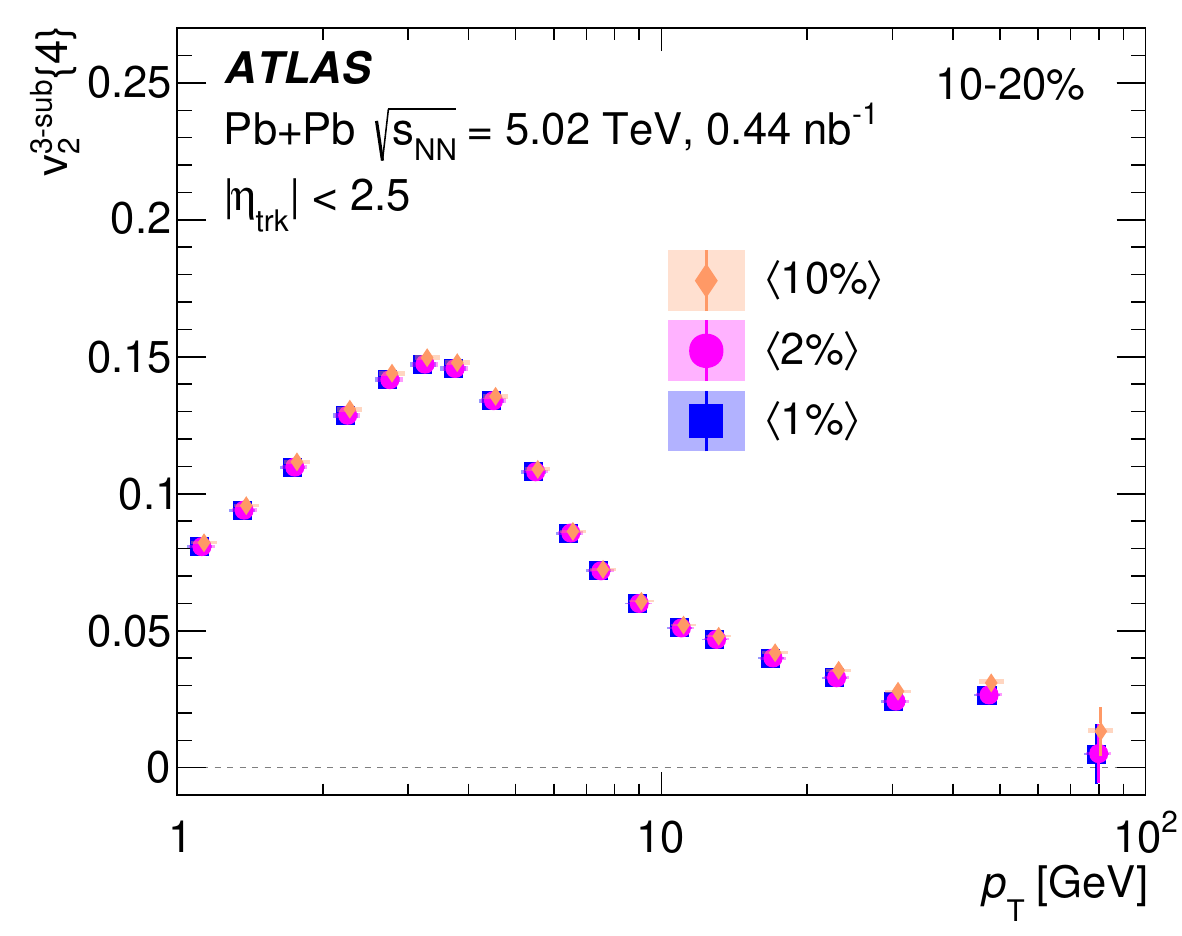}
}
\hfill
\centering
\subfloat[]{
\centering
\includegraphics[width=0.48\textwidth]{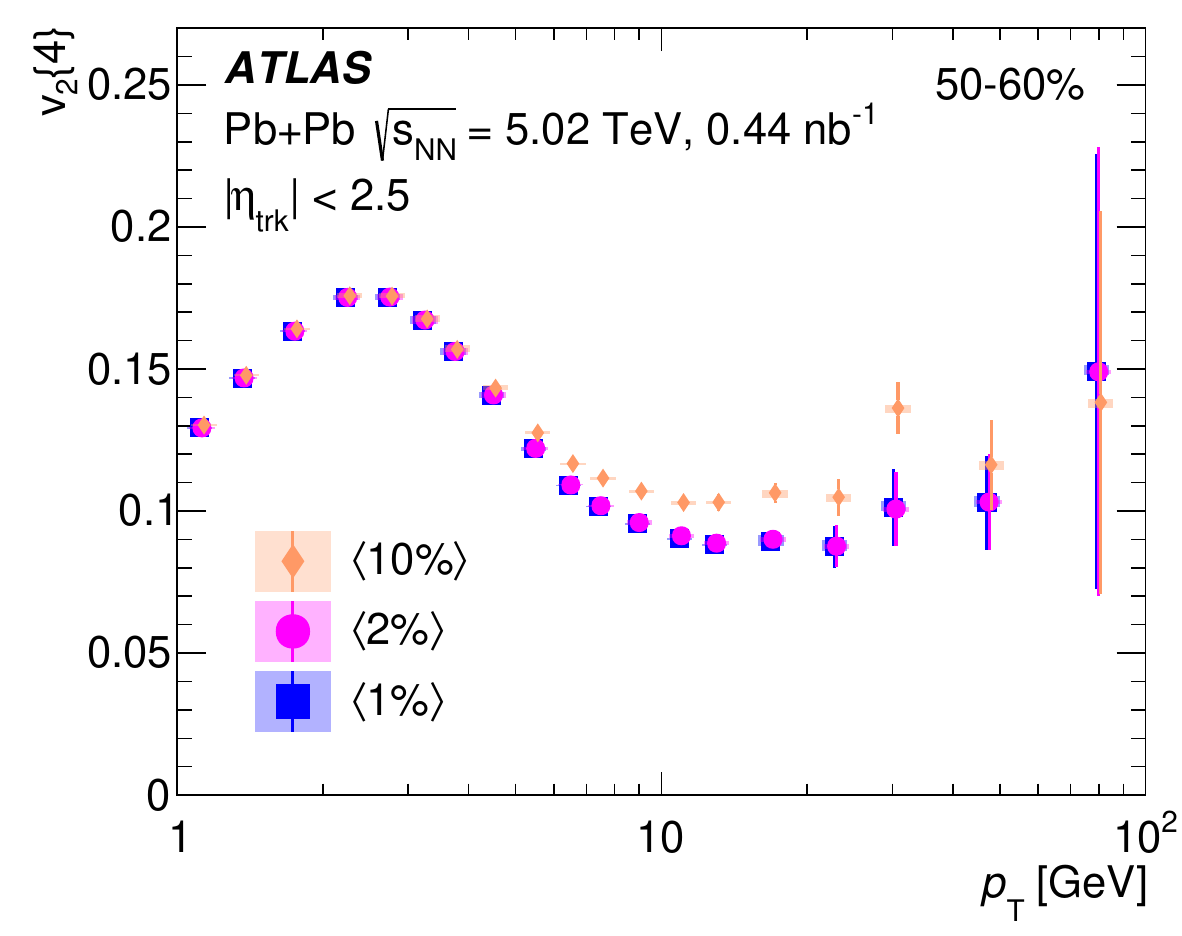}
}
\subfloat[]{
\centering
\includegraphics[width=0.48\textwidth]{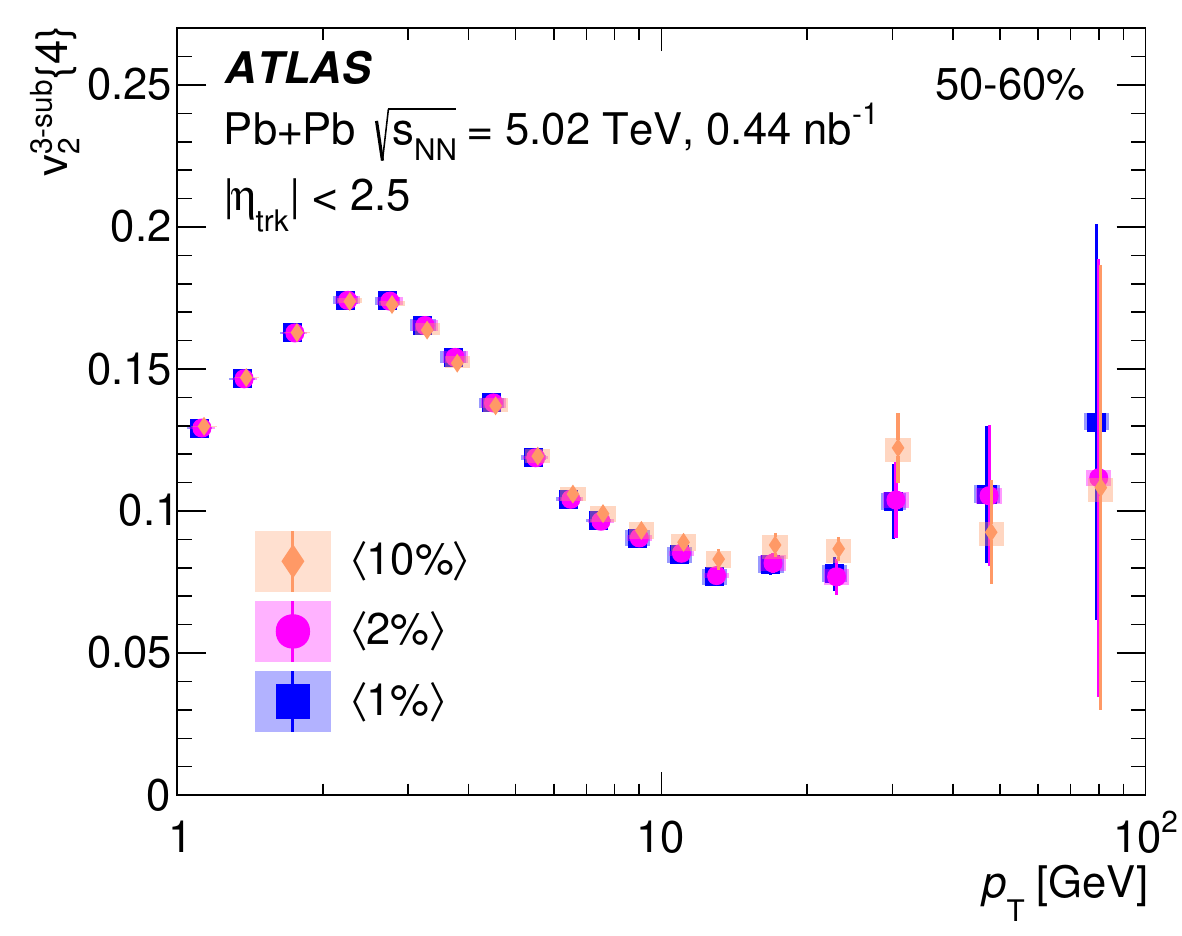}
}
\hfill
\caption{Comparison between the \ec{1}, \ec{2}, and \ec{10} event combination procedures in the centrality interval 10--20\% for (a) \vnm{2}{4} and (b) \vnmsub{2}{4}{\text{3-sub}}, and in the centrality interval 50--60\% for (c) \vnm{2}{4} and (d) \vnmsub{2}{4}{\text{3-sub}}. The statistical uncertainties are shown as error bars and the systematic uncertainties are shown as boxes.}
\label{fig:comp_mpc_evtclassv2}
\end{figure*}

\begin{figure*}[ht!]
\centering
\subfloat[]{
\centering
\includegraphics[width=0.48\textwidth]{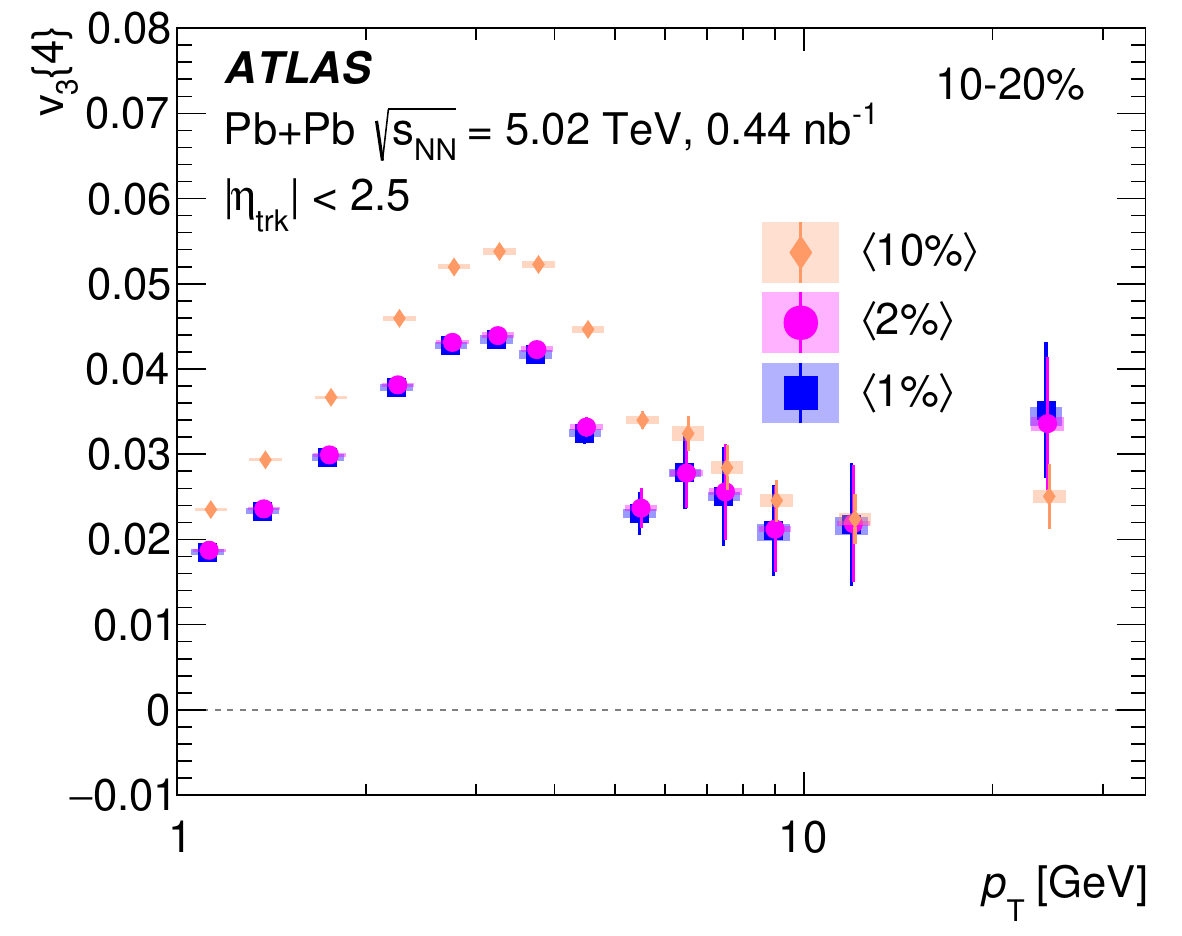}
}
\subfloat[]{
\centering
\includegraphics[width=0.48\textwidth]{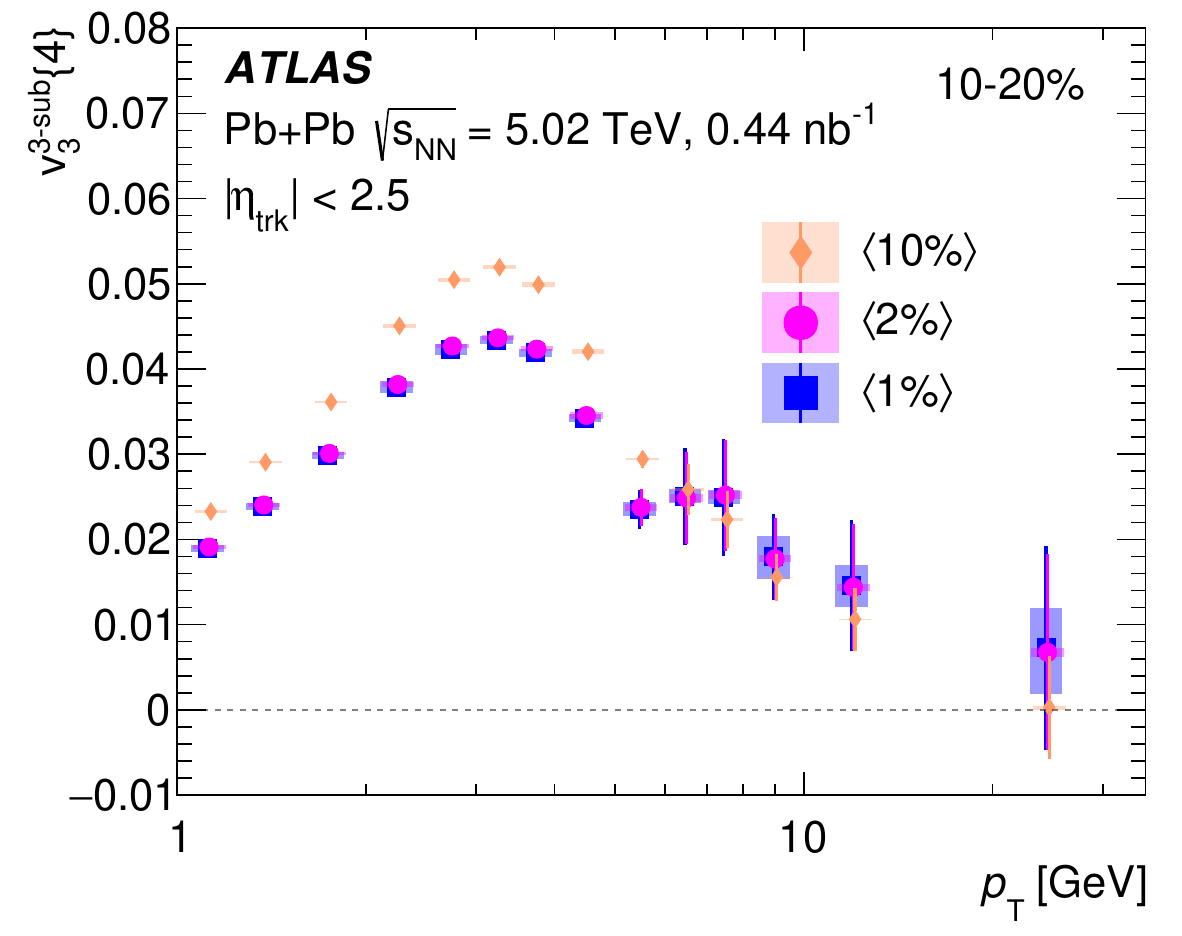}
}
\hfill
\centering
\subfloat[]{
\centering
\includegraphics[width=0.48\textwidth]{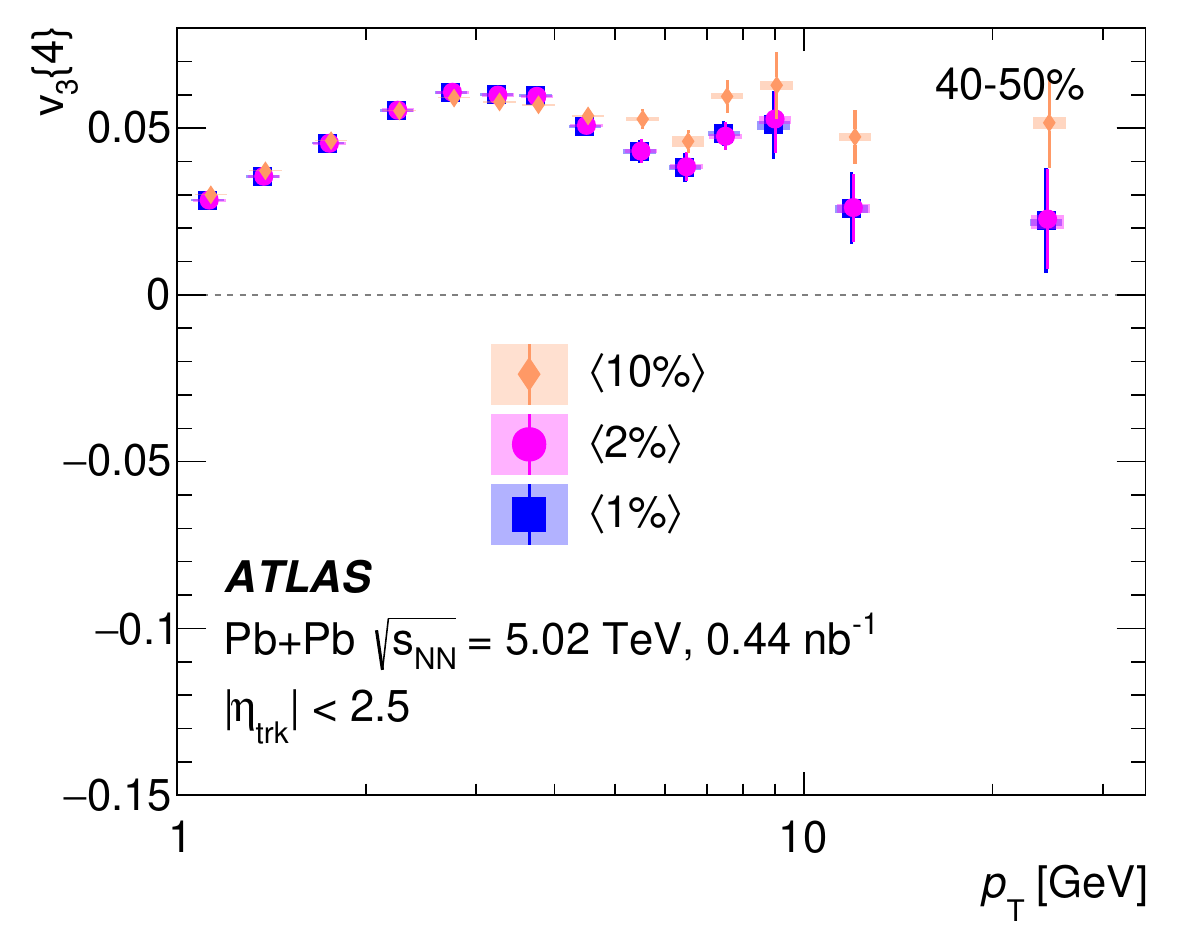}
}
\subfloat[]{
\centering
\includegraphics[width=0.48\textwidth]{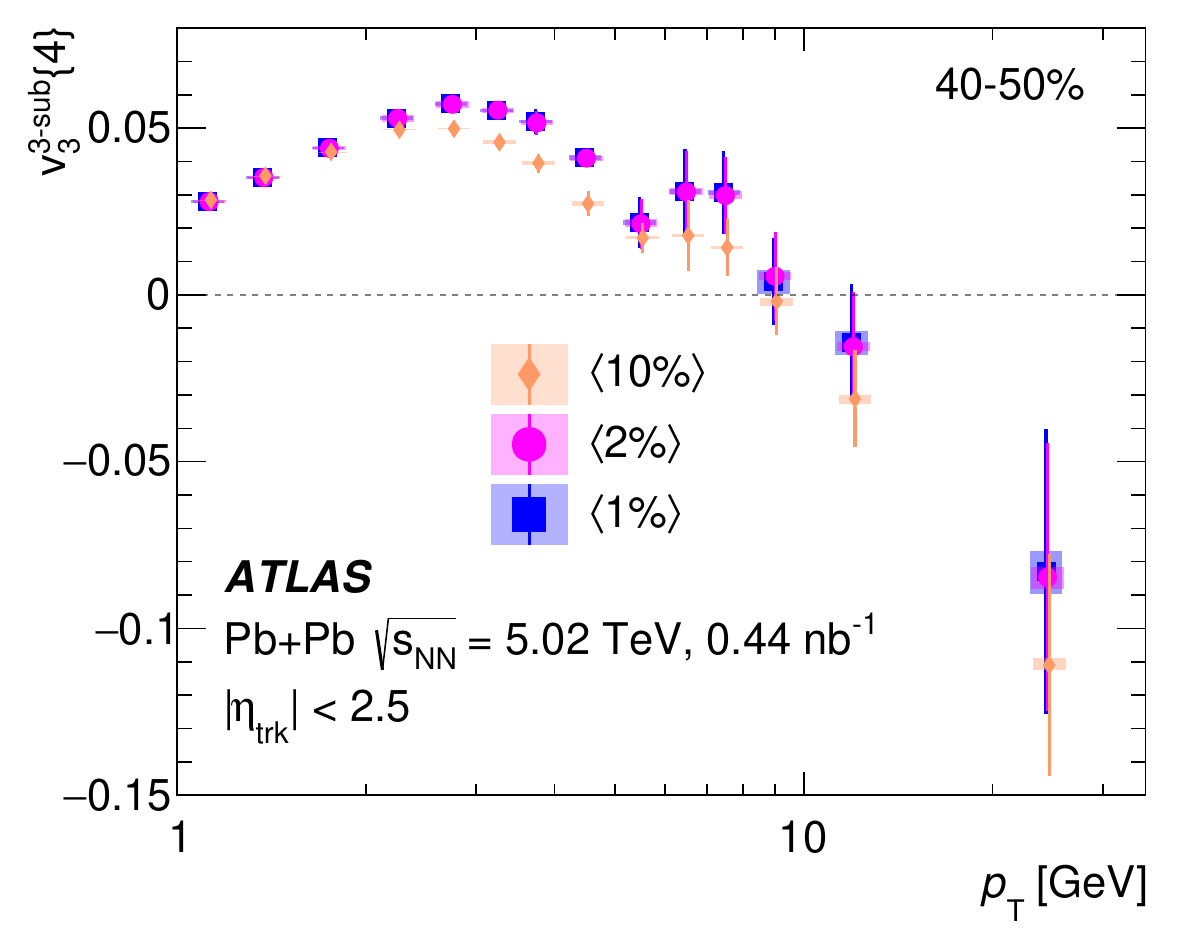}
}
\hfill
\caption{Comparison between the \ec{1}, \ec{2}, and \ec{10} event combination procedures in the centrality interval 10--20\% for (a) \vnm{3}{4} and (b) \vnmsub{3}{4}{\text{3-sub}}, and in the centrality interval 40--50\% for (c) \vnm{3}{4} and (d) \vnmsub{3}{4}{\text{3-sub}}. The statistical uncertainties are shown as error bars and the systematic uncertainties are shown as boxes.}
\label{fig:comp_mpc_evtclassv3}
\end{figure*}

The \vn{n} values measured with four-particle cumulants are sensitive to the mean value as well as fluctuations of the underlying distributions,
which includes contributions from both nonflow effects and initial-state geometry \cite{Borghini:2001vi}.
To better understand how the event-by-event distributions of nonflow effects and the initial-state geometry contribute to the measurements of \vn{n}, comparisons of \vnm{n}{4} using different event combination procedures are
shown in Figures~\ref{fig:comp_mpc_evtclassv2} and~\ref{fig:comp_mpc_evtclassv3}.
In all centrality and \pT\ ranges measured, the \ec{2} results converge to those obtained with the default \ec{1} procedure, indicating that centrality-resolution-related fluctuations are negligible in these measurements when using the \ec{2} or \ec{1} procedure. Therefore, the \ec{1} procedure results serve as an appropriate baseline for comparison.
In the low-\pt\ region, the event combination procedure affects the fluctuation terms but not the mean value of the underlying distribution,
whereas in the high-\pt\ region, computing cumulants directly in a wider centrality interval also biases the mean of measurements toward more central events.

Figure~\ref{fig:comp_mpc_evtclassv2} shows that for \vn{2} the impact of event combination procedure is dependent on centrality, \pt, and less pronounced in three-subevent cumulants.
In both of the quoted centrality intervals, the \ec{10} procedure shows a greater discrepancy from the \ec{1} procedure for \vnm{2}{4} than for \vnmsub{n}{4}{\text{3-sub}}, and the difference between \vnm{2}{4} and \vnmsub{n}{4}{\text{3-sub}} is more pronounced in peripheral events.
With the application of three-subevent cumulants, the discrepancy between event combination procedures is vanishingly small.
This discrepancy shows a centrality and \pT\ dependence, as well as a sensitivity to the three-subevent cumulants similar to that observed for nonflow effects,
suggesting the four-particle cumulant \vn{2} is primarily sensitive to fluctuations in nonflow effects.
Similar observations have been made in studies of low-\pt\ flow~\cite{ATLAS:2019peb,Jia:2017hbm}.

Figure~\ref{fig:comp_mpc_evtclassv3} shows a comparison between different event combination procedures for \vnm{3}{4} and \vnmsub{3}{4}{\text{3-sub}}.
It can be seen that the discrepancy between \ec{10} and \ec{1} is much more substantial than \vn{2}
in the quoted centrality intervals for both the standard and the three-subevent cumulants, the latter of which have been shown to suppress short-range nonflow effects in Figure~\ref{fig:3-sub}.
These observations show that \vn{3} values are more sensitive to fluctuations in the initial geometry than \vn{2}.
It has been found in the measurements of reference flow that the fluctuations of \vn{3} are more significant relative to the mean in comparison to \vn{2}~\cite{ATLAS:2013xzf}.
Results from Figures~\ref{fig:comp_mpc_evtclassv2} and~\ref{fig:comp_mpc_evtclassv3} %
suggest that event-by-event fluctuations are more pronounced in the triangularity of the initial geometry than the ellipticity.
The discrepancy among event combination procedures decreases in the 40--50\% centrality interval, where the centrality dependence of \vnm{3}{4} has been observed to flatten~\cite{ATLAS:2019peb}.
Different orderings of the \vn{3} values in the event combination procedure are observed between \vnm{3}{4} and \vnmsub{3}{4}{\text{3-sub}}, particularly at high \pT.
This can arise from the suppression of short-rage nonflow correlations and residual back-to-back dijet contributions in the three-subevent $Q$ cumulants.

\begin{figure*}[ht!]
\centering
\subfloat[]{
\centering
\includegraphics[width=0.48\textwidth]{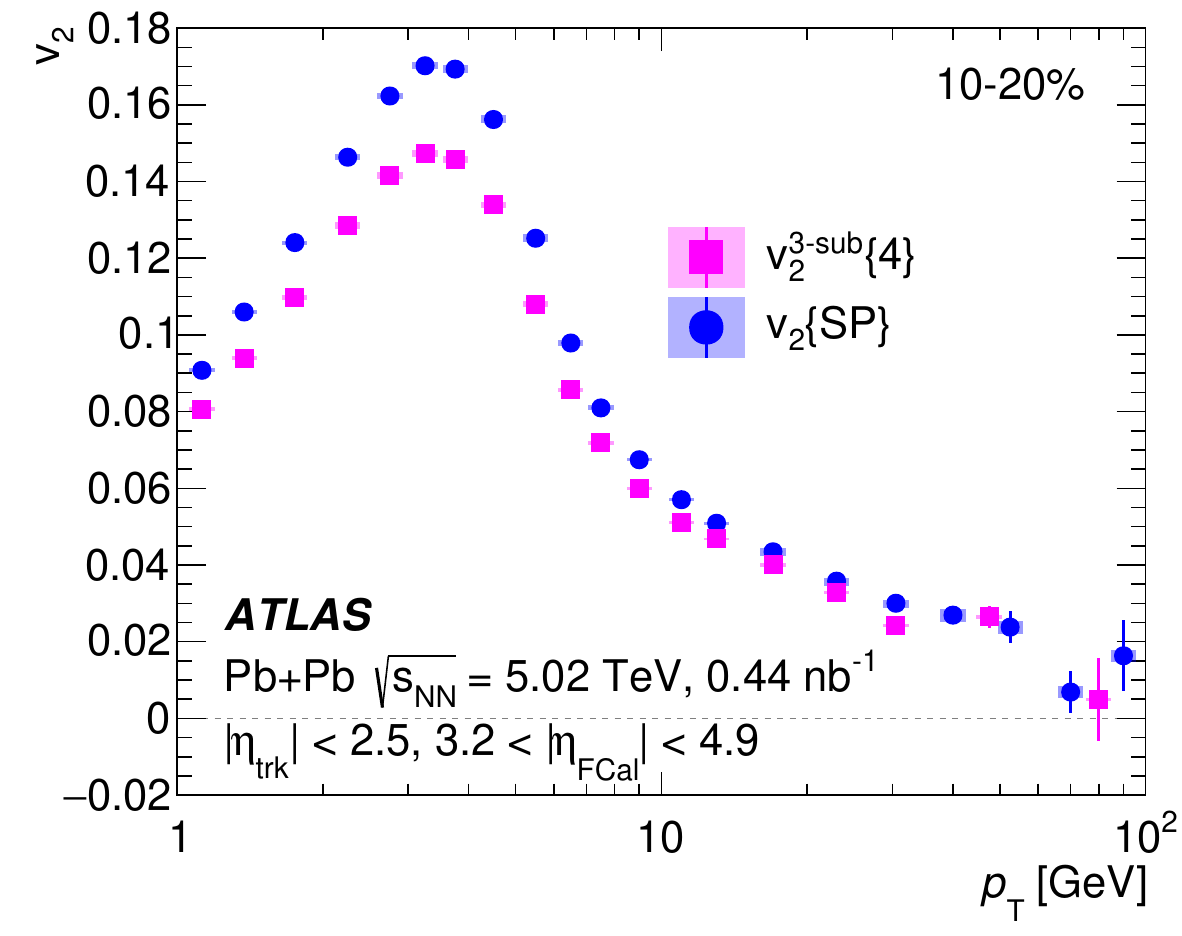}
}
\subfloat[]{
\centering
\includegraphics[width=0.48\textwidth]{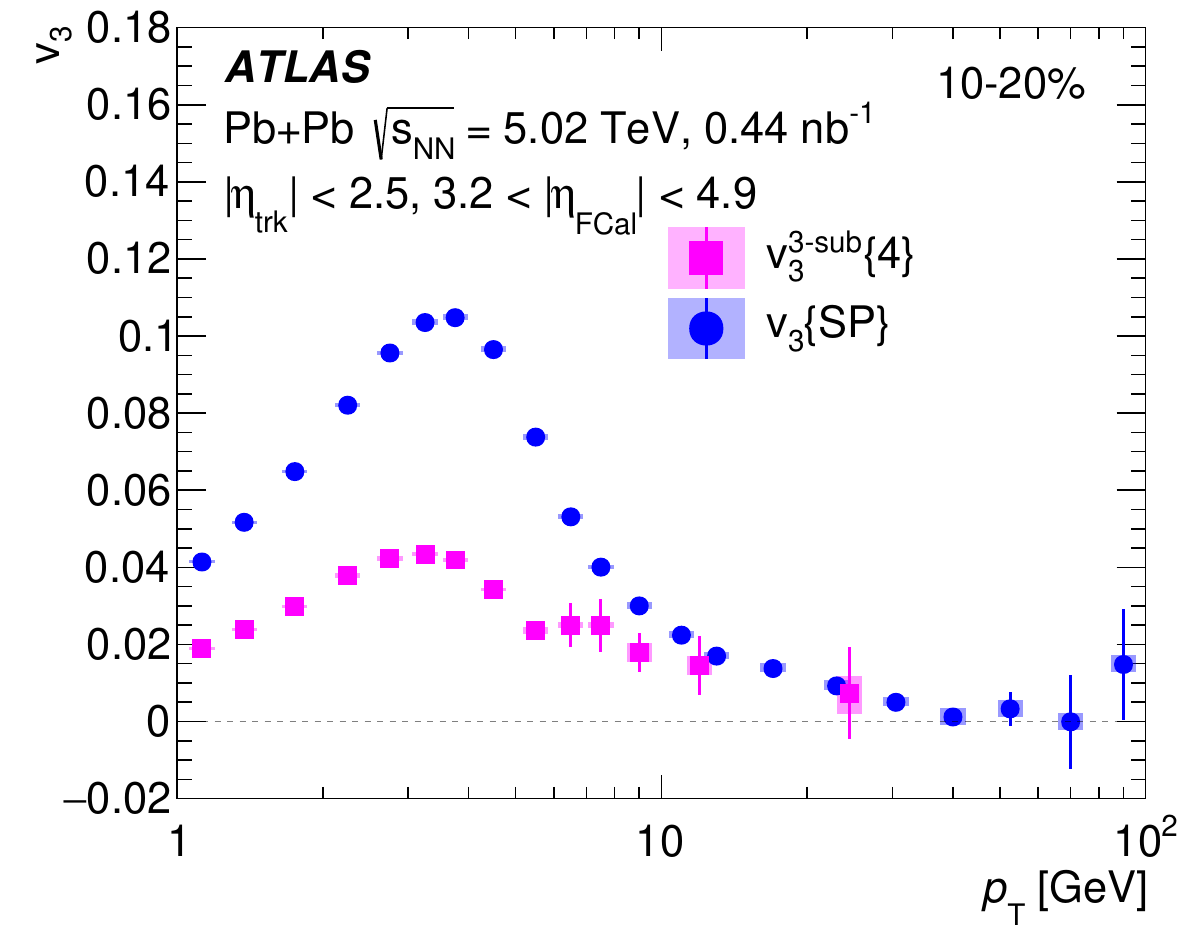}
}
\hfill
\centering
\subfloat[]{
\centering
\includegraphics[width=0.48\textwidth]{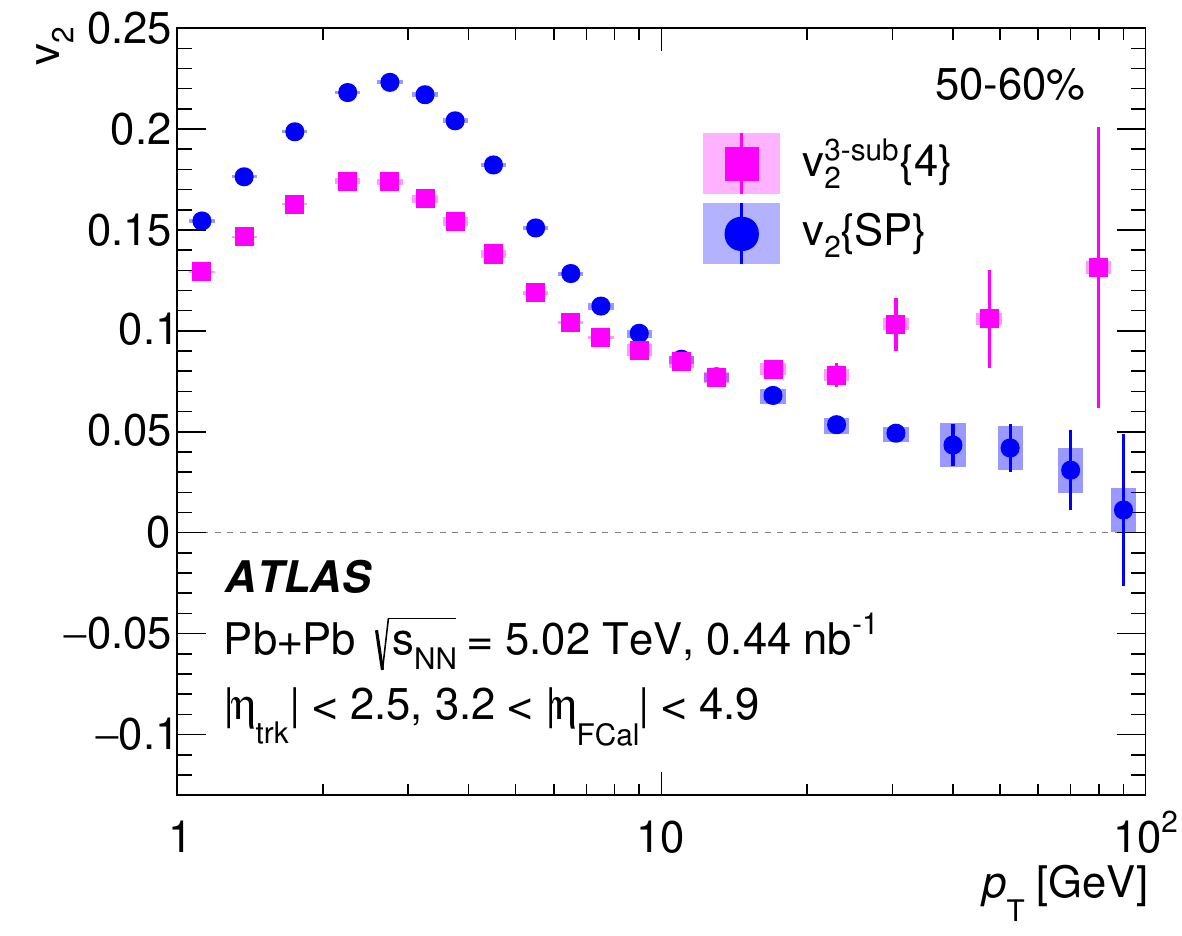}
}
\subfloat[]{
\centering
\includegraphics[width=0.48\textwidth]{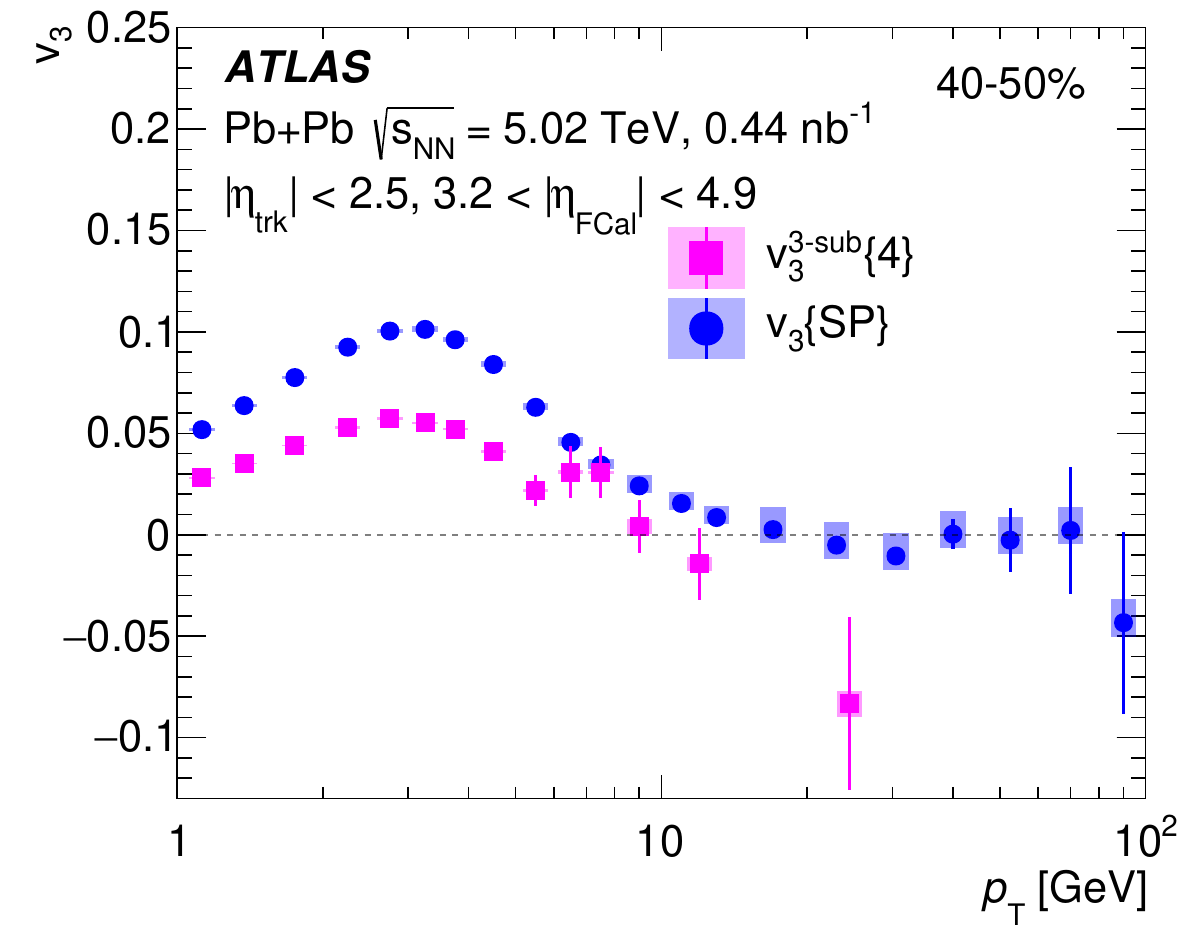}
}
\hfill
\caption{Comparison between the \vnsp{n} results and \vnmsub{n}{4}{\text{3-sub}} in centrality intervals 10--20\% for (a) $n=2$ and (b) $n=3$, and (c) in centrality intervals 50--60\% for $n=2$ and (d) in centrality intervals 40--50\% for $n=3$. The statistical uncertainties are shown as error bars and the systematic uncertainties are shown as boxes.}
\label{fig:comp_sp_mpc}
\end{figure*}

Figure~\ref{fig:comp_sp_mpc} shows a comparison between the SP method and the three-subevent $Q$-cumulant MPC method using the default \ec{1} event combination procedure.
The \vnmsub{3}{4}{\text{3-sub}} results are truncated at a \pt\ of 35~\GeV\ due to limited statistics.
Values of \vnsp{n} are greater than \vnmsub{n}{4}{\text{3-sub}}, similar to the comparison observed in \vn{n} measurements in \pbpb\ collisions at $\sqrtsnn=2.76$~\TeV\ \cite{ATLAS:2014qxy}.
At $\pt>10$~\GeV, the comparison between the two methods contains information about how hard scattered partons
respond to the event-by-event distribution of the initial-state QGP geometry.
Different behaviors are observed for different centrality intervals.
In central events, the relation $\vnsp{n} > \vnmsub{n}{4}{\text{3-sub}}$ holds up to \pt\ of approximately $15$~\GeV\ for both \vn{2} and \vn{3}.
Whereas in peripheral events, different trends are observed between \vn{2} and \vn{3} toward the high \pT~region.
In the centrality interval 50--60\%, the two methods converge with increasing \pt\ and then intersect around approximately \SI{15}{\GeV} for \vn{2}, above which the two methods diverge with $\vnmsub{2}{4}{\text{3-sub}}>\vnsp{2}$, whereas for \vn{3}, the methods diverge from each other for $\pt > 10$~\GeV, and always exhibiting $\vnsp{3}>\vnmsub{3}{4}{\text{3-sub}}$.
This difference between \vn{2} and \vn{3} suggests residual nonflow dijet contributions, which have greater impact in peripheral events.
It should be noted that multiple effects could contribute to the observed trends in peripheral events, such as other nonflow correlations, the effects of the event combination procedure, and event plane longitudinal decorrelation.

\FloatBarrier

\newpage

% End of text imported from the .//ANA-HION-2023-04-PAPER-result.tex input file

%

\FloatBarrier

%
%
%

%
%
%
%

% The next lines are included from the .//ANA-HION-2023-04-PAPER-conclusions.tex input file
\newpage
\section{Conclusions}
\label{sec:conclusion}

This paper describes the measurements of azimuthal anisotropies of charged particles in \pbpb\ collisions at $\sqrtsnn=5.02$~\TeV, using a dataset corresponding to an integrated luminosity of \PEBlum\ collected with the ATLAS detector at the LHC in 2018.
The azimuthal anisotropy coefficients, \vn{2} and \vn{3}, are measured
using the scalar-product and multiparticle cumulant methods over the charged particles with \pt\ up to 400~\GeV\ in several centrality intervals spanning the 0--60\% most central events.
Additionally, different pseudorapidity ranges and strategies for \vn{n} values are compared.

The \pt-differential \vnsp{n} values are measured over a pseudorapidity range $|\eta| < 2.5$ and a \pt\ range of 1--400~\GeV\ in the 0--60\% most central collisions.
In mid-central collisions, the \vnsp{2} values are found to remain positive up to \pt\ of 100~\GeV\ with a value of 1--2\%; the \vnsp{3} values are found to be positive up to \pt\ of approximately 25~\GeV.
These results are consistent with previous measurements from the LHC experiments, and the increased luminosity and implemented data taking innovations resulted in a larger dataset that has allowed for improved precision at high \pt. Values of \vnsp{n} are also measured across two additional pseudorapidity ranges: $|\eta| < 1.1$ and $1.1 < |\eta| < 2.5$, over a \pt\ range of 1--100~\GeV.
In the low-\pt\ region, a larger pseudorapidity gap between the correlated flow vectors decreases both \vnsp{2} and \vnsp{3} values, which can arise from decreased nonflow effects and the increased longitudinal decorrelation.
By contrast, at high \pt, a larger pseudorapidity gap between the correlated flow vectors decreases \vnsp{2} but increases \vnsp{3} values, suggesting nonflow dijet contributions, especially in peripheral events.

Using the MPC method, the \pt-differential \vnm{2}{4} values are measured over a pseudorapidity range of $|\eta| < 2.5$ and a \pt\ up to 100~\GeV\ in the most central 0--60\% events.
For $\pt>10$~\GeV, it is observed that the three-subevent technique suppresses short-range nonflow contributions.
Meanwhile, nonflow dijet contributions are still observed with the three-subevent technique.
Measurements of \vnm{2}{4} and \vnm{3}{4} using two additional event combination procedures, \ec{2} and \ec{10}, were also performed and compared to results using the default \ec{1} procedure to understand the dominating source of flucutations in \vn{n} measurements.
It is observed that for \vnm{2}{4}, the event combination procedure affects nonflow contributions primarily,
whereas, for \vnm{3}{4}, initial geometry contributions are more significantly affected.

The comparison between the SP and MPC methods shows different trends between \vn{2} and \vn{3} in peripheral events.
For \vn{2}, the difference between the SP and the MPC methods decreases toward zero up to \pt\ of 15~\GeV, then flips sign for $\pT>15$~\GeV. For \vn{3}, however, this difference continues to increase with \pT\ with the same sign. The contrasting behavior between \vn{2} and \vn{3} indicates remaining nonflow dijet contributions, which contribute to the SP and the MPC methods differently.

These results provide comprehensive information on the mechanism and fluctuations of azimuthal anisotropies of hard-scattered particles in the \pbpb\ collision system. The positive \vn{2} and \vn{3} values observed suggest that the energy loss of the hard-scattered partons
is influenced by the event-by-event distribution of the initial geometry of the QGP, and these values can be used to constrain the path-length dependence of jet quenching.
The comparison studies presented provide key insights on the different contributions of the various nonflow sources, as well as revealing the different sources of fluctuations that dominate in \vn{2} and \vn{3}.

% End of text imported from the .//ANA-HION-2023-04-PAPER-conclusions.tex input file

%
%
%
%
%
\section*{Acknowledgements}
%

%

% The next lines are included from the .//acknowledgements/Acknowledgements.tex input file
%
%

%
%

We thank CERN for the very successful operation of the LHC and its injectors, as well as the support staff at
CERN and at our institutions worldwide without whom ATLAS could not be operated efficiently.

The crucial computing support from all WLCG partners is acknowledged gratefully, in particular from CERN, the ATLAS Tier-1 facilities at TRIUMF/SFU (Canada), NDGF (Denmark, Norway, Sweden), CC-IN2P3 (France), KIT/GridKA (Germany), INFN-CNAF (Italy), NL-T1 (Netherlands), PIC (Spain), RAL (UK) and BNL (USA), the Tier-2 facilities worldwide and large non-WLCG resource providers. Major contributors of computing resources are listed in Ref.~\cite{ATL-SOFT-PUB-2025-001}.

We gratefully acknowledge the support of ANPCyT, Argentina; YerPhI, Armenia; ARC, Australia; BMWFW and FWF, Austria; ANAS, Azerbaijan; CNPq and FAPESP, Brazil; NSERC, NRC and CFI, Canada; CERN; ANID, Chile; CAS, MOST and NSFC, China; Minciencias, Colombia; MEYS CR, Czech Republic; DNRF and DNSRC, Denmark; IN2P3-CNRS and CEA-DRF/IRFU, France; SRNSFG, Georgia; BMFTR, HGF and MPG, Germany; GSRI, Greece; RGC and Hong Kong SAR, China; ICHEP and Academy of Sciences and Humanities, Israel; INFN, Italy; MEXT and JSPS, Japan; CNRST, Morocco; NWO, Netherlands; RCN, Norway; MNiSW, Poland; FCT, Portugal; MNE/IFA, Romania; MSTDI, Serbia; MSSR, Slovakia; ARIS and MVZI, Slovenia; DSI/NRF, South Africa; MICIU/AEI, Spain; SRC and Wallenberg Foundation, Sweden; SERI, SNSF and Cantons of Bern and Geneva, Switzerland; NSTC, Taipei; TENMAK, T\"urkiye; STFC/UKRI, United Kingdom; DOE and NSF, United States of America.

Individual groups and members have received support from BCKDF, CANARIE, CRC and DRAC, Canada; CERN-CZ, FORTE and PRIMUS, Czech Republic; COST, ERC, ERDF, Horizon 2020, ICSC-NextGenerationEU and Marie Sk{\l}odowska-Curie Actions, European Union; Investissements d'Avenir Labex, Investissements d'Avenir Idex and ANR, France; DFG and AvH Foundation, Germany; Herakleitos, Thales and Aristeia programmes co-financed by EU-ESF and the Greek NSRF, Greece; BSF-NSF and MINERVA, Israel; NCN and NAWA, Poland; La Caixa Banking Foundation, CERCA Programme Generalitat de Catalunya and PROMETEO and GenT Programmes Generalitat Valenciana, Spain; G\"{o}ran Gustafssons Stiftelse, Sweden; The Royal Society and Leverhulme Trust, United Kingdom.

In addition, individual members wish to acknowledge support from Armenia: Yerevan Physics Institute (FAPERJ); CERN: European Organization for Nuclear Research (CERN DOCT); Chile: Agencia Nacional de Investigaci\'on y Desarrollo (FONDECYT 1230812, FONDECYT 1240864); China: Chinese Ministry of Science and Technology (MOST-2023YFA1605700, MOST-2023YFA1609300), National Natural Science Foundation of China (NSFC - 12175119, NSFC 12275265, NSFC-12075060); Czech Republic: Czech Science Foundation (GACR - 24-11373S), Ministry of Education Youth and Sports (ERC-CZ-LL2327, FORTE CZ.02.01.01/00/22\_008/0004632), PRIMUS Research Programme (PRIMUS/21/SCI/017); EU: H2020 European Research Council (ERC - 101002463); European Union: European Research Council (BARD No. 101116429, ERC - 948254, ERC 101089007), Horizon 2020 Framework Programme (MUCCA - CHIST-ERA-19-XAI-00), European Union, Future Artificial Intelligence Research (FAIR-NextGenerationEU PE00000013), Italian Center for High Performance Computing, Big Data and Quantum Computing (ICSC, NextGenerationEU); France: Agence Nationale de la Recherche (ANR-20-CE31-0013, ANR-21-CE31-0013, ANR-21-CE31-0022, ANR-22-EDIR-0002); Germany: Baden-Württemberg Stiftung (BW Stiftung-Postdoc Eliteprogramme), Deutsche Forschungsgemeinschaft (DFG - 469666862, DFG - CR 312/5-2); Italy: Istituto Nazionale di Fisica Nucleare (ICSC, NextGenerationEU), Ministero dell'Università e della Ricerca (NextGenEU 153D23001490006 M4C2.1.1, NextGenEU I53D23000820006 M4C2.1.1, NextGenEU I53D23001490006 M4C2.1.1); Japan: Japan Society for the Promotion of Science (JSPS KAKENHI JP22H01227, JSPS KAKENHI JP22H04944, JSPS KAKENHI JP22KK0227, JSPS KAKENHI JP23KK0245); Norway: Research Council of Norway (RCN-314472); Poland: Ministry of Science and Higher Education (IDUB AGH, POB8, D4 no 9722), Polish National Agency for Academic Exchange (PPN/PPO/2020/1/00002/U/00001), Polish National Science Centre (NCN 2021/42/E/ST2/00350, NCN OPUS 2023/51/B/ST2/02507, NCN OPUS nr 2022/47/B/ST2/03059, NCN UMO-2019/34/E/ST2/00393, NCN \& H2020 MSCA 945339, UMO-2020/37/B/ST2/01043, UMO-2021/40/C/ST2/00187, UMO-2022/47/O/ST2/00148, UMO-2023/49/B/ST2/04085, UMO-2023/51/B/ST2/00920); Portugal: Foundation for Science and Technology (FCT); Spain: Generalitat Valenciana (Artemisa, FEDER, IDIFEDER/2018/048), Ministry of Science and Innovation (MCIN \& NextGenEU PCI2022-135018-2, MICIN \& FEDER PID2021-125273NB, RYC2019-028510-I, RYC2020-030254-I, RYC2021-031273-I, RYC2022-038164-I); Sweden: Carl Trygger Foundation (Carl Trygger Foundation CTS 22:2312), Swedish Research Council (Swedish Research Council 2023-04654, VR 2018-00482, VR 2021-03651, VR 2022-03845, VR 2022-04683, VR 2023-03403), Knut and Alice Wallenberg Foundation (KAW 2018.0458, KAW 2019.0447, KAW 2022.0358); Switzerland: Swiss National Science Foundation (SNSF - PCEFP2\_194658); United Kingdom: Leverhulme Trust (Leverhulme Trust RPG-2020-004), Royal Society (NIF-R1-231091); United States of America: U.S. Department of Energy (ECA DE-AC02-76SF00515), Neubauer Family Foundation.

%
%

% End of text imported from the .//acknowledgements/Acknowledgements.tex input file

%
\clearpage

\clearpage
% The next lines are included from the .//ANA-HION-2023-04-PAPER-appendix.tex input file
\appendix
\part*{Appendix}
\addcontentsline{toc}{part}{Appendix}
\section{Measurements and comparisons of \vn{n} in additional centrality intervals}
\label{sec:appendix}

\begin{figure*}[!ht]
\centering
\subfloat[]{
\centering
\includegraphics[width=0.48\textwidth]{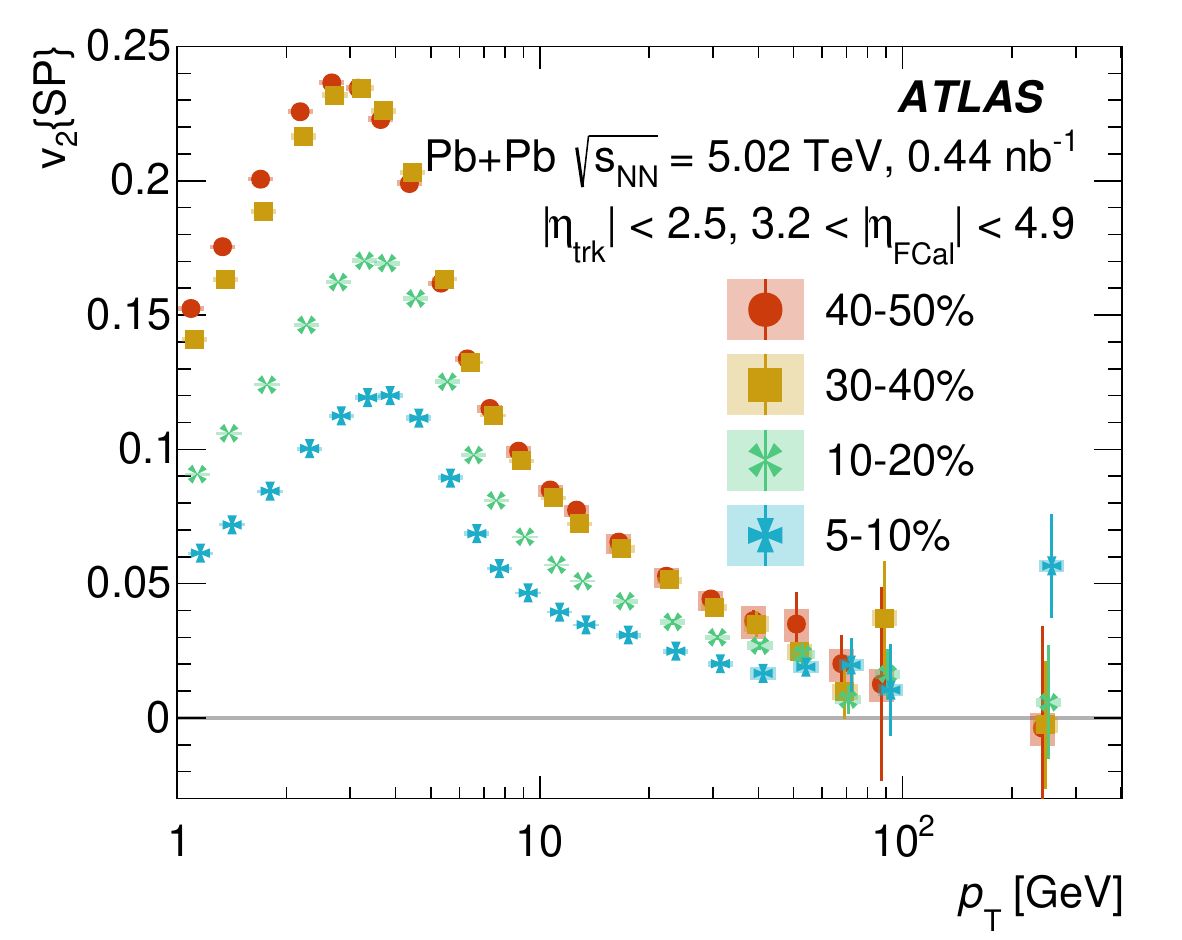}
}
\subfloat[]{
\centering
\includegraphics[width=0.48\textwidth]{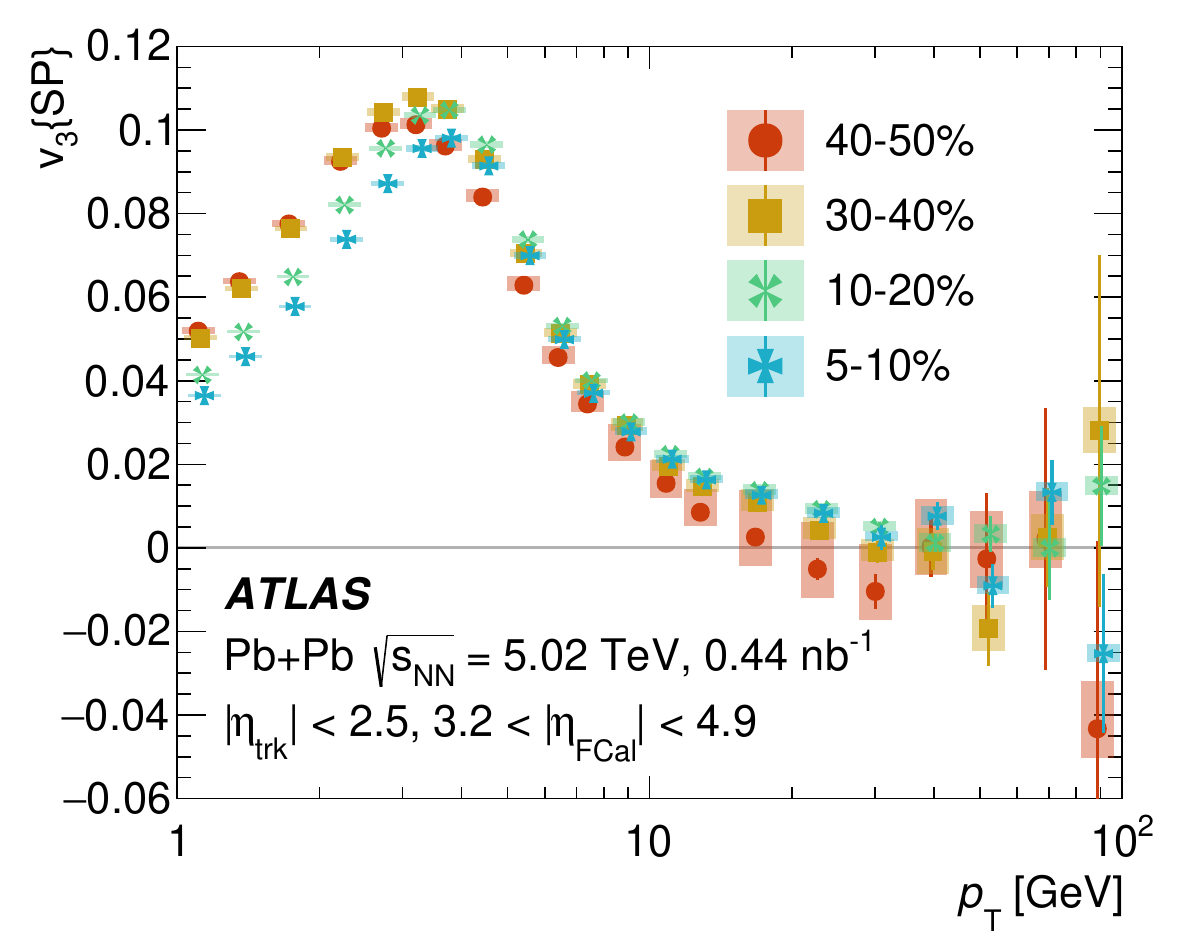}
}
\hfill
\caption{The (a) \vnsp{2} and (b) \vnsp{3} values as a function of charged-particle \pt\
for the centrality intervals 5--10\%, 10--20\%, 30--40\%, and 40--50\%.
The statistical uncertainties are shown as error bars and the systematic uncertainties
are shown as boxes.
}
\label{fig:vn_sys_sp_aux}
\end{figure*}

\begin{figure*}[!ht]
\centering
\subfloat[]{
\centering
\includegraphics[width=0.48\textwidth]{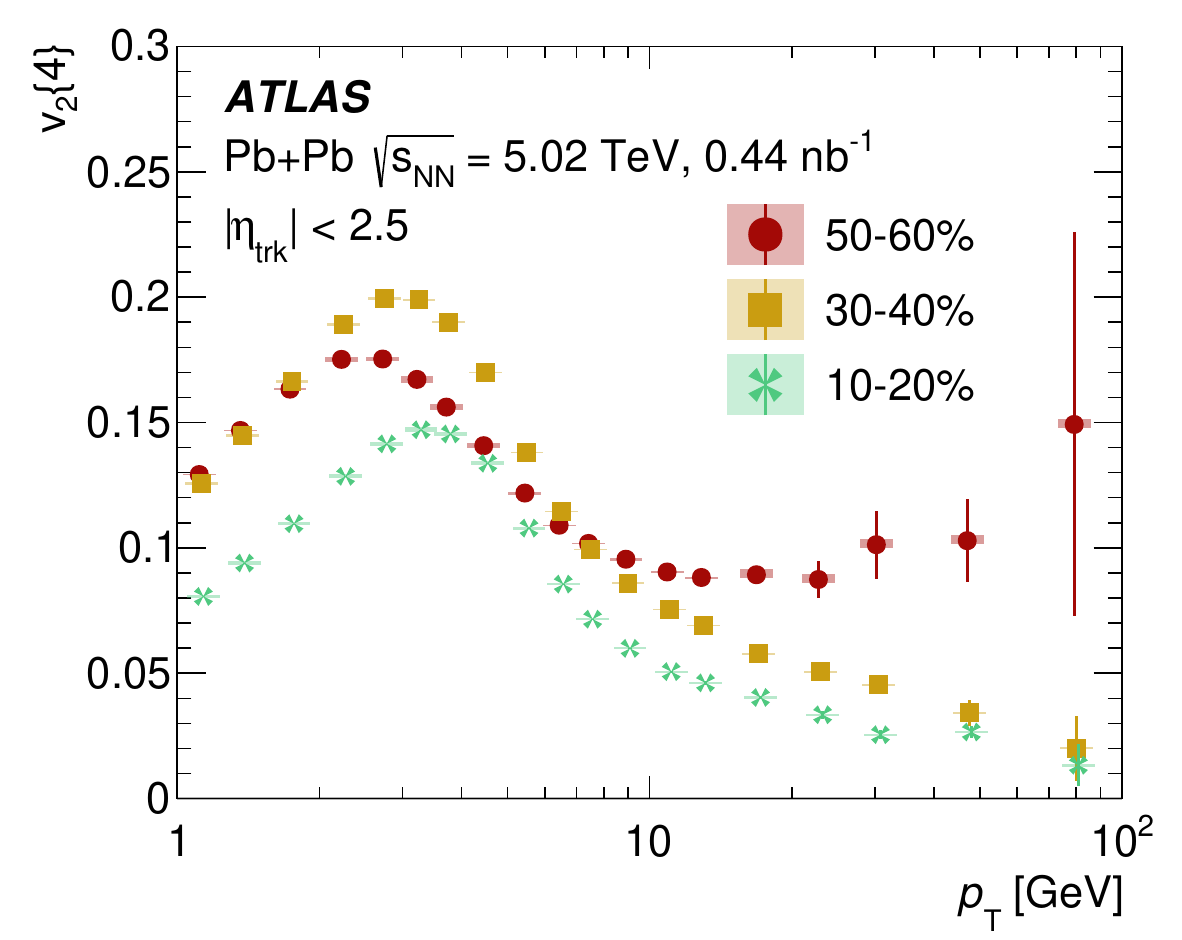}
}
\subfloat[]{
\centering
\includegraphics[width=0.48\textwidth]{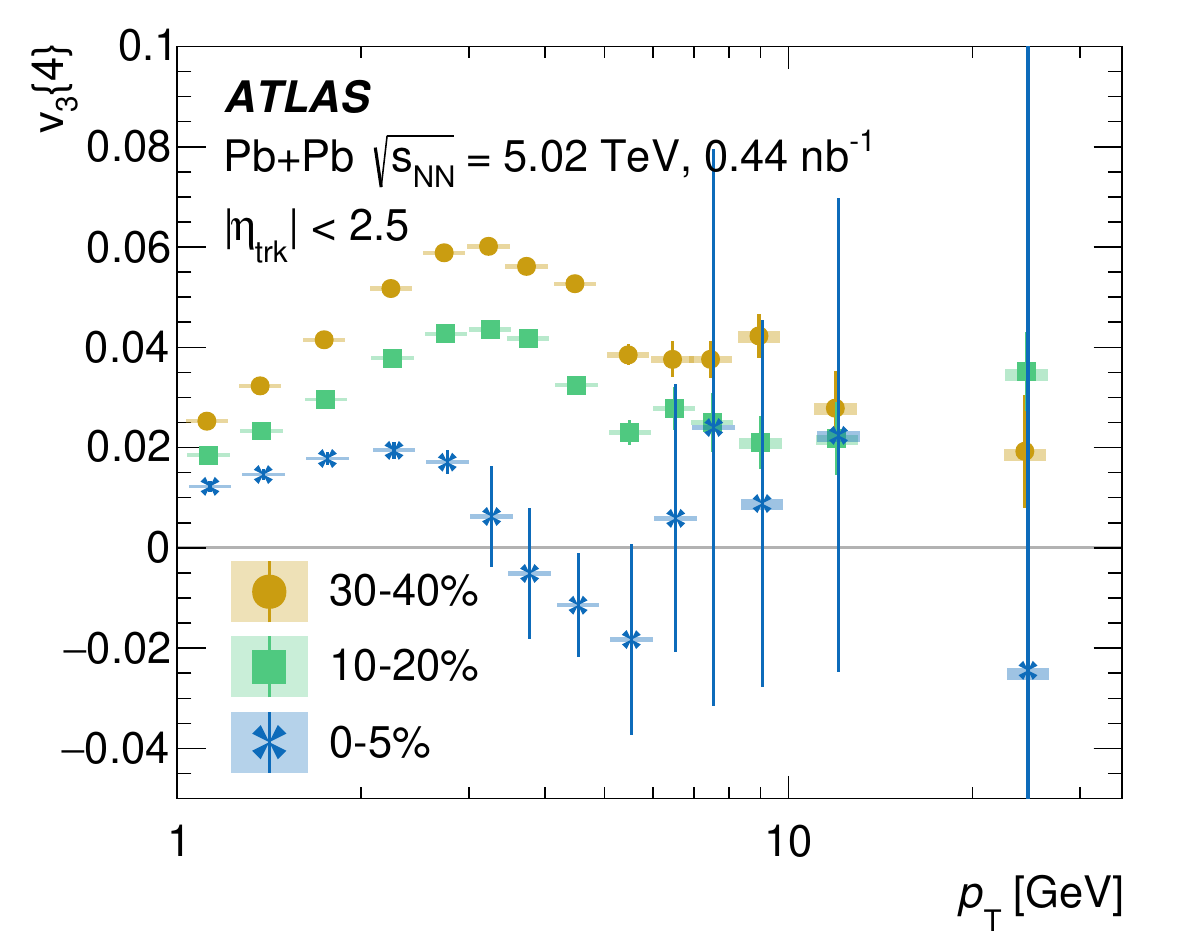}
}
\hfill
\caption{The (a) \vnm{2}{4} and (b) \vnm{3}{4} values as a function of charged-particle \pt\ for different centrality intervals. For \vnm{2}{4}, 10--20\%, 30--40\% and 50--60\% are shown, and for \vnm{3}{4}, 0--5\%, 10--20\%, and 30--40\% are shown.
The statistical uncertainties are shown as error bars and the systematic uncertainties
are shown as boxes.
}
\label{fig:vn_sys_mpc_aux}
\end{figure*}

This appendix includes additional centrality intervals for the measurements presented in Section~\ref{sec:result}.
For comparison purposes, the selected centrality intervals are all of 10\% width, except in Figures~\ref{fig:vn_sys_sp_aux} and~\ref{fig:vn_sys_mpc_aux}, where the 5\% wide centrality intervals are also shown.
These additional centrality intervals supplement the transitioning of trends from the central 10--20\% interval to the peripheral 40--50\% or 50--60\% interval shown in Section~\ref{sec:result}.

\begin{figure*}[!ht]
\centering
\hfill
\subfloat[]{
\centering
\includegraphics[width=0.48\textwidth]{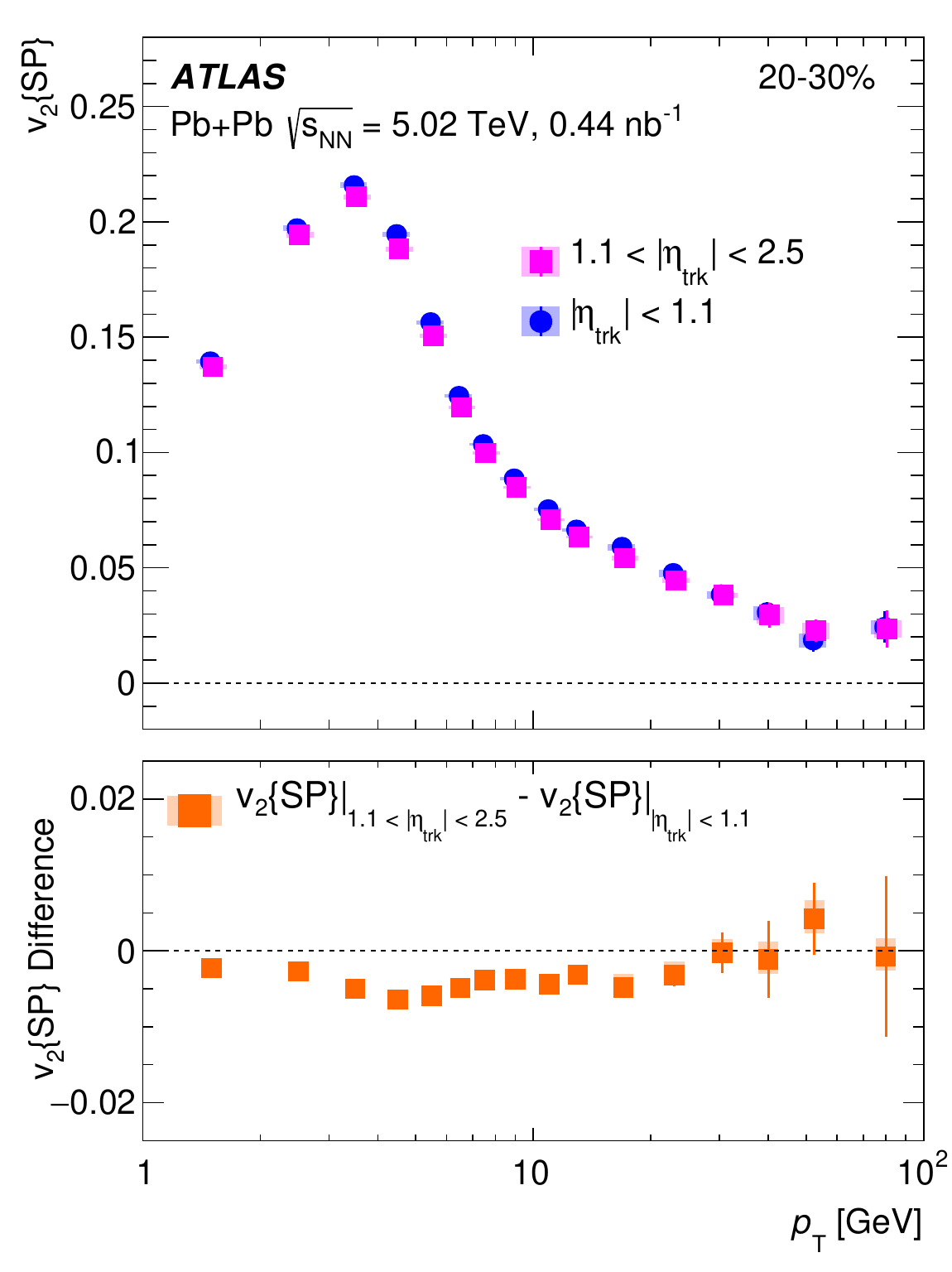}
}
\subfloat[]{
\centering
\includegraphics[width=0.48\textwidth]{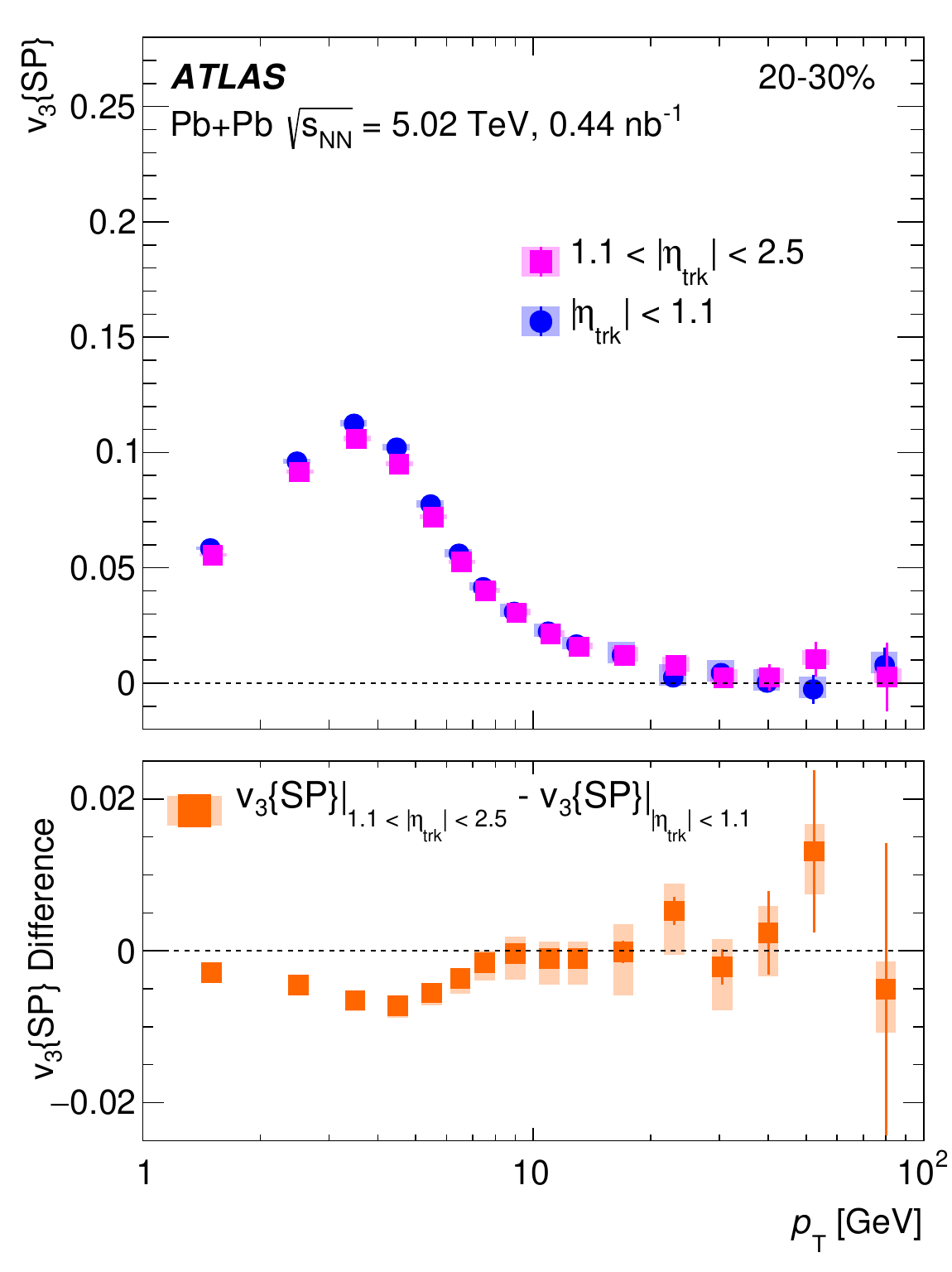}
}
\hfill
\caption{The (a) \vnsp{2} and (b) \vnsp{3} values as a function of charged-particle \pt\
in the centrality interval 20--30\% for different pseudorapidity ranges.
The bottom panel shows the difference in the \vnsp{3} measured between the two ranges.
The statistical uncertainties are shown as error bars and the systematic uncertainties are shown as boxes.
}
\label{fig:eta_dep_auxab}
\end{figure*}

\begin{figure*}
\centering
\hfill
\subfloat[]{
\centering
\includegraphics[width=0.48\textwidth]{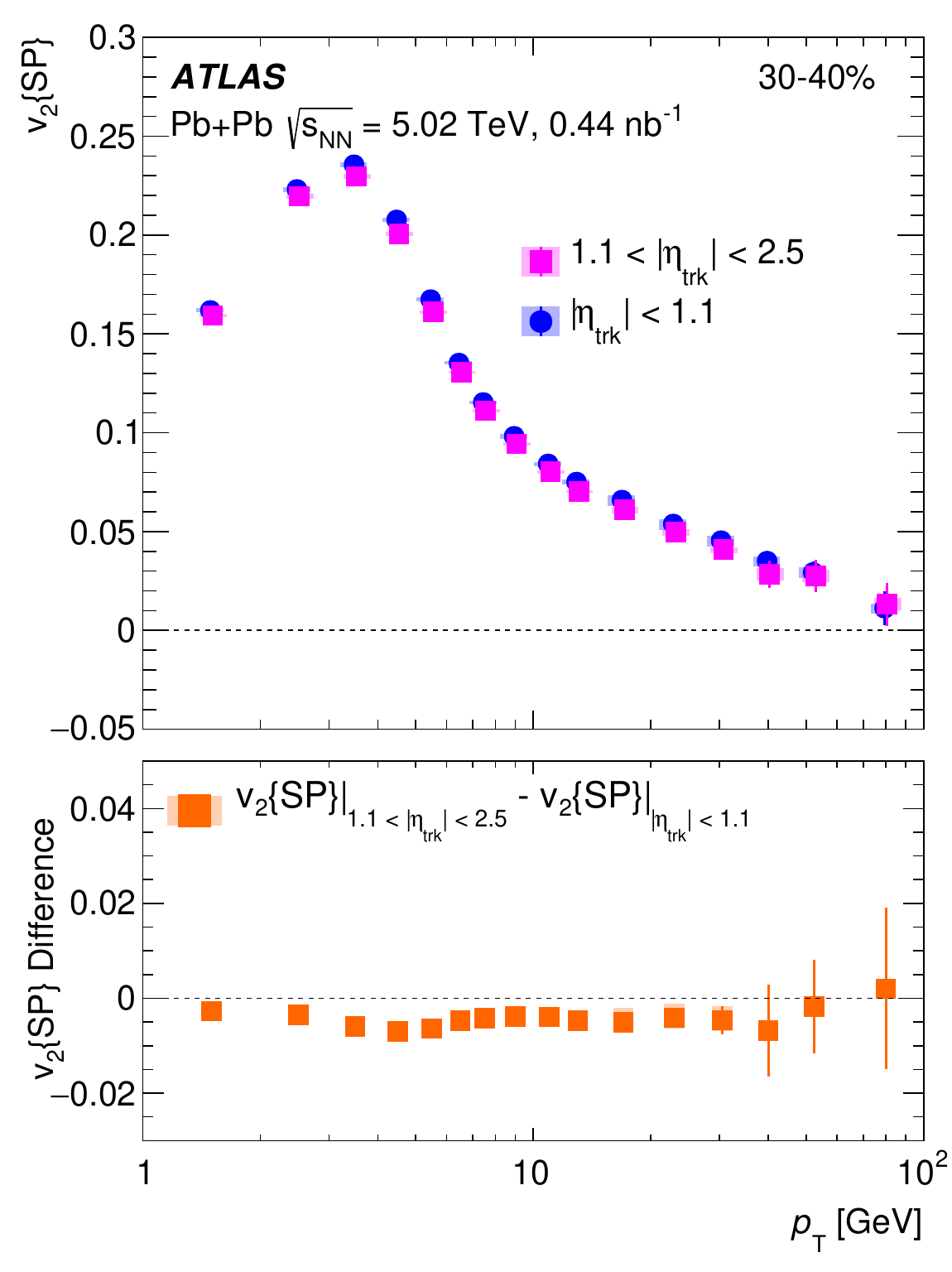}
}
\subfloat[]{
\centering
\includegraphics[width=0.48\textwidth]{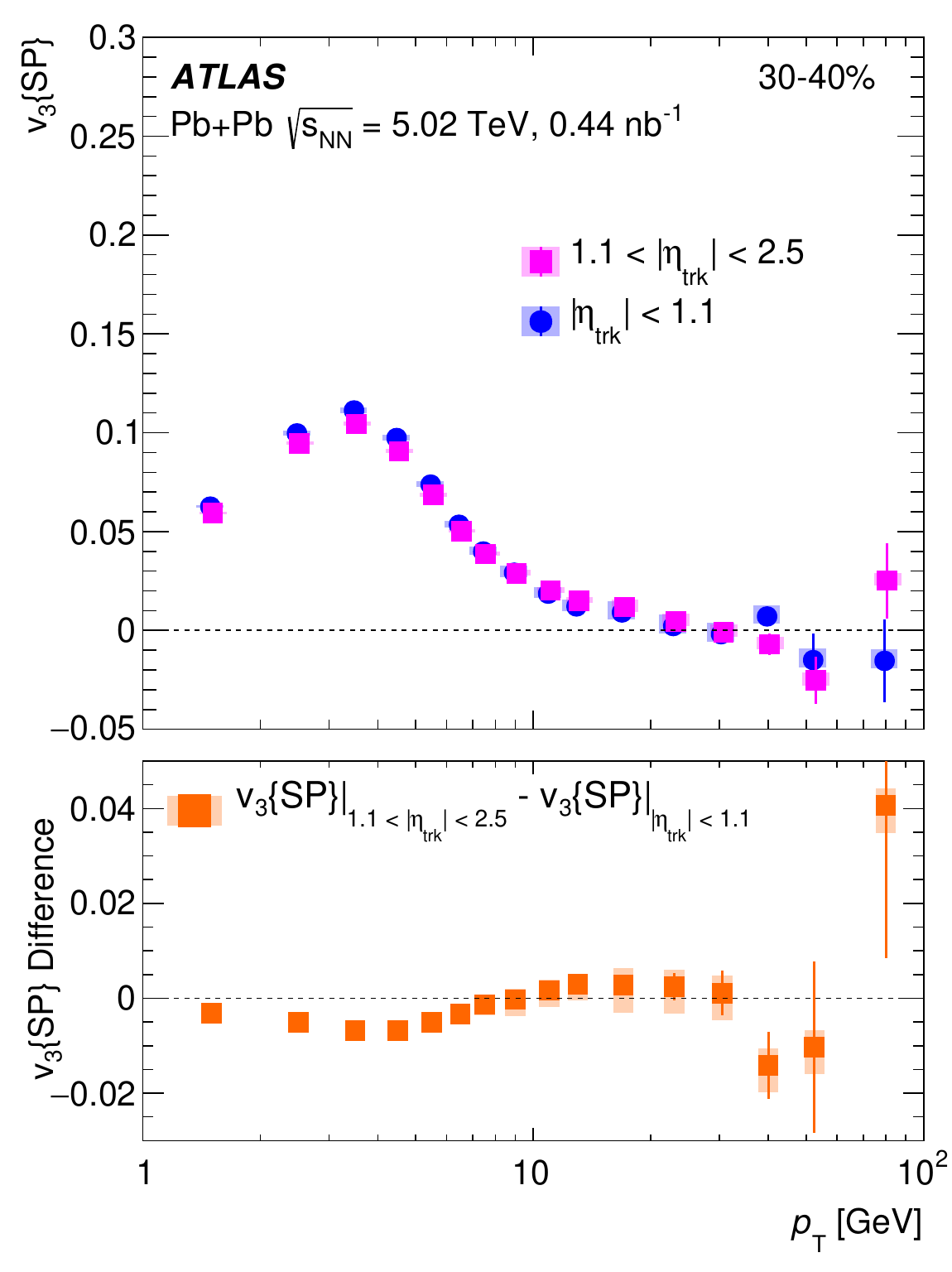}
}
\hfill
\caption{The (a) \vnsp{2} and (b) \vnsp{3} values as a function of charged-particle \pt\
in the centrality interval 30--40\% for different pseudorapidity ranges.
The bottom panel shows the difference in the \vnsp{3} measured between the two ranges.
The statistical uncertainties are shown as error bars and the systematic uncertainties are shown as boxes.
}
\label{fig:eta_dep_auxb}
\end{figure*}

Figure~\ref{fig:vn_sys_sp_aux} shows additional centrality intervals for \vnsp{n} results as a function of \pt, where the same centrality and \pt\ trends shown in Figure~\ref{fig:vn_sys_sp} can be observed.
Figures~\ref{fig:eta_dep_auxab} and~\ref{fig:eta_dep_auxb} show the comparison of different pseudorapidity ranges for \vnsp{2} and \vnsp{3} results in two mid-central intervals.
In these comparisons, the difference between pseudorapidity ranges does not show as strong a dependence on \pt\ for \vnsp{2} as it does for \vnsp{3}.

\begin{figure*}[ht!]
\centering
\subfloat[]{
\centering
\includegraphics[width=0.48\textwidth]{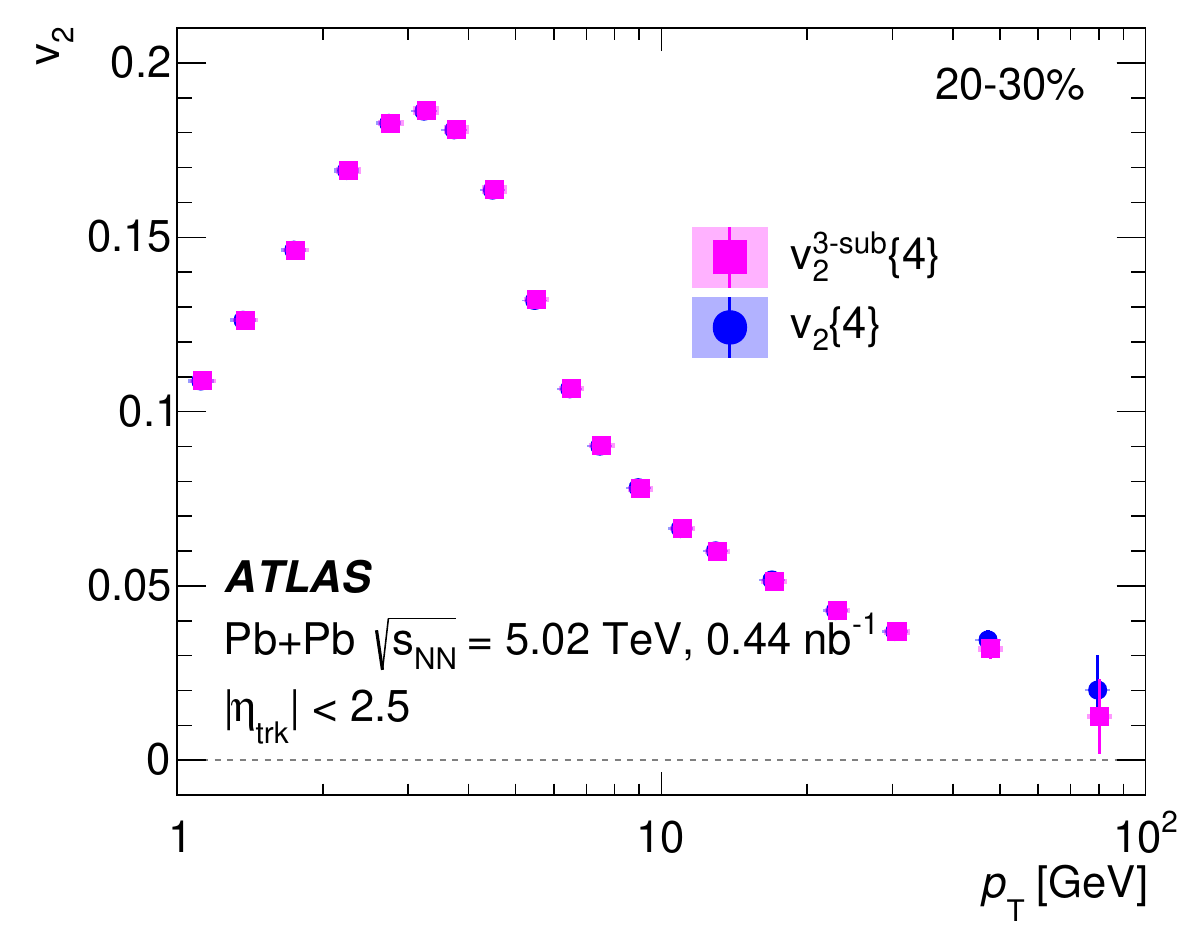}
}
\subfloat[]{
\centering
\includegraphics[width=0.48\textwidth]{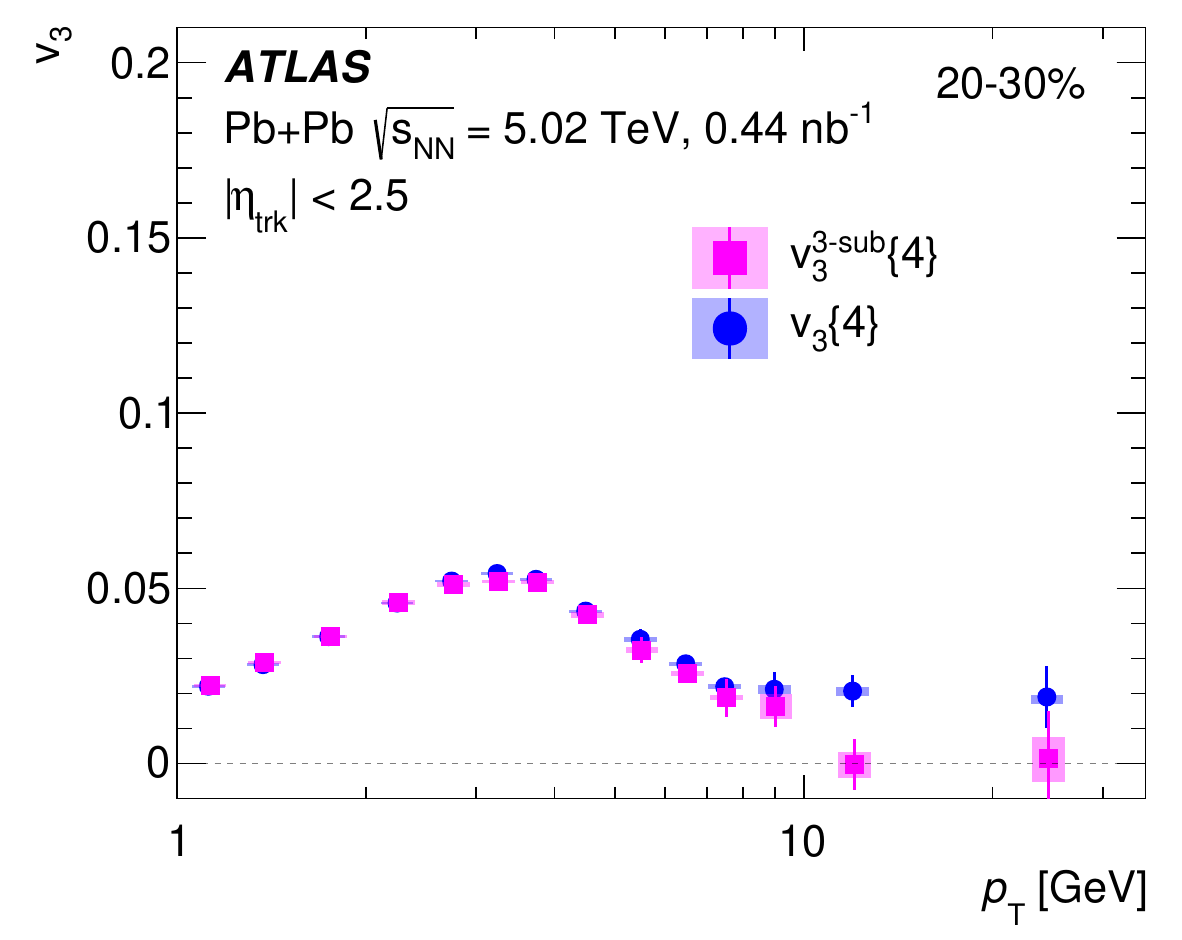}
}
\hfill
\centering
\subfloat[]{
\centering
\includegraphics[width=0.48\textwidth]{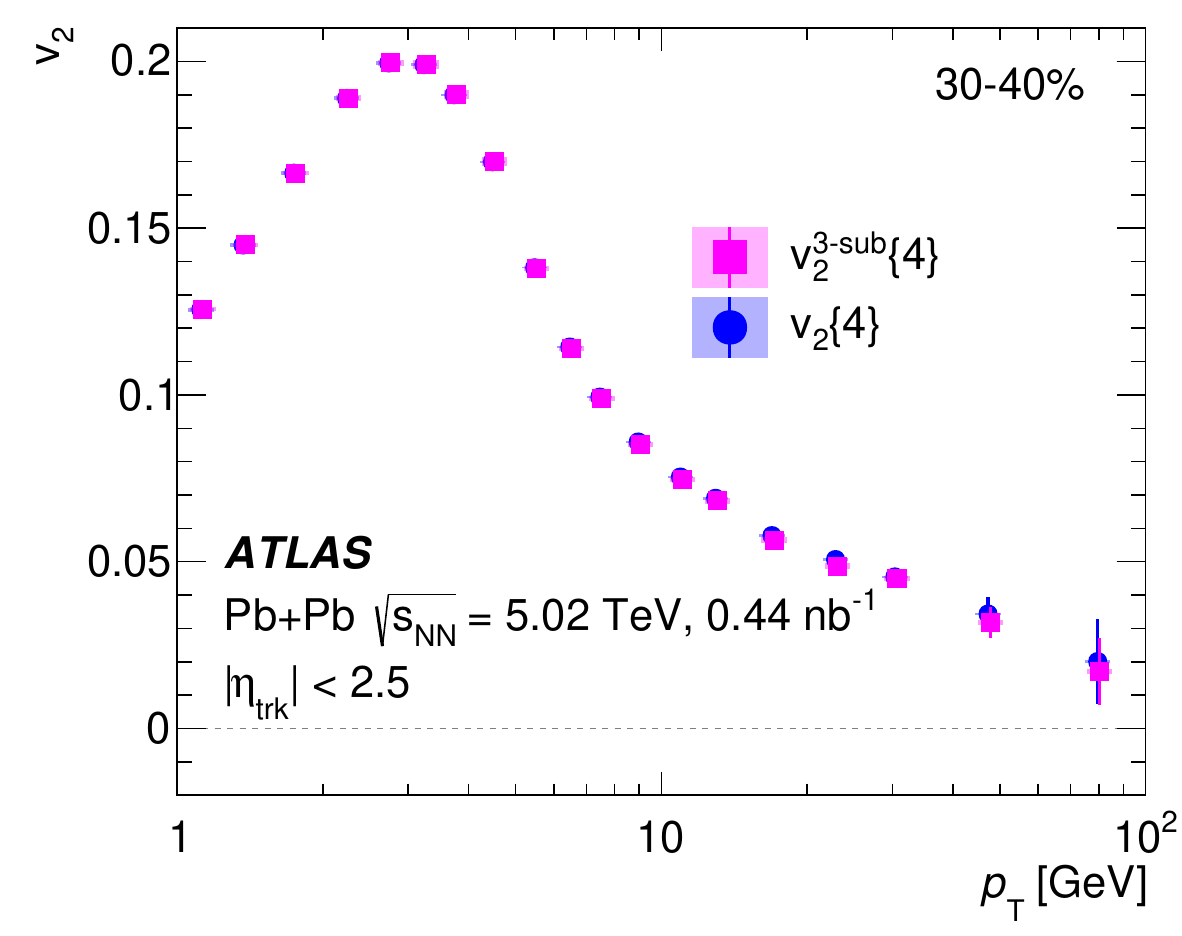}
}
\subfloat[]{
\centering
\includegraphics[width=0.48\textwidth]{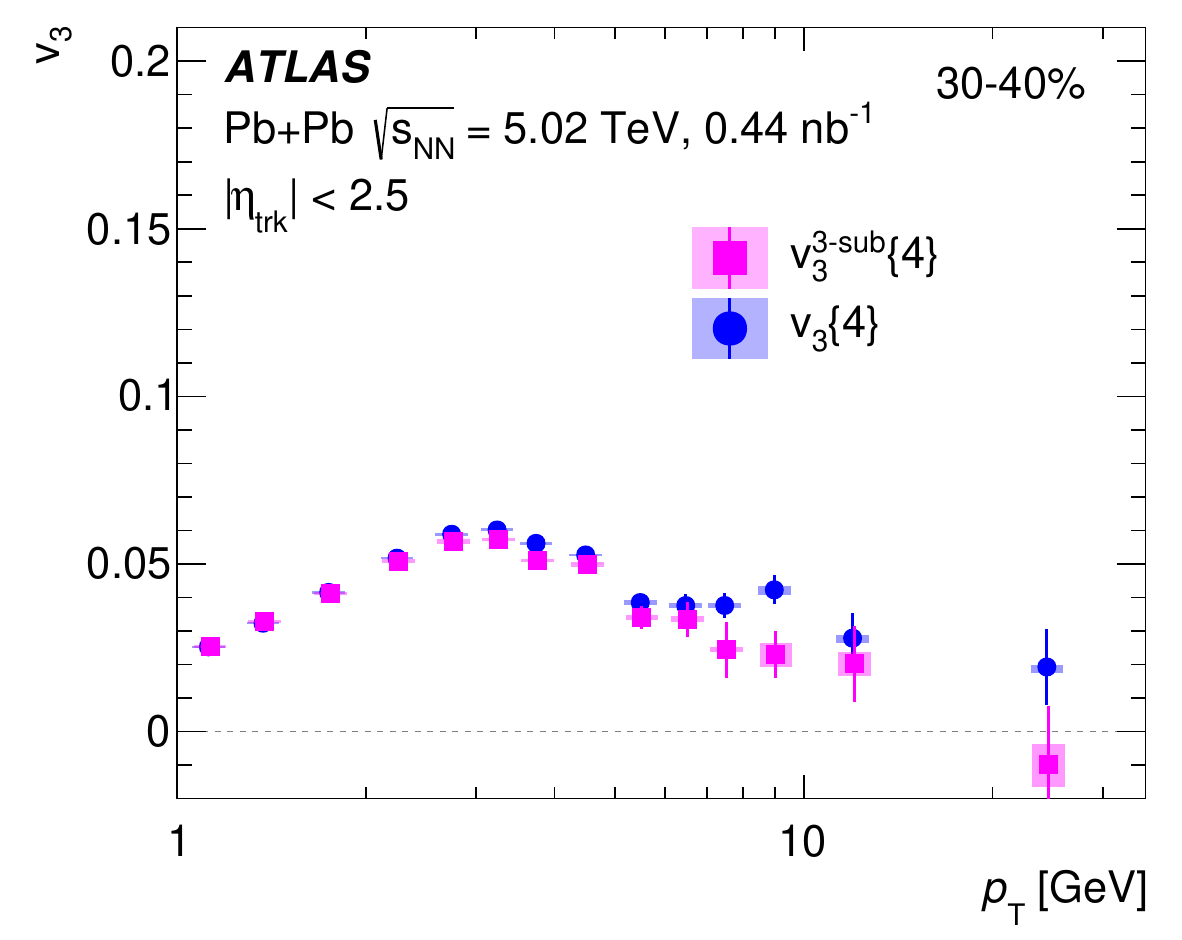}
}
\hfill
\caption{\vnm{n}{4} comparison between standard and three-subevent $Q$ cumulants in the centrality interval 20--30\% for (a) \vnm{2}{4} and (b) \vnm{3}{4}, and in the centrality interval 30--40\% for (c) \vnm{2}{4} and (d) \vnm{3}{4}. The statistical uncertainties are shown as error bars and the systematic uncertainties are shown as boxes.}
\label{fig:3-sub_aux}
\end{figure*}

\begin{figure*}[ht!]
\centering
\subfloat[]{
\centering
\includegraphics[width=0.48\textwidth]{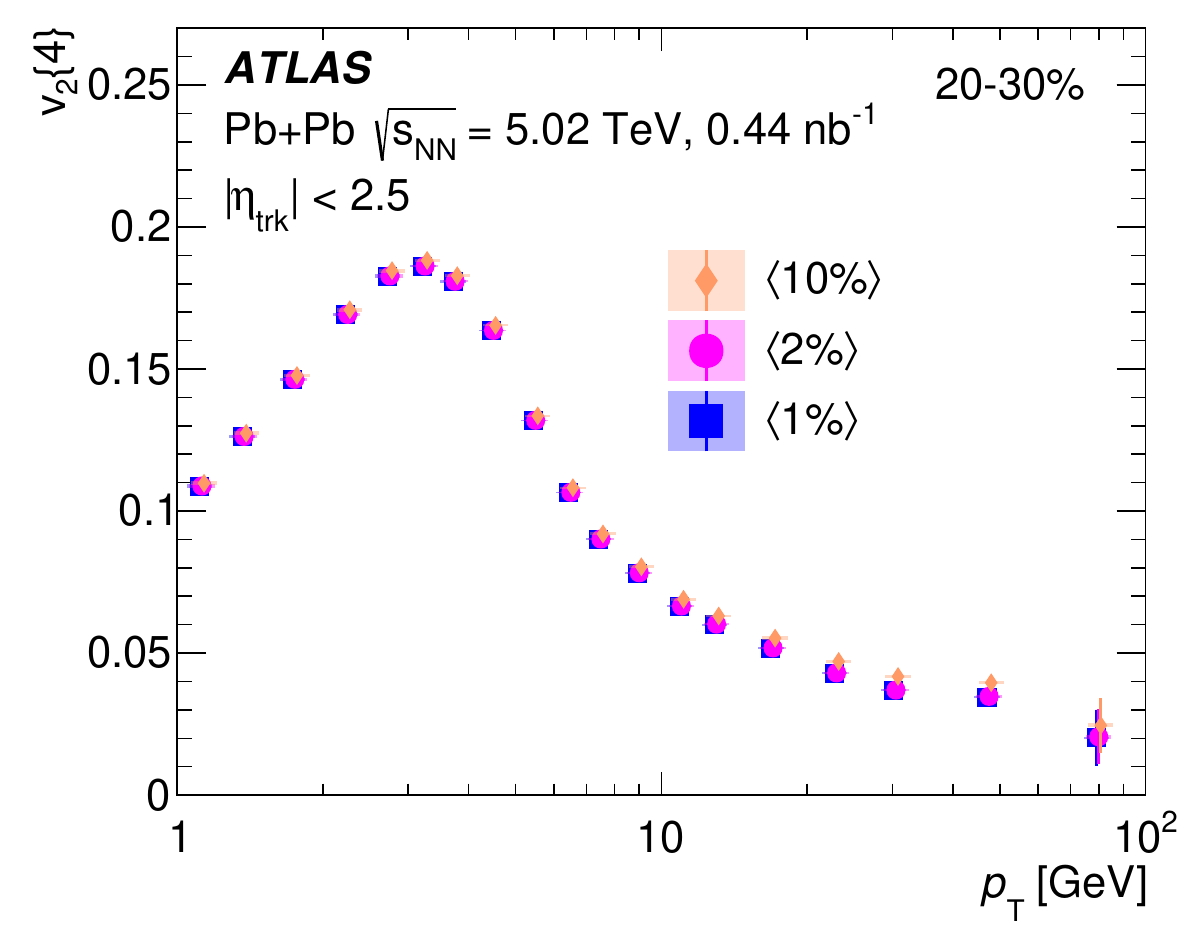}
}
\subfloat[]{
\centering
\includegraphics[width=0.48\textwidth]{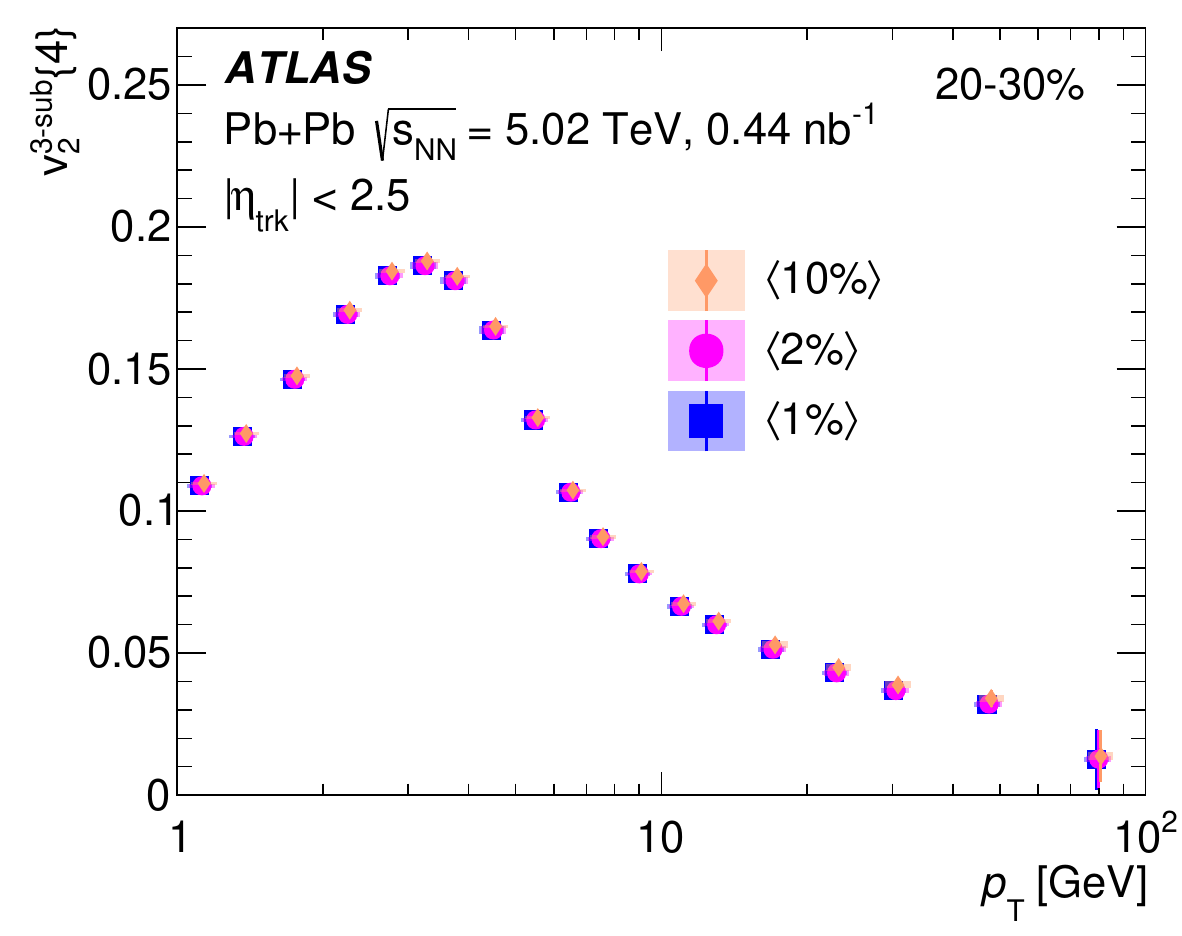}
}
\hfill
\centering
\subfloat[]{
\centering
\includegraphics[width=0.48\textwidth]{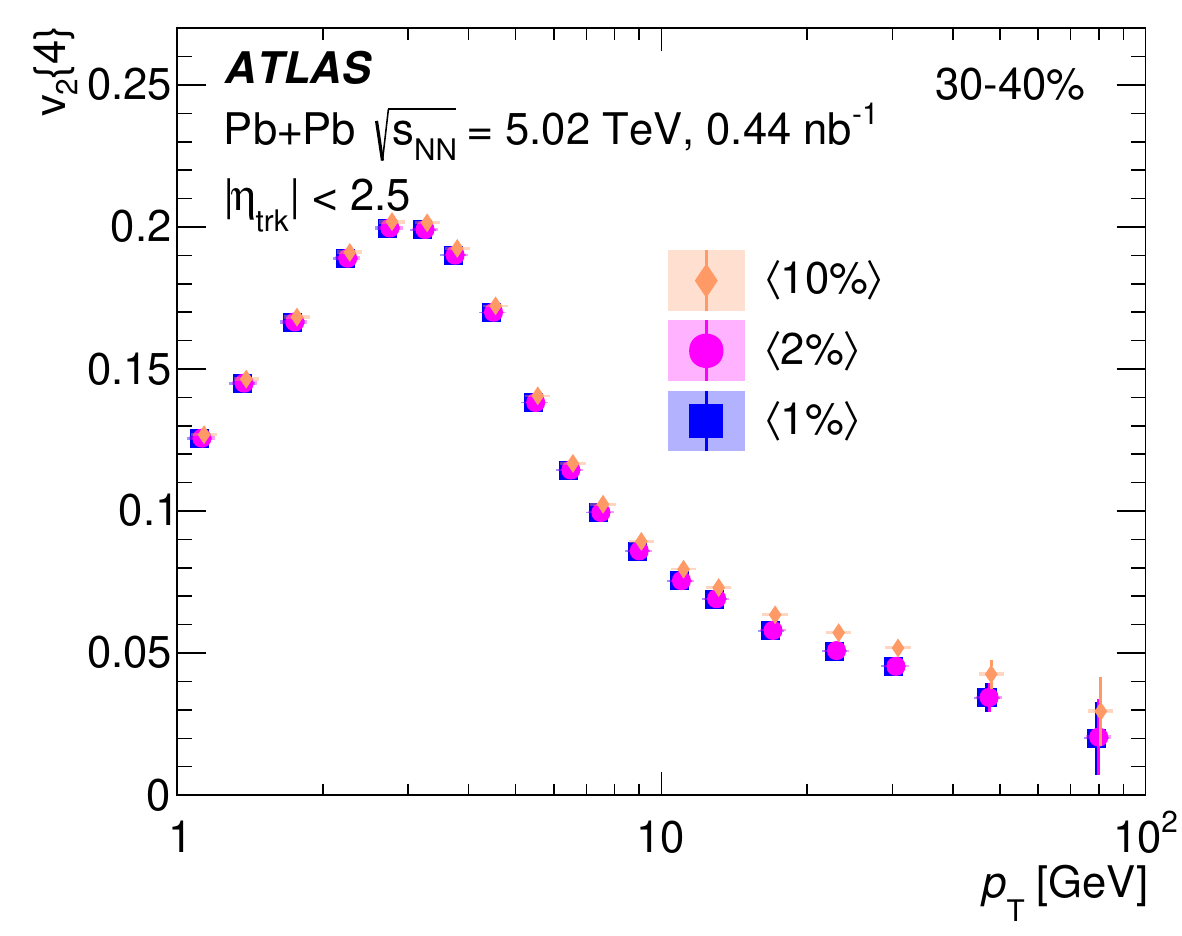}
}
\subfloat[]{
\centering
\includegraphics[width=0.48\textwidth]{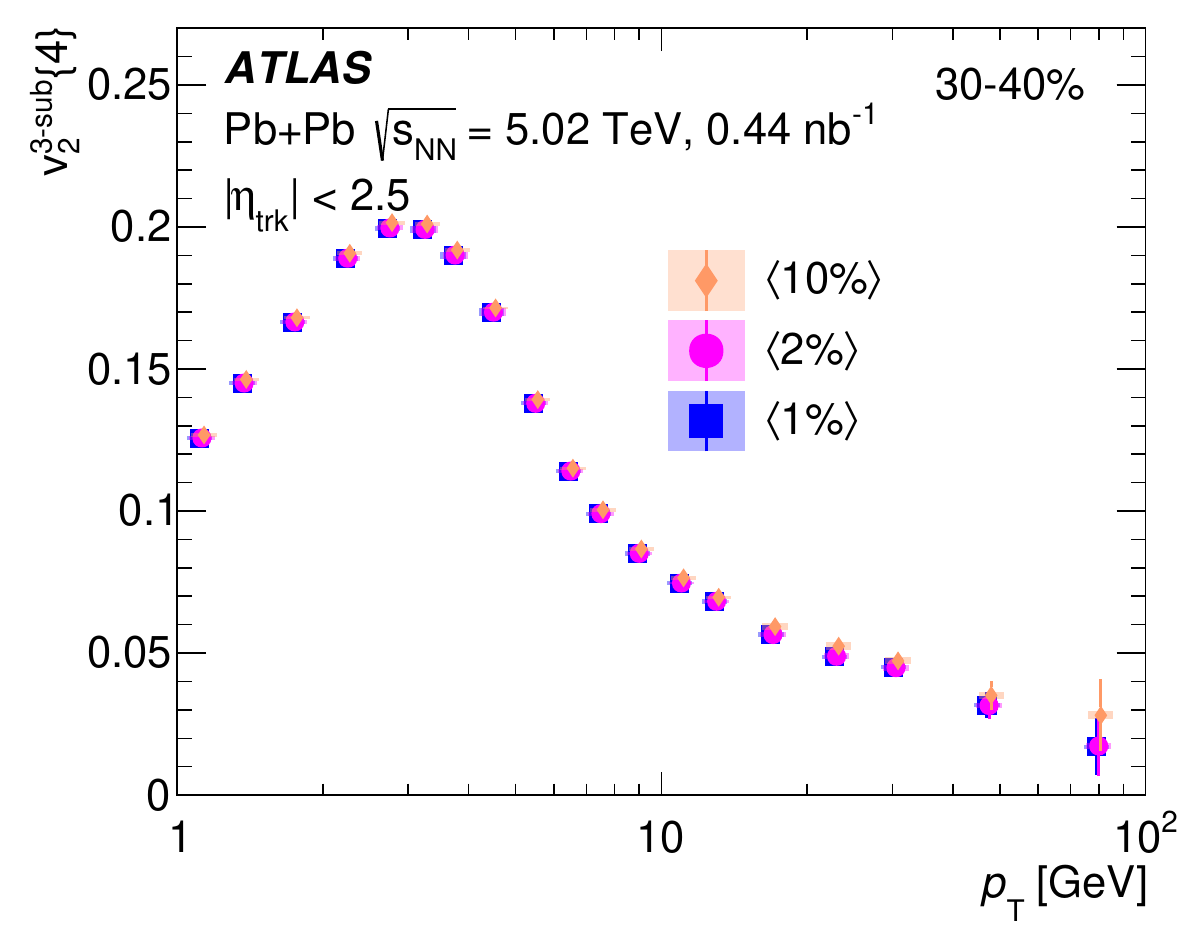}
}
\hfill
\caption{Comparison between \ec{1}, \ec{2} and \ec{10} event combination procedures in the centrality intervals 20--30\% for (a) \vnm{2}{4} and (b) \vnmsub{2}{4}{\text{3-sub}}, and in the centrality interval 30--40\% for (c) \vnm{2}{4} and (d) \vnmsub{2}{4}{\text{3-sub}}. The statistical uncertainties are shown as error bars and the systematic uncertainties are shown as boxes.}
\label{fig:evtclassv2}
\end{figure*}

\begin{figure*}[ht!]
\centering
\subfloat[]{
\centering
\includegraphics[width=0.48\textwidth]{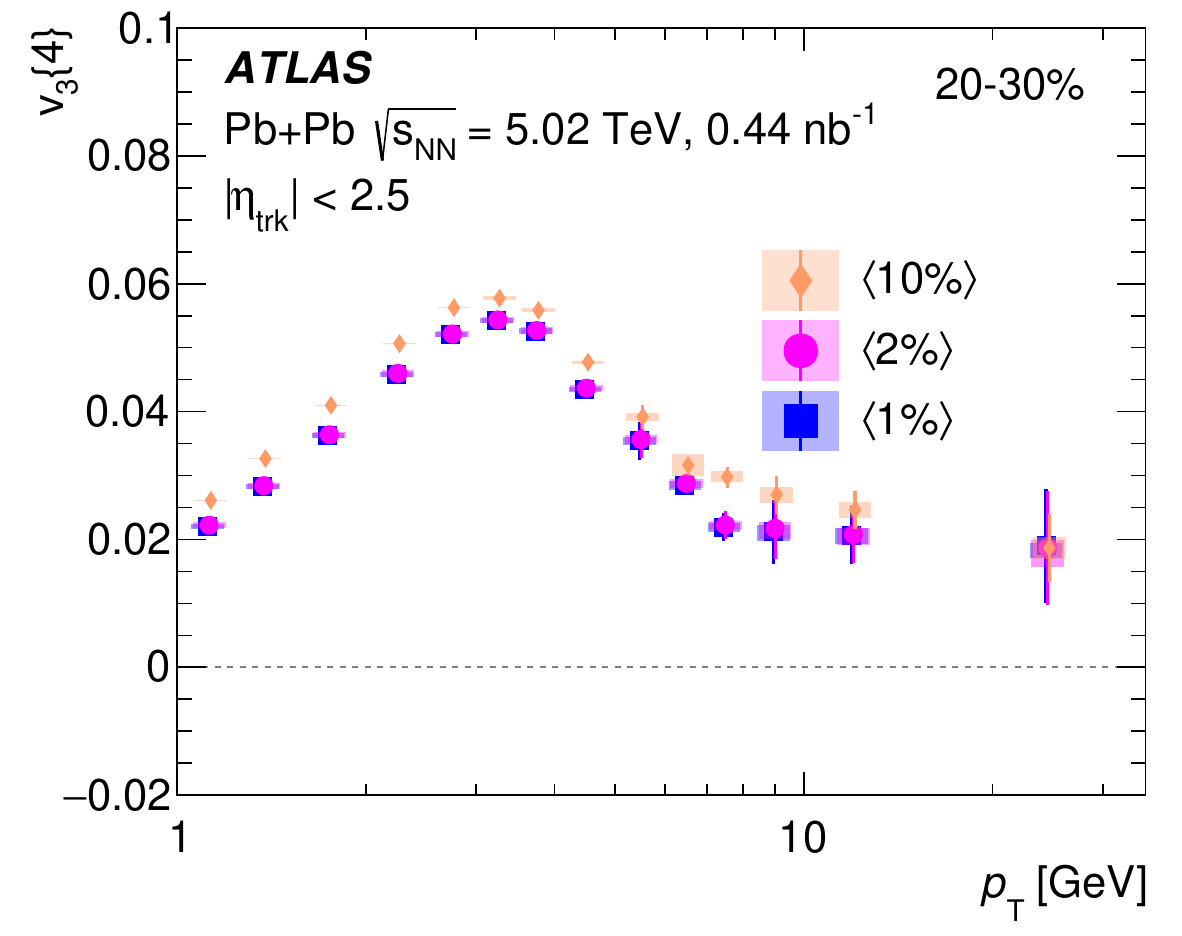}
}
\subfloat[]{
\centering
\includegraphics[width=0.48\textwidth]{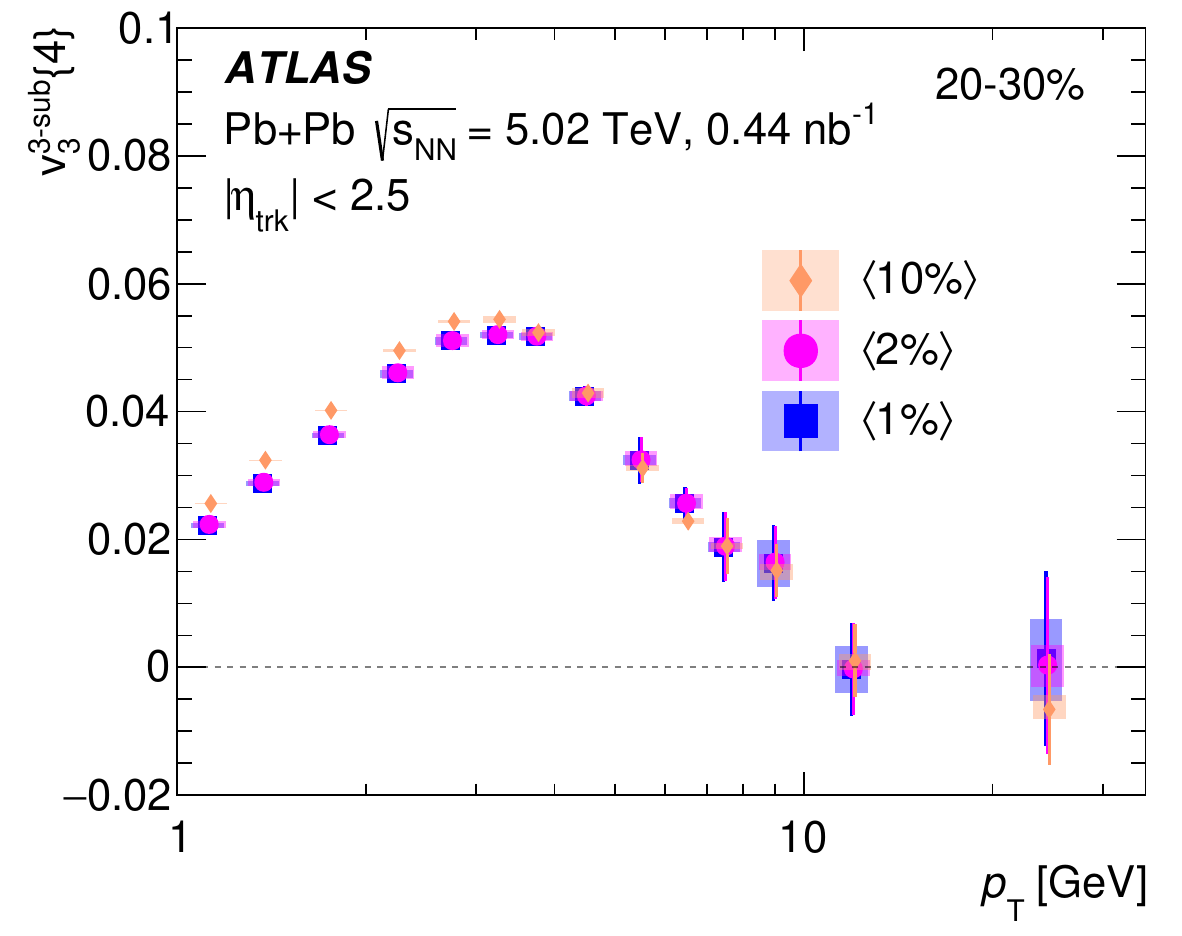}
}
\hfill
\centering
\subfloat[]{
\centering
\includegraphics[width=0.48\textwidth]{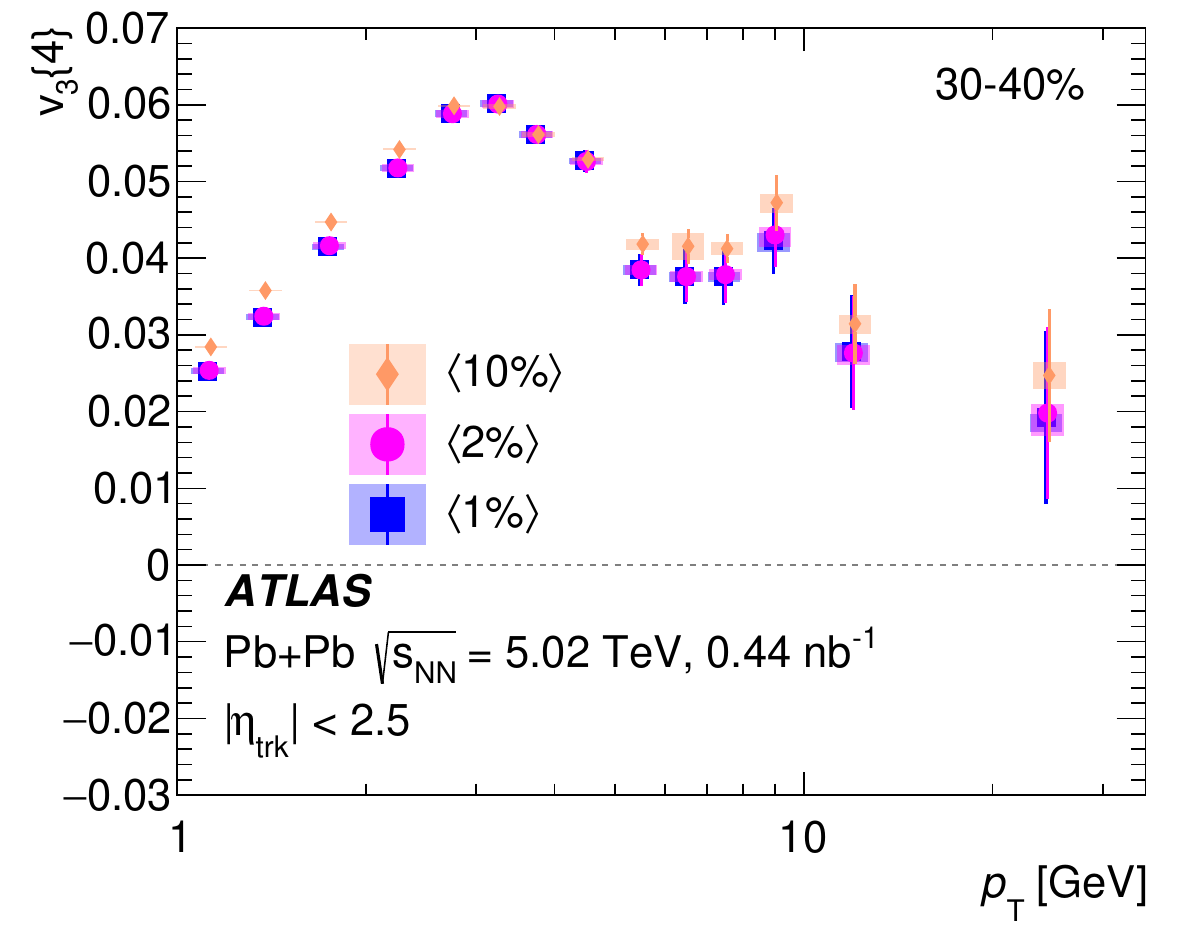}
}
\subfloat[]{
\centering
\includegraphics[width=0.48\textwidth]{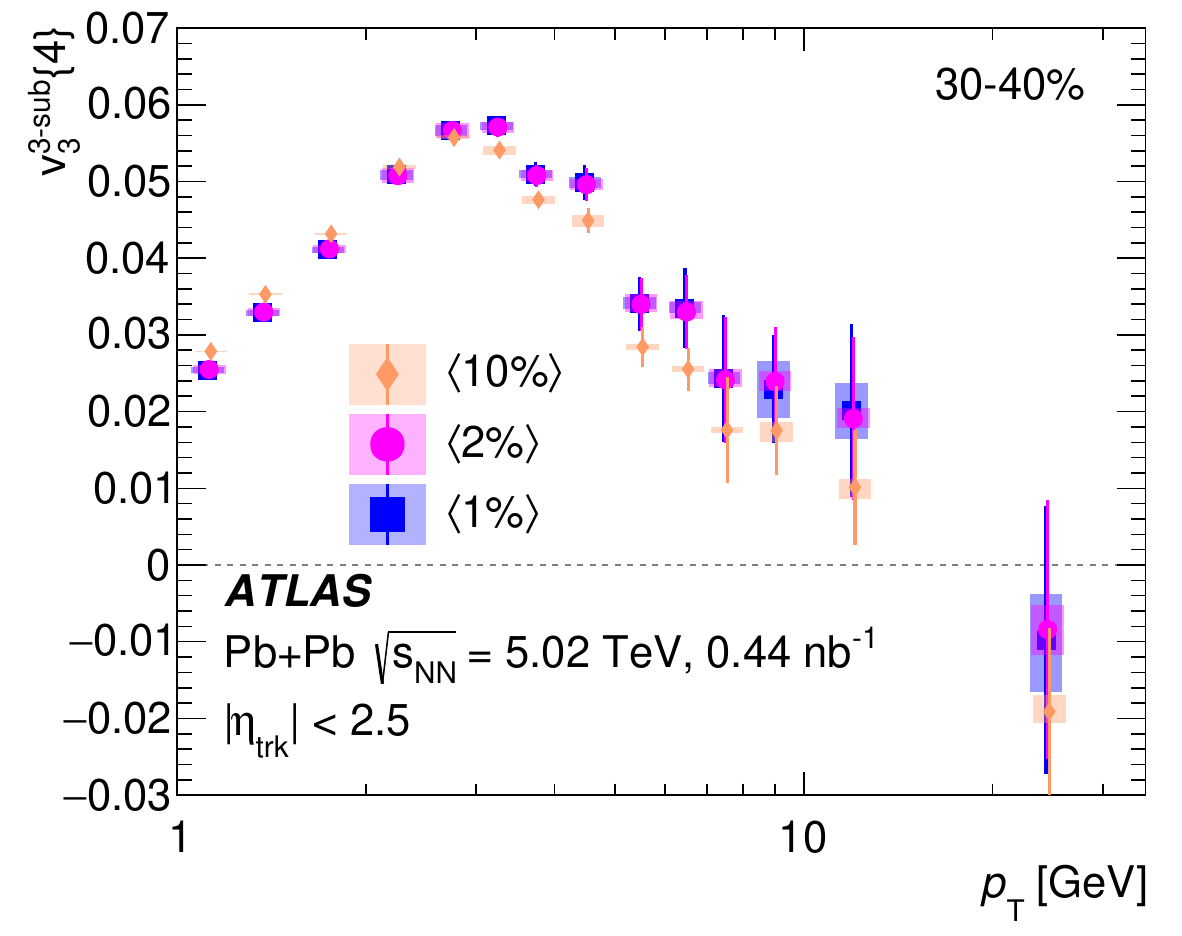}
}
\hfill
\caption{Comparison between \ec{1}, \ec{2} and \ec{10} event combination procedures in the centrality interval 20--30\% for (a) \vnm{3}{4} and (b) \vnmsub{3}{4}{\text{3-sub}}, and in the centrality interval 30--40\% for (c) \vnm{3}{4} and (d) \vnmsub{3}{4}{\text{3-sub}}. The statistical uncertainties are shown as error bars and the systematic uncertainties are shown as boxes.}
\label{fig:evtclassv3}
\end{figure*}

\FloatBarrier
Figure~\ref{fig:vn_sys_mpc_aux} shows additional centrality intervals for \vnm{n}{4} results with standard $Q$ cumulants using the \ec{1} event combination procedure as a function of \pt, where a  slight increase at high \pt\ is observed for \vnm{2}{4} in the 50--60\% centrality interval.
Figure~\ref{fig:3-sub_aux} shows the comparison of standard versus three-subevent cumulants in \vnm{n}{4} for mid-central events, where a stronger suppression is observed in \vn{3} than \vn{2}.
Figures~\ref{fig:evtclassv2} and~\ref{fig:evtclassv3} show the event combination procedure comparisons for \vnm{2}{4} and \vnm{3}{4}, respectively, for both the standard and three-subevent cumulants.
The event combination procedure discrepancy is much smaller in \vn{2} than in \vn{3}, and is well suppressed by the three-subevent cumulants. In contrast, \vnmsub{3}{4}{\text{3-sub}} still shows a significant discrepancy between different event combination procedures.

% End of text imported from the .//ANA-HION-2023-04-PAPER-appendix.tex input file
\clearpage

\printbibliography
\clearpage
% ATLAS Collaboration author list
% Reference date of HION-2023-04 is 2024-10-29
% Author list last updated on date 06-JUL-25
% Data extracted on 06-Jul-2025 for paper reference HION-2023-04
% at 11:06am
 
\begin{flushleft}
\hypersetup{urlcolor=black}
{\Large The ATLAS Collaboration}

\bigskip

\AtlasOrcid[0000-0002-6665-4934]{G.~Aad}$^\textrm{\scriptsize 105}$,
\AtlasOrcid[0000-0001-7616-1554]{E.~Aakvaag}$^\textrm{\scriptsize 17}$,
\AtlasOrcid[0000-0002-5888-2734]{B.~Abbott}$^\textrm{\scriptsize 124}$,
\AtlasOrcid[0000-0002-0287-5869]{S.~Abdelhameed}$^\textrm{\scriptsize 120a}$,
\AtlasOrcid[0000-0002-1002-1652]{K.~Abeling}$^\textrm{\scriptsize 56}$,
\AtlasOrcid[0000-0001-5763-2760]{N.J.~Abicht}$^\textrm{\scriptsize 50}$,
\AtlasOrcid[0000-0002-8496-9294]{S.H.~Abidi}$^\textrm{\scriptsize 30}$,
\AtlasOrcid[0009-0003-6578-220X]{M.~Aboelela}$^\textrm{\scriptsize 46}$,
\AtlasOrcid[0000-0002-9987-2292]{A.~Aboulhorma}$^\textrm{\scriptsize 36e}$,
\AtlasOrcid[0000-0001-5329-6640]{H.~Abramowicz}$^\textrm{\scriptsize 158}$
% ...
\bigskip
\\

% ...
$^{17}$Department for Physics and Technology, University of Bergen, Bergen; Norway.\\
% ...
$^{30}$Physics Department, Brookhaven National Laboratory, Upton NY; United States of America.\\
% ...
$^{36}$$^{(a)}$Facult\'e des Sciences Ain Chock, Universit\'e Hassan II de Casablanca;$^{(b)}$Facult\'{e} des Sciences, Universit\'{e} Ibn-Tofail, K\'{e}nitra;$^{(c)}$Facult\'e des Sciences Semlalia, Universit\'e Cadi Ayyad, LPHEA-Marrakech;$^{(d)}$LPMR, Facult\'e des Sciences, Universit\'e Mohamed Premier, Oujda;$^{(e)}$Facult\'e des sciences, Universit\'e Mohammed V, Rabat;$^{(f)}$Institute of Applied Physics, Mohammed VI Polytechnic University, Ben Guerir; Morocco.\\
% ...
$^{46}$Physics Department, Southern Methodist University, Dallas TX; United States of America.\\
% ...
$^{50}$Fakult\"{a}t Physik , Technische Universit{\"a}t Dortmund, Dortmund; Germany.\\
% ...
$^{56}$II. Physikalisches Institut, Georg-August-Universit\"{a}t G\"ottingen, G\"ottingen; Germany.\\
% ...
$^{105}$CPPM, Aix-Marseille Universit\'e, CNRS/IN2P3, Marseille; France.\\
% ...
$^{120}$$^{(a)}$New York University Abu Dhabi, Abu Dhabi;$^{(b)}$United Arab Emirates University, Al Ain; United Arab Emirates.\\
% ...
$^{124}$Homer L. Dodge Department of Physics and Astronomy, University of Oklahoma, Norman OK; United States of America.\\
% ...
$^{158}$Raymond and Beverly Sackler School of Physics and Astronomy, Tel Aviv University, Tel Aviv; Israel.\\
% ...
$^{a}$ Also Affiliated with an institute covered by a cooperation agreement with CERN.\\
$^{b}$ Also at An-Najah National University, Nablus; Palestine.\\
$^{c}$ Also at Borough of Manhattan Community College, City University of New York, New York NY; United States of America.\\
$^{d}$ Also at Center for High Energy Physics, Peking University; China.\\
$^{e}$ Also at Center for Interdisciplinary Research and Innovation (CIRI-AUTH), Thessaloniki; Greece.\\
$^{f}$ Also at CERN, Geneva; Switzerland.\\
% ...
$^{*}$ Deceased

\end{flushleft}

% Created with Glance <Atlas.Glance@cern.ch>

%

%
%
%

\end{document}